\documentclass[fleqn,usenatbib]{mnras} 

\usepackage{newtxtext,newtxmath}         

\usepackage[T1]{fontenc}
\usepackage{ae,aecompl}

\usepackage{array}
\usepackage{mathtools}

\newcolumntype{?}{!{\vrule width 1.5pt}}
\usepackage{afterpage}

\usepackage{multirow}
\usepackage{graphicx}	
\usepackage{amsmath}	
\usepackage{amssymb}	
\usepackage[usenames]{color}
\usepackage{bm}		
\usepackage{booktabs,makecell}
\usepackage{adjustbox}
\usepackage{natbib}
\usepackage[dvipsnames]{xcolor}

\usepackage[switch,pagewise]{lineno}

\newcommand{\redmagic}{\textit{redMaGiC}}

\newcommand{\bk}{\textbf{k}}

\newcommand{\tj}[6]{ \begin{pmatrix}
       #1 & #2 & #3 \\
       #4 & #5 & #6 
    \end{pmatrix}}

\newcommand{\avg}[1]{\langle #1 \rangle}

\definecolor{purple}{RGB}{150,0,200}

\usepackage{lineno}

\title[Cosmology with Mass Maps moments]{Dark Energy Survey Year 3 results:  cosmology with moments of weak lensing mass maps - validation on simulations}

\makeatletter
\def \blfootnote{\xdef\@thefnmark{}\@footnotetext}
\makeatother

\author[M. Gatti, et al.]{
\parbox{\textwidth}{
\Large 
M.~Gatti$^{1\star}$,  
C.~Chang$^{2,3}$, 
O.~Friedrich$^{4}$,  
B.~Jain$^{5}$, 
D.~Bacon$^{6}$, 
M.~Crocce$^{7,8}$, 
J.~DeRose$^{9,10}$,
I.~Ferrero$^{11}$, 
P.~Fosalba$^{7,8}$, 
E.~Gaztanaga$^{7,8}$, 
D.~Gruen$^{9,10,12}$, 
I.~Harrison$^{13}$, 
N.~Jeffrey$^{14}$, 
N.~MacCrann$^{15,16}$, 
T.~McClintock$^{17}$, 
L.~Secco$^{5}$,
L.~Whiteway$^{14}$, 
T.~M.~C.~Abbott$^{18}$, 
S.~Allam$^{19}$, 
J.~Annis$^{19}$, 
S.~Avila$^{20}$, 
D.~Brooks$^{14}$, 
E.~Buckley-Geer$^{19}$, 
D.~L.~Burke$^{10,12}$, 
A.~Carnero~Rosell$^{21,22}$, 
M.~Carrasco~Kind$^{23,24}$, 
J.~Carretero$^{1}$, 
R.~Cawthon$^{25}$, 
L.~N.~da Costa$^{22,26}$, 
J.~De~Vicente$^{21}$, 
S.~Desai$^{27}$, 
H.~T.~Diehl$^{19}$, 
P.~Doel$^{14}$, 
T.~F.~Eifler$^{28,29}$, 
J.~Estrada$^{19}$, 
S.~Everett$^{30}$, 
A.~E.~Evrard$^{31,32}$, 
J.~Frieman$^{19,3}$, 
J.~Garc\'ia-Bellido$^{20}$, 
D.~W.~Gerdes$^{31,32}$, 
R.~A.~Gruendl$^{23,24}$, 
J.~Gschwend$^{22,26}$, 
G.~Gutierrez$^{19}$, 	
D.~J.~James$^{30}$, 
M.~D.~Johnson$^{24}$, 
E.~Krause$^{28}$, 
K.~Kuehn$^{33,34}$, 
M.~Lima$^{35,22}$, 
M.~A.~G.~Maia$^{22,26}$, 
M.~March$^{5}$, 
J.~L.~Marshall$^{36}$, 
P.~Melchior$^{37}$, 
F.~Menanteau$^{23,24}$, 
R.~Miquel$^{38, 1}$, 
A.~Palmese$^{19,3}$, 
F.~Paz-Chinch\'{o}n$^{23,24}$, 
A.~A.~Plazas$^{37}$, 
C.~S{\'a}nchez$^{5}$, 
E.~Sanchez$^{21}$, 
V.~Scarpine$^{19}$, 	
M.~Schubnell$^{32}$, 
S.~Santiago$^{7,8}$, 
I.~Sevilla-Noarbe$^{21}$, 
M.~Smith$^{39}$, 
M.~Soares-Santos$^{40}$, 
E.~Suchyta$^{41}$, 
M.~E.~C.~Swanson$^{24}$, 
G.~Tarle$^{32}$, 
D.~Thomas$^{6}$, 
M.~A.~Troxel$^{42}$, 
J.~Zuntz$^{43}$, 
\begin{center} (DES Collaboration) \end{center}
}
}

\vspace{0.2cm}

\date{\today}

\begin{document}     
\label{firstpage} 
\pagerange{\pageref{firstpage}--\pageref{lastpage}}%
\maketitle        	 

\begin{abstract}
We present a simulated cosmology analysis using the second and third moments of the weak lensing mass (convergence) maps. The analysis is geared towards the third year (Y3) data from the Dark Energy Survey (DES), but the methodology can be applied to other weak lensing data sets. The second moment, or variances, of the convergence as a function of smoothing scale contains information similar to standard shear 2-point statistics. The third moment, or the skewness, contains additional non-Gaussian information. We present the formalism for obtaining the convergence maps from the measured shear and for obtaining the second and third moments of these maps given partial sky coverage. We estimate the covariance matrix from a large suite of numerical simulations. We test our pipeline through a simulated likelihood analyses varying 5 cosmological parameters and 10 nuisance parameters. Our forecast shows that the combination of second and third moments provides a 1.5 percent constraint on $S_8 \equiv \sigma_8 (\Omega_{\rm m}/0.3)^{0.5}$ for DES Y3 data. This is 20 percent better than an analysis using a simulated DES Y3 shear 2-point statistics, owing to the non-Gaussian information captured by the inclusion of higher-order statistics. The methodology developed here can be applied to current and future weak lensing datasets to make use of information from non-Gaussianity in the cosmic density field and to improve constraints on cosmological parameters.



\end{abstract}


\blfootnote{$^{\star}$ E-mail: mgatti@ifae.es}
\setcounter{footnote}{1}




\section{Introduction}


A map of the mass distribution of the Universe, or the large-scale structure (LSS), contains a vast amount of cosmological information. A given cosmological model predicts the spatial statistics of the mass distribution as well as its evolution over time. One of the cleanest ways to probe the mass distribution in the Universe is through weak (gravitational) lensing. Gravitational lensing refers to the phenomenon that light rays from distant galaxies bend as they travel through space-time, causing distortion of the observed galaxy images. This is because the space-time is perturbed by mass distribution between the galaxy and the observer according to General Relativity \citep{Einstein1936}. Weak lensing is the regime where this perturbation is small; its effect is usually much smaller than the noise on a single galaxy basis, and the signal is extracted statistically using a very large ensembles of galaxies. As lensing is a purely gravitational effect it requires fewer assumptions about galaxy formation physics compared to cosmological probes that use galaxies directly, such as galaxy clustering \citep[for a review of weak gravitational lensing see e.g.,][]{Bartelmann2001}.

A key element of a weak lensing analysis is to have a large number of galaxies with well-measured shapes. This means that we need 1) cosmological surveys that collect photons from as many galaxies as possible, and 2) well-controlled systematic errors in the shape measurement of these galaxies. Motivated by the potential cosmological power of weak lensing, photometric galaxy surveys targeted at weak lensing science have been operating over the past two decades. Today, unprecedented large galaxy surveys such as the Dark Energy Survey \citep[DES,][]{Flaugher2005}, the Hyper Suprime-Cam (HSC) Subaru Strategic Program \citep{Aihara2018}, the Kilo-Degree Survey \citep[KiDS,][]{deJong2013} are all pushing the limits of weak lensing measurements. 

Most of the current weak lensing analyses have focussed on tomographic 2-point correlation measurements  \citep[e.g.][]{Troxel2017,Hildebrandt2017,Hikage2019}. With the past two decades of work, the theoretical modelling of the shear 2-point correlation function has matured significantly. Although there is still active research on, for example, the modelling of the small scales and of non-linear lensing corrections, the baseline theory of shear 2-point correlation function is reasonably robust. State-of-the-art datasets from the first year (Y1) of the DES currently give the tightest constraints from galaxy surveys on the Universe's clustering amplitude under a $\Lambda$CDM cosmology, $S_8 \equiv \sigma_8 \sqrt{\Omega_{\rm m}/0.3} = 0.782^{+0.027}_{-0.027}$ \citep{Troxel2017}. The parameter $S_{8}$, which is a combination of $\sigma_{8}$ (the Universe's density variance on a scale of 8 Mpc) and $\Omega_{\rm m}$ (the density of the total matter today) is designed to be approximately the parameter most constrained by weak lensing observations. We note that these constraints are at a level similar to those provided by the cosmic microwave background (CMB) from the \emph{Planck} satellite, $S_8 = 0.841^{+0.027}_{-0.025}$, when marginalising over neutrino mass \citep{Troxel2017}.

However, there is much more information stored in the matter fields beyond what can be captured by 2-point statistics. Two-point correlation functions only capture the Gaussian information stored in the field, while it is well known that the probability distribution function (PDF) of the galaxy density contrast in the late Universe has a 1-point distribution that is approximated better as log-normal than Gaussian \citep{Hubble1934,Coles1991,Wild2005}. Over the years, efforts have been made to explore statistics beyond 2-point for cosmology. These include 3-point correlation functions and bi-spectrum \citep{Takada2003, Takada2004, Semboloni2011, Fu2014}, shear peak statistics \citep{Kratochvil2010, Liu2015, Kacprzak2016}, higher moments of the weak lensing convergence field \citep{VanWaerbeke2013,Petri2015,Chang2018}, the PDF of the weak lensing convergence field \citep{Patton2016}, density-split statistics \citep{Friedrich2018, Gruen2018}, Minkowski functionals \citep{Kratochvil2012,Petri2015} and the Minimum Spanning Tree (MST, \citealt{Naidoo2019}). For some of these summary statistics (peak statistics, Minkowski functionals), one major challenge is that no analytic theoretical prediction of the target statistics exist and cosmological constraints must come from a large number of numerical simulations that span a range of cosmological parameters. In addition, these simulations also need to be closely matched to data and it is not clear what the requirements are for the matching between simulation and data \citep[though there exist some work in systematically addressing this question, e.g.][]{Bruderer2016,Kacprzak2019}. With the increasingly large datasets, the demand on simulations for these statistics become increasingly hard to meet. For the other statistics where analytical forms exist (3-point function, higher moments, PDF, density split statistics), most of the exploration work has been carried out with idealised simulations that in many respects do not represent the survey data.  
One of the reasons for this is that once one moves beyond 2-point statistics, the measurements and modelling becomes more complicated. This means that the noise and systematic effects propagate non-trivially.

In this paper, we focus on using the second and third moments of the weak lensing convergence field to constrain cosmology. In particular, this work validates the methodology using simulations and is targeted for the third year (Y3) of DES data. A companion paper applying this framework to the DES Y3 data will follow. First studied in \citet{Jain1997}, the moments of the weak lensing convergence field is one of the simpler high-order statistics both in terms of the measurement and in terms of the theoretical modelling. Several papers (e.g., \citealt{Gaztanaga1998,Fosalba2008, VanWaerbeke2013,Pujol2016,Chang2018}) have performed various moments measurements on simulations and/or data and compared the results with theoretical predictions, although this information was not then used to place constraints on cosmological parameters. In \citet{Vafaei2010}, the authors studied the tradeoff between different survey strategies in CFHTLenS for combined two and three-point statistics using simulations. They concluded that combining two and three-point statistics of the convergence field could increase the cosmological constraints by 10-20 per cent, in the case of CFHTLens data. In \citet{Petri2015}, the authors used a set of simulations with different cosmological parameters to study how the moments of the convergence field can help constrain cosmology. They included up to the fourth moments and showed that the constraints improve up to three times compared to the power spectrum-only constraints.

We build on the previous work and make several improvements. First, we use an analytic framework to incorporate the effect of masking, adapting a well-tested pseudo angular power spectrum estimation formalism (pseudo-C$_{\ell}$ in the following). Second, we include several systematic effects that are commonly accounted for in shear 2-point correlation function measurements and are key to obtaining unbiased cosmological constraints: namely, shear calibration bias, photometric redshift calibration uncertainty and intrinsic alignment. Third, we test how robust our statistics are to small-scales, higher  order lensing corrections such as reduced shear and source clustering, and to the effect of small-scales baryonic physics. Finally, we test our framework with two different sets of simulations (simple log-normal simulations and full N-body simulations that match the characteristics of the dataset of interest), each suited for specific purposes. Although the simulations and analysis choices here are specific to the DES Y3 data, we note that the general approach in this paper can be easily transferred to a different dataset.      

The paper is organised as follows. In \S~\ref{sect:model} we describe how we generate the weak lensing convergence maps from a shear catalog and how the second and third moments of this convergence map can be modelled, taking into account the effect of the mask as well as other systematics. In \S~\ref{sect:sims} we describe the characteristics and purpose of the two set of simulations used in this work. We test the validity of our modelling with simulations in \S~\ref{sect:testing} and determine the regime where our model can correctly predict the second and third moments. In \S~\ref{sect:like_tt} we derive the final components needed for a cosmology analysis: the covariance matrix, the scale cuts, and the likelihood. We describe also a fast emulator for evaluating the theory prediction for the cosmology inference. In \S~\ref{sect:cosmology} we determine the final fiducial scale cuts by examining how the cosmological constraints are biased as a function of scale cuts, and we forecast the cosmological constraints for DES Y3 and Y5 data. We summarise our findings in \S~\ref{sect:summary}.   
\section{Map making and theoretical modelling}
\label{sect:model}

In order to extract cosmological information from weak lensing convergence maps, we need to first construct the convergence map $\kappa$ from the more directly observed weak lensing shear $\gamma$. The theoretical modelling of the moments measured from the convergence map depends on the particular procedure one took to construct the map. As such, we first describe in \S~\ref{sect:Map Making} our map construction procedure and next introduce in \S~\ref{sect:theo} the theoretical model of our moment measurements.

\subsection{Map making}\label{sect:Map Making}

We implement a full-sky, spherical harmonics approach to obtain an estimate of the convergence field $\kappa$ from the estimated shear $\boldsymbol{\gamma}$ \citep{Castro2005,Leistedt2017,Wallis2017}. Such a full-sky formalism has been applied to both DES SV and Y1 data \citep{Wallis2017, Chang2018}. In \citet{Wallis2017}, the authors show that the convergence maps constructed using various flat-sky projection schemes could introduce up to 10 per cent error in the estimation of the curl-free modes (E-modes) of the convergence and up to 20 per cent for divergence-free modes (B-modes) of the convergence for an area approximately the DES Y3 footprint (5000 deg$^2$). As a result, it is necessary that we use this full-sky formalism in this work.


At any position in comoving space ($\chi$,$\theta$,$\phi$), one can relate the lensing potential $\psi$ to the local Newtonian potential $\Psi$ along the line-of-sight:


\begin{equation}
\label{eq:born}
\psi (\chi,\theta,\phi) = \frac{2}{c^2} \int_0^{\chi} d \chi' \frac{f_k (\chi -\chi')}{f_k (\chi)f_k (\chi')} \Psi (\chi',\theta,\phi),
\end{equation}
where $f_k$ assumes values of $\rm sin \chi$, $\chi$, $\rm sinh \chi$ for a closed ($k = 1$), flat ($k = 0$) and open ($k = -1$) universe respectively. Eq.~\ref{eq:born} implicitly assumes the Born approximation (i.e, the photons move along the unperturbed geodesics when computing
their deflection angle). The lensing potential in Eq.~\ref{eq:born} can be related to convergence $\kappa$ and shear $\gamma$ following \cite{Castro2005}:

\begin{equation}
\label{eq:kappa}
  \kappa = \frac{1}{4} (\eth \bar{\eth} +\bar{\eth} \eth) \psi ,
\end{equation}

\begin{equation}
\label{eq:gamma}
  \gamma = \gamma^1 + i \gamma^2 = \frac{1}{2} \eth \eth  \psi ,
\end{equation}
where $\eth$ and $\bar{\eth}$ are the raising and lowering operators acting on spin-weighted spherical harmonics defined in, e.g., \cite{Castro2005}. Expanding $\psi(\chi,\theta,\phi)$ in spherical harmonics leads to:

\begin{equation}
\psi(\chi,\theta,\phi) = \sum_{\ell m} \psi_{\ell m}(\chi)_0 Y_{\ell m}(\theta,\phi)
\end{equation}
\begin{equation}
\psi_{\ell m} (\chi)  = \int d\Omega \psi_{\ell m}(\chi ,\theta ,\phi)_0Y_{\ell m}^{*}(\theta ,\phi), 
\end{equation}
where $_0 Y_{\ell m}(\theta,\phi)$ are the spin-0 spherical harmonic basis set and $\psi_{\ell m} (\chi)$ the harmonic coefficients at a given comoving distance.
Analogously, we can expand $\kappa$ and $\gamma$:

\begin{equation}
\kappa = \kappa_{E} + i \kappa_{B} = \sum_{\ell m} ({\kappa}_{E,\ell m} + i {\kappa}_{B,\ell m} )_0Y_{\ell m},
\end{equation}
 
\begin{equation}
\label{eq:gammanomask}
  \gamma = \gamma^1 + i \gamma^2 = 2 \sum_{\ell m} {\gamma}_{\ell m} {}_{2} Y_{\ell m},
\end{equation}
with $_2Y_{\ell m}$ spin-2 spherical harmonics. We note that the convergence field has been divided into curl-free E-modes and divergence-free B-modes. One can relate the shear signal to the convergence field as follow:

\begin{equation}
{\kappa}_{E,\ell m} + i {\kappa}_{B,\ell m} = - \frac{1}{2} \ell  (\ell +1) \Psi_{\ell m},
\end{equation}

\begin{center}
\begin{equation}
{\gamma}_{\ell m} = \hat{\gamma}_{E,\ell m} + i {\gamma}_{B,\ell m} = \frac{1}{2}[\ell (\ell +1 )(\ell -1)(\ell +2)]^{1/2}\Psi_{\ell m}
\end{equation}
\end{center}

\begin{equation}
\label{eq:KS}
{\gamma}_{\ell m} =  -\sqrt{\frac{(\ell +2)(\ell -1)}{\ell (\ell +1)}} ({\kappa}_{E,\ell m} + i {\kappa}_{B,\ell m}).
\end{equation}

The shear field needs first to be decomposed into spherical harmonics; then E and B modes of the convergence field follows from applying Eq.~\ref{eq:KS}.
Curl-free E-modes carry most of the cosmological signal. Divergence-free B-modes can arise due to non-linear lensing corrections (such as deflection along the first-order Born approximation), clustering of the lenses and reduced shear corrections \citep{Schneider1998,Schneider2002b,Krause2009}. These effects are assumed to be small for current stage III weak lensing surveys (e.g., DES, KIDS, HSC) and will be neglected in the rest of the paper (see also $\S$~\ref{sect:HR2}). Biases in the shear measurement pipeline or object selection biases can also produce B-modes that can affect the parameters inference by few percent \citep{Hoekstra2004,Asgari2018}. Finally, partial sky coverage can induce mode mixing, producing spurious B-modes in the reconstructed convergence maps due to E-mode leakage. This will be the only source of B-modes that we will take into account here (see $\S$~\ref{sect:theo}).

To apply Eq.~\ref{eq:KS} to data, we need an estimate of the shear field. In practice, the shear field cannot be directly measured. The observable is the reduced shear:
\begin{equation}
\label{eq:reduced_mm}
g = \frac{\gamma}{1-\kappa}.
\end{equation}

Since galaxies have an intrinsic shape, what we actually measure is the ellipticity, or shape of the galaxy, which is a noisy estimate of the reduced shear:

\begin{equation}
\epsilon = \frac{g+\epsilon_{int}+\epsilon_{m}}{1+g(\epsilon_{int}+\epsilon_{m})},
\end{equation}
where $ \epsilon_{int}$ is the intrinsic shape of the galaxy, and $\epsilon_{m}$ the shape measurement noise. The latter two quantities should average to zero for large number of galaxies (though in $\S$ \ref{systematics} we explain how to handle shear measurement biases). Moreover, in the weak lensing regime, $\gamma, \kappa \ll 1$, so the observed shape results in a noisy estimator for the shear field $\epsilon \approx \gamma + \epsilon_{int} +\epsilon_{m}$\footnote{We note that we have ignored intrinsic alignment so far. In $\S$ \ref{systematics} we explain how to include intrinsic alignment in our formalism.}. The fewer the galaxies, the noisier the estimate of $\gamma$. This means also that our estimate of the convergence field will be noisy:

\begin{equation}
\kappa_{E,{\rm obs}} = \kappa_{E, {\rm true}} + \kappa_{E,{\rm true}}
\end{equation}

\begin{equation}
\kappa_{B,{\rm obs}} = \kappa_{B,{\rm true}} + \kappa_{B,{\rm true}}
\end{equation}

The contribution of the noise to the convergence field can be estimated by randomly rotating the shape of the galaxies and applying the full-sky spherical harmonics approach to obtain the convergence \citep{VanWaerbeke2013,Chang2018}. As the random rotation should completely erase the cosmological contribution, the resulting convergence signal will just contain noise and should average to 0 (but with a non-negligible variance). 

It follows that when estimating second and third moments from noisy convergence maps it is necessary to properly de-noise the measured moments. Following \cite{VanWaerbeke2013}:
%
%
%
%
\begin{multline}
\label{eq:deno2}
\langle\hat{\kappa}^2\rangle^{i,j} = \langle  \kappa^2\rangle^{i,j} -  \langle\kappa \kappa_{{\rm rand}}\rangle^{i,j} -  \langle  \kappa_{\rm rand} \kappa \rangle^{i,j} -
 \langle \kappa^2_{{\rm rand}}\rangle^{i,j},
\end{multline}
\begin{multline}
\label{eq:deno3}
\langle \hat{\kappa}^3\rangle^{i,j,k} = \langle \kappa^3\rangle^{i,j,k} -  \langle \kappa^3_{{\rm rand}}\rangle^{i,j,k} - \\ \left[\langle  \kappa^2_{{\rm rand}}\kappa  \rangle^{i,j,k} - \langle  \kappa\kappa^2_{{\rm rand}}  \rangle^{i,j,k} + {\rm cycl.} \right].
\end{multline}
In the above equations, the term $\langle \kappa^2_{{\rm rand}}\rangle^{i,j}$  is the noise-only contribution to the second moment of the tomographic bins $i,j$; for $i \neq j$ it vanishes. The map $\kappa_{{\rm rand}}$ represents the estimate of the shape noise contribution to the convergence map; it is estimated by randomly rotating the galaxy shapes. The intrinsic ellipticity distribution of observed galaxies is not expected to be perfectly Gaussian, but by the central limit theorem, it would be the correct distribution in the limit of large numbers of galaxies averaged in the pixels of the convergence map \citep{Jeffrey2018}. If this holds, also the term $\langle \kappa^3_{{\rm rand}}\rangle^{i,j,k}$ (which is the noise-only contribution to the third moment of the tomographic bin  $i,j,k$) would vanish. Additional checks will need to be performed on DES Y3 data, as we do not include potential sources of noise inhomogeneities (e.g. astrophysical or observational systematics) in this work. Finally, under the hypothesis of no correlation between the convergence field and the shape noise, mixed terms should be consistent with zero (we will check this hypothesis in $\S$ \ref{sect:testing}). 


The above theoretical derivation describes how to obtain the convergence maps from an estimate of the shear field. The method we implement in this paper does not assume any prior knowledge of the convergence field to be reconstructed. There exist methods, however, which implicitly or explicitly assume priors that improve the map reconstructions over a range of metrics (e.g. \citealt{Jeffrey2018,Mawdsley2019,Jeffrey2019}). Some of these methods will be explored in a future DES Y3 Mass Maps paper (in prep.). We are not considering these methods here: in this paper, the observables (convergence moments) are modelled from theory, and including the effects of such priors on the maps moments will be difficult. On the other hand, these alternative methods are valuable when N-body simulations are used to model the observables (e.g., \citealt{Petri2015,Fluri2018}).

We do not describe the detailed map making procedure here as it has been detailed in previous DES papers (e.g., \citealt{Chang2018}). To summarize, we construct the maps using \texttt{HEALPIX} pixelisation \citep{GORSKI2005}. The first step in the reconstruction of the mass map involves making pixelised ellipticity (or shear estimate) maps $\epsilon_1$ and $\epsilon_2$ from a shear catalogue. These are obtained by averaging the two components of the shape estimate over all the galaxies belonging to a given \texttt{HEALPIX} pixel. We use NSIDE = 1024, corresponding to a pixel size of 3.44 arcmin. Next, we perform the spin transformation which converts the ellipticity maps into a curl-free E-mode convergence map $\hat{\kappa}_E$ and a divergence-free B-mode convergence map $\hat{\kappa}_B$. We use the \texttt{HEALPIX} functions \texttt{MAP2ALM} to decompose the shear field in spherical harmonic space obtaining the coefficients $\hat{\gamma}_{E,\ell m}$, $\hat{\gamma}_{B,\ell m}$ and calculate $\hat{\kappa}_{E,\ell m}$, $\hat{\kappa}_{B,\ell m}$  following Eq. \ref{eq:KS}. Finally, we use the \texttt{HEALPIX} function \texttt{ALM2MAP} to convert these coefficients back to the real space $\kappa_E$ and  $\kappa_B$ maps. 

\newpage

\subsection{Theoretical modelling }\label{sect:theo}
We adopt the theoretical model for second and third moments (variance and skewness) of the convergence field using a 
non-linear extension of cosmological perturbation theory \citep{VanWaerbeke2001,Scoccimarro2001,Bernardeau2002}.

As we are interested in highlighting the features of our convergence field at different angular scales, we smooth our recovered convergence fields using a top-hat filter at different angular scales. We note that the choice of the type of the filter is rather arbitrary, e.g., \citealt{VanWaerbeke2013} used a Gaussian filter. We chose a top-hat filter to facilitate the analytical evaluation of third moments (see Appendix~\ref{sect:Skewness}). A top-hat filter $W$ in harmonic space of smoothing length $\theta_0$ is defined as:


\begin{equation}
\label{eq:filter}
W_{l}(\theta_0) = \frac{P_{l-1}(cos(\theta_0))-P_{l+1}(cos(\theta_0))}{(2l+1)(1-cos(\theta_0))},
\end{equation}
where $P_{\ell}$ are Legendre polynomials of order $l$.
The variance of matter contrast $\delta$ smoothed by such a filter at a given comoving distance $\chi$ is:

\begin{equation}
\label{eq:dd2}
\langle \delta^2_{\theta_0,\rm NL} \rangle (\chi) = \sum_{\ell} \frac{2\ell+1}{4 \pi} P_{\rm NL} (\ell/\chi,\chi) F_{\ell}^2 W_{\ell}(\theta_0)^2,
\end{equation}
where $F_{\ell}$ is the pixel window function\footnote{$F_{\ell}$  is modelled using the pixel window function provided by \texttt{HEALPIX}.} and $P_{\rm NL} (\ell/\chi,\chi)$ the non linear power spectrum. For the latter we used \texttt{HALOFIT} as detailed in \cite{Takahashi2014} and assumed in the fiducial DES Y3 analysis.

For the smoothed version of skewness of the matter contrast, at leading order in perturbation theory it reads:

\begin{equation}
\label{eq:dd3}
\langle \delta^3_{\theta_0, \rm NL} \rangle (\chi) = S_3 [\langle \delta^2_{\theta_0,\rm NL} \rangle (\chi)]^2,
\end{equation}
where $S_3$ is the reduced skewness parameter. The analytical derivation of the reduced skewness parameter is performed to leading order, which is linear in the power spectrum, but as such predictions perform well even in the mildly non-linear regime ($k \approx 0.1 h^{ - 1}$ Mpc \citealt{Bernardeau2002}), we assume their validity when a non-linear power spectrum (the \texttt{HALOFIT} from \citealt{Takahashi2014}) is used to compute the variance. We also implement a refinement of the treatment of the skewness at small scales based on N-body, cold dark matter only simulations  \citep{Scoccimarro2001,GilMarin2012}. In particular, we implement the analytical fitting formulae from \cite{Scoccimarro2001}. The modelling choices for the third moments are validated in \S~\ref{sect:validate_others}. We also include in our error budget (see \S~\ref{sect:covariance}) the scatter between different small scales models to grasp the modelling uncertainty. For the analytical expression of the reduced skewness parameter see Appendix \ref{sect:Skewness}.

The above equations, together with a galaxy-matter bias model, could be used to predict the second and third moments of the galaxy density field (for a possible application of the moments of the galaxy density contrast distribution see, e.g., \citealt{Salvador2019}). Here, however, we are interested in the second and third moments of the convergence field, for which a galaxy-matter bias model is not needed. Under the Limber approximation \citep{Takada2002}, one obtains:

\begin{equation}
\label{eq:ksm2}
\langle \kappa^2_{\theta_0}\rangle^{i,j}  = \int d\chi \frac{q^i(\chi) q^j(\chi)}{\chi^2} \langle \delta^2_{\theta_0} \rangle (\chi),
\end{equation}
\begin{equation}
\label{eq:ksm3}
\langle \kappa^3_{\theta_0}\rangle^{i,j,k}  = \int d\chi \frac{q^i(\chi) q^j(\chi)q^k(\chi)}{\chi^4} \langle \delta^3_{\theta_0} \rangle (\chi).
\end{equation}
$i,j,k$ refers to different tomographic bins. We have dropped the subscript $_{\rm NL}$ for brevity. The term $q^i$ represents the lensing kernel and reads:
\begin{equation}
\label{eq:wlk}
q^i(\chi) = \frac{3H_0^2\Omega_{\rm m}}{2c^2}\frac{\chi}{a(\chi)} \int_{\chi}^{\chi_h} d\chi' n^i(z(\chi')) dz/d\chi'\frac{\chi'-\chi}{\chi},
\end{equation}
where $H_0$ is the Hubble constant at present time, $\Omega_{\rm m}$ the matter density, $c$ the speed of light, $n^i(z)$ the normalised redshift distribution of a given tomographic bin, and $a(\chi)$ the scale factor.

We note that the variance and skewness of the convergence field have differing dependencies on the parameters $\Omega_{\rm m}$ and $\sigma_8$ \citep{Bernardeau1997, Seljak1996}.

\subsubsection{Effects of masking}
One of the problems in estimating the convergence field from the observed shapes is that we observe only a portion of the sky. This means that the reconstruction will suffer edge effects, due to the convolution with a  window function representing the survey footprint. Some 
methods deal with mask effects at the level of map making \citep{Pires2009, Mawdsley2019}, whereas in this work, we will account for the mask effects in our theoretical predictions using a pseudo-$C_\ell$ formalism \citep{Brown2005,Hikage2011}.

The pseudo-$C_\ell$ formalism correctly recovers the shear power spectrum estimated from the shear field in the case of partial sky coverage. It also predicts mode mixing (that is, part of the E-modes leaks into B-modes and vice-versa). In particular, if we define:

\begin{equation}
C_{\ell }^{EE} = \frac{1}{2\ell +1}\sum_m | \hat{\gamma}_{E,\ell m} |^2,
\end{equation}
\begin{equation}
C_{\ell }^{EB} = \frac{1}{2\ell +1}\sum_m  \hat{\gamma}_{E,\ell m}  \hat{\gamma}_{B,\ell m}^* ,
\end{equation}
\begin{equation}
C_{\ell }^{BB} = \frac{1}{2\ell +1}\sum_m | \hat{\gamma}_{B,\ell m} |^2,
\end{equation}
one can write the masked (pseudo) spectra as the convolution of the true spectra with a mixing matrix:
\begin{equation}
\hat{\textbf{C}}_{\ell} = \sum_{\ell'} \textbf{M}_{\ell \ell'} \textbf{C}_{\ell'},
\end{equation}
where have we introduced the vector $\textbf{C}_{\ell} \equiv (C_{\ell}^{EE},C_{\ell}^{EB},C_{\ell}^{BB})$. The mode-mode coupling matrix $\textbf{M}$ is expressed in terms of $M^{EE,EE}_{\ell \ell'}$, $M^{BB,BB}_{\ell \ell'}$, $M^{EB,EB}_{\ell \ell'}$, $M^{EE,BB}_{\ell \ell'}$. The mixing matrices contain information about the survey geometry; analytical expressions for the mixing matrices in terms of the window function can be found in \cite{Hikage2011} and in Appendix \ref{sect:MM}. The pseudo-$C_\ell$ formalism can be incorporated in Eq. \ref{eq:ksm2} as:
\begin{multline}
\label{eq:ksmM2}
\langle \kappa^2_{\theta_0}\rangle^{i,j, EE/BB}  = \int d\chi \frac{q^i(\chi) q^j(\chi)}{\chi^2} \times \\ \sum_{\ell} \frac{2\ell+1}{4 \pi} f_{\ell}^{-1} W_{\ell}(\theta_0)^2 \sum_{\ell'}  M_{\ell \ell'}^{EE/BB ,EE}P_{NL} (\ell'/\chi,\chi) F_{\ell'}^2 f_{\ell'}.
\end{multline}
In the above equations, the factor $f_{\ell} = [(\ell+2)(\ell-1)]/[\ell(\ell+1)]$ accounts for the fact that the mixing matrix is applied to the shear field rather than to the convergence field directly. Depending on the mixing matrix used ($M_{\ell \ell'}^{EE ,EE} $ or $M_{\ell \ell'}^{BB ,EE} $), with Eq.~\ref{eq:ksmM2} we can predict the variance of both E and B modes of the recovered convergence field.
As for the the skewness, if we neglect the contribution of the masking to the skewness parameter, we can write:
\begin{equation}
\label{eq:ksmM3}
\langle \kappa^3_{\theta_0}\rangle^{i,j,k,EE/BB}  = \int d\chi \frac{q^i(\chi) q^j(\chi)q^k(\chi)}{\chi^3} S_3 [\langle \delta^2_{\theta_0,NL}\rangle^{EE/BB}  (\chi)]^2.
\end{equation}
%
We will check the validity of assuming the mask effect to be small on $S3$ against simulations in $\S$~\ref{sect:testing}.

\subsubsection{Systematic effects}\label{systematics}

Astrophysical and measurement systematic effects are modelled through nuisance parameters and are included in our likelihood (see \S~\ref{sect:like}). We marginalise over all the nuisance parameters when estimating the cosmological parameters. Values and priors are summarised in Table~\ref{table1}.

\begin{table}
\tiny
\caption {Cosmological, systematic and astrophysical parameters. We report the boundaries for both Flat and Gaussian priors. For Gaussian priors we also report the mean and the 1 $\sigma$ in the prior column. }
\centering
\begin{adjustbox}{width=0.4\textwidth}
\begin{tabular}{|c|c|c|}

 \hline
\textbf{Parameter} & \textbf{Range}& \textbf{Prior}\\
 \hline
{$\Omega_{\rm m}$} &$0.1 ... 0.9$& Flat\\
{$\sigma_{\rm 8}$} &$0.4 ... 1.3$& Flat \\
{$h_{\rm 100}$} &$55 ... 90$& Flat\\ 
{$n_{\rm s}$} &$0.87 ... 1.07$& Flat\\ {$\Omega_{\rm b}$} &$0.03 ... 0.07$& Flat \\
 \hline
{$m_1-m_2\times10^2$} &$-10.0 ... 10.0$& 0.0 $\pm$ 2.3 \\
{$\Delta z_1\times10^2$} &$-10.0 ... 10.0$& 0.0 $\pm$ 1.6 \\
{$\Delta z_2\times10^2$} &$-10.0 ... 10.0$& 0.0 $\pm$ 1.3 \\
{$\Delta z_3\times10^2$} &$-10.0 ... 10.0$& 0.0 $\pm$ 1.1 \\
{$\Delta z_4\times10^2$} &$-10.0 ... 10.0$& 0.0 $\pm$ 2.2 \\

 \hline
{$A_{\rm IA,0}$} &$-5.0 ... 5.0$& Flat \\
{$\alpha_{\rm IA}$} &$-5.0 ... 5.0$& Flat \\
 \hline
\end{tabular}
\end{adjustbox}
\label{table1}
\end{table}

\textit{Photometric redshift uncertainties}. Photometric redshift uncertainties are parametrized through a shift $\Delta z$ in the mean of the redshift distribution:

\begin{equation}
n^i(z) = \hat{n}^i(z+\Delta z ),
\end{equation}
where $\hat{n}^i$ is the original estimate of the redshift distribution coming from photometric redshift code. We assume DES Y1 priors for the shift parameters.

\textit{Multiplicative shear biases}. Biases coming from the shear measurement pipeline are modelled through a multiplicative parameter $1+m^i$. Such parameter depends on the tomographic bin considered and affects our moments in the following way:

\begin{equation}
\langle \kappa^2_{\theta_0}\rangle^{i,j} \rightarrow (1+m^i) (1+m^j)\langle \kappa^2_{\theta_0}\rangle^{i,j},
\end{equation}
\begin{equation}
\langle \kappa^3_{\theta_0}\rangle^{i,j,k} \rightarrow (1+m^i) (1+m^j) (1+m^k)\langle \kappa^3_{\theta_0}\rangle^{i,j,k}.
\end{equation}
Gaussian priors are assumed for each of the $m^i$.

\textit{Intrinsic galaxy alignments}. IA is modelled according to the non-linear alignment (NLA) model \citep{Hirata2004,Bridle2007}. It can be incorporated in the modelling by modifying the lensing kernel:

\begin{equation}
q^i(\chi) \rightarrow q^i(\chi) - A(z(\chi)) \frac{n^i(z(\chi))}{\avg{n^i}}\frac{dz}{d\chi}.
\end{equation}

The NLA model is usually used in the context of 2-point correlation statistics, but the above equation generalises it to the third moments case as well.
The amplitude of the IA contribution can be written as a power-law:

\begin{equation}
A(z) = A_{\rm IA,0} \left(\frac{1+z}{1+z_0} \right)^{\alpha_{\rm IA}} \frac{c_1 \rho_{m,0}}{D(z)},
\end{equation}
with $z_0= 0.62$, $c_1\rho_{m,0}=0.0134$ \citep{Bridle2007,methodpaper} and $D(z)$ the linear growth factor. We marginalise over $A_{\rm IA,0}$ and $\alpha_{\rm IA}$ assuming flat priors.

\section{Simulations}
\label{sect:sims}

\begin{figure}
\begin{center}
\includegraphics[width=0.45 \textwidth]{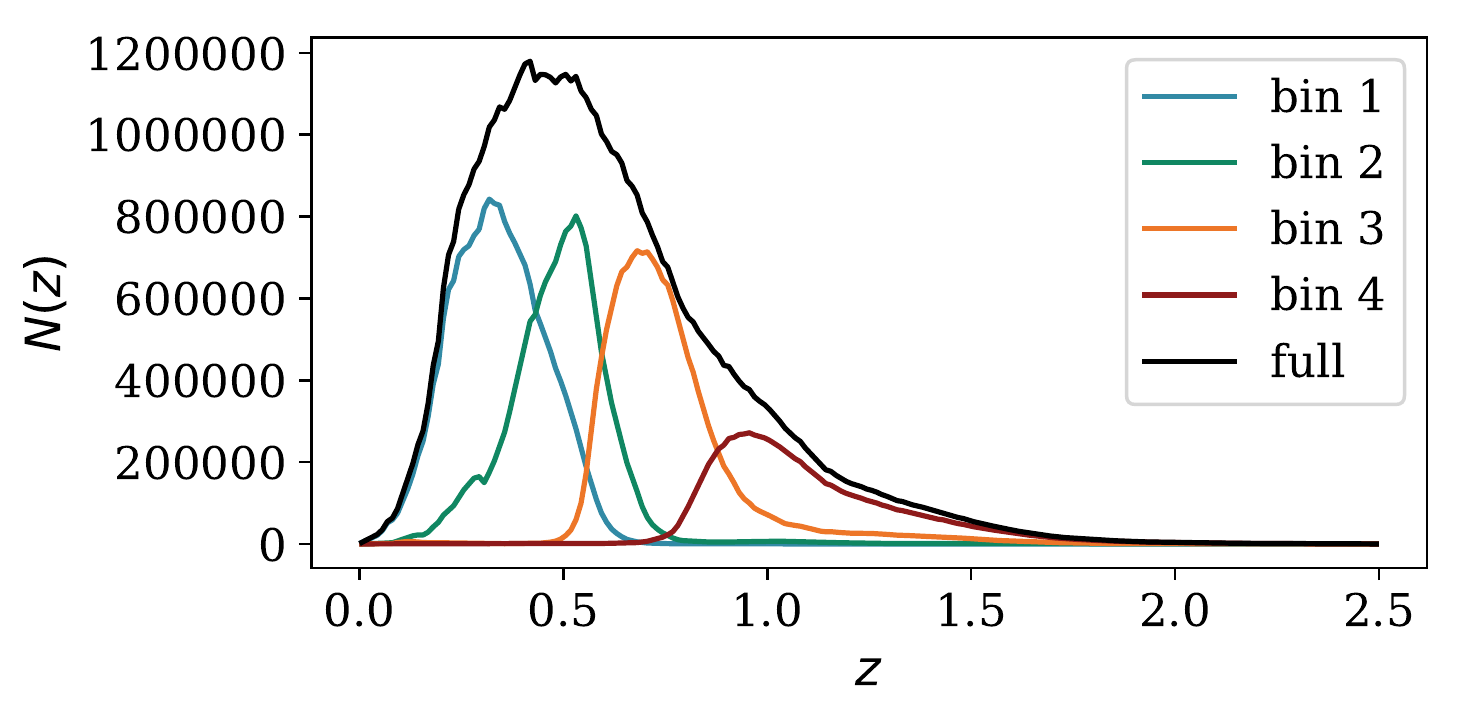}
\end{center}
\caption{Redshift distributions of the 4 tomographic WL bins (and the full sample), from a fiducial DES simulated sample \citep{DeRose2018}. A bin width of $\Delta_z = 0.01$ has been used for the histograms.}
\label{fig:redshift}
\end{figure}

Two different sets of simulations are used to validate our measurement. These simulations differ in the complexity of the physics included, and are used to validate different parts of our methodology. In particular, we make use of:

\begin{itemize}
    \item \texttt{FLASK} simulations \citep{Xavier2016}. These are log-normal realisations, and are used to produce a large number of realisations (of the order of 1000) of the shear and convergence fields. They require input power spectra for their predictions, so they cannot be used to test the modelling of the second and the third moments, as they are key ingredients to run the simulations. We use them to model the covariance matrices of our measurements and to test the modelling of mask effects.
    

    \item \cite{Takahashi2017} mocks. We use 100 full-sky gravitational lensing convergence and shear maps obtained from full N-body simulations and a ray-tracing algorithm described in \cite{Takahashi2017}. We use these to validate the theoretical modelling of second and third moments over a large number of simulations. We also use them to check the effect of non-linear lensing corrections in our modelling.

\end{itemize}
Below we provide a more in-depth descriptions of each of the simulations.

\subsection{\texttt{FLASK} simulations}\label{sect:FLASK}
The \texttt{FLASK} software \citep{Xavier2016} allows one to rapidly generate full-sky, log-normal realisations of a given field (in our case, the convergence field). In particular, \texttt{FLASK} assumes the convergence field to be described by a zero-mean shifted log-normal distribution, where the parameters of the log-normal probability distribution function (PDF) are chosen to match the variance and skewness of the input. The lognormal approximation is usually adopted for the density field \citep{Hubble1934,Coles1991,Wild2005} and is not expected to exactly hold for the convergence field, as it is a weighted projection of the mass density field along the line of sight. Tests on numerical simulations showed a lognormal PDF to be a reasonable model (e.g \citealt{Taruya2002, Hilbert2011}), although generalised lognormal PDFs have been shown to improve the fit at the tails of the distribution \citep{Das2006,Takahashi2011,Joachimi2011}. Observational evidences from DES SV \citep{Clerkin2017} finds that at intermediate scales between 10 and 20 arcmin, the convergence distributions are more lognormal than Gaussian (at larger scales noise dominates). We show in $\S$ \ref{sect:scale_cuts} that relying on the lognormal approximation to build our covariance matrix does not bias the recovery of the cosmological parameters.

The software requires as inputs a set of auto and cross power spectra and a log-normal shift parameter. This latter parameter is a combination of the variance and skewness \citep{Xavier2016} and it is computed from theory and fixed to the value at no smoothing\footnote{Formally, this means that the third moment computed in \texttt{FLASK} should match theoretical predictions only at no smoothing. Slight variations can occur with a non zero smoothing as the convergence field is not perfectly log-normal. The second moment should agree at every smoothing scale as the full power spectrum is provided.}. We generated theoretical predictions for the power spectra of the convergence field for four tomographic bins of our WL source sample. We used the true redshift distributions of the WL sample as measured in a fiducial DES simulated sample (\texttt{Buzzard}, \citealt{DeRose2018}). Redshift distributions are shown in Fig.~\ref{fig:redshift}. We fixed the cosmology of our input power spectra to be $\Omega_{\rm m} = 0.286$, $\sigma_8 = 0.82$, $\Omega_b = 0.047$, $n_s = 0.96 $, $h_{100} = 0.7$\footnote{The values of the cosmological parameters used to compute the covariance are slightly different than the ones of the mocks used to validate the modelling of second and third moments. These values have been chosen to facilitate the comparison with other simulated cosmological analysis for DES Y3.}. We generated 1000 realisations of the convergence fields in the form of \texttt{HEALPIX} maps \citep{GORSKI2005} with NSIDE = 1024. This resolution is chosen based on the expected number density of the DES Y3 WL sample. For each of the realisations, we cut out a DES Y3 footprint. We assign shape noise to each pixel of the shear fields based on the shape noise properties of the DES Y3 WL sample and using \texttt{FLASK} realisations of the WL sample density field. The final number densities of each bin are respectively 1.38, 1.36, 1.35, 0.86 gal/arcmin$^2$, while the $\sigma_e$ (the standard deviation of the two components for the measured galaxy shapes) are 0.29, 0.29, 0.29, 0.30. We use such \texttt{FLASK} mocks to validate our modelling of the mask effects and to generate covariance matricies for our measurements.

\subsection{Takahashi N-body simulation}

The simulations are a set of 108 full-sky 
lensing convergence and shear maps obtained for a range of redshifts between z = 0.05 and 5.3 at intervals of 150 $h^{ - 1}$ Mpc comoving distance. 

Initial conditions were generated using 2LPTIC \citep{Crocce2006} and the N-body run using L-GADGET2 \citep{springel2005}, consistent with WMAP 9 year results \citep{Hinshaw2013}: $\Omega_{\rm m} = 0.279$, $\sigma_8 = 0.82$, $\Omega_b = 0.046$, $n_s = 0.97 $, $h = 0.7$.

The simulations begin with 14 boxes with side lengths L = 450, 900, 1350, ..., 6300 $h^{ - 1}$ Mpc in steps of 450 $h^{ - 1}$ Mpc, with six independent copies at each box size and 2048$^3$ particles per box. Snapshots are taken at the redshift corresponding to the lens planes at intervals of 150 $h^{ - 1}$ Mpc comoving distance. The authors checked that the agreement of the average matter power spectra with the revised \texttt{HALOFIT} \citep{Takahashi2012} was within $5$ per cent for $k < 1$ $h$ Mpc$^{ - 1}$ at $z < 1$, for $k < 0.8$ $h$ Mpc$^{ - 1}$ at $z < 3$, and for $k < 0.5$ $h$ Mpc$^{ - 1}$ at $z < 7$. In this paper we also check the validity of the simulations for the angular scales of interest in \S\ref{sect:validate_others}.
Weak lensing quantities were estimated using the multiple plane ray-tracing algorithm \texttt{GRayTrix} \citep{Hamana2015}, and shear and convergence \texttt{HEALPIX} maps with resolution NSIDE = 4096 or larger are provided. Halos are identified in the simulation using the public code \texttt{ROCKSTAR} \citep{Behroozi2013}. The simulations do not come with a galaxy catalogue. For each of the 108 realisations, we produced convergence maps for the 4 WL tomographic bins by properly stacking the convergence snapshots taking into account the redshift distributions of the bins. We used the same redshift distribution as that used in the \texttt{FLASK} simulations.


\section{Testing the modelling with simulations}\label{sect:testing}

\begin{figure*}
\begin{center}
\includegraphics[width=0.95 \textwidth]{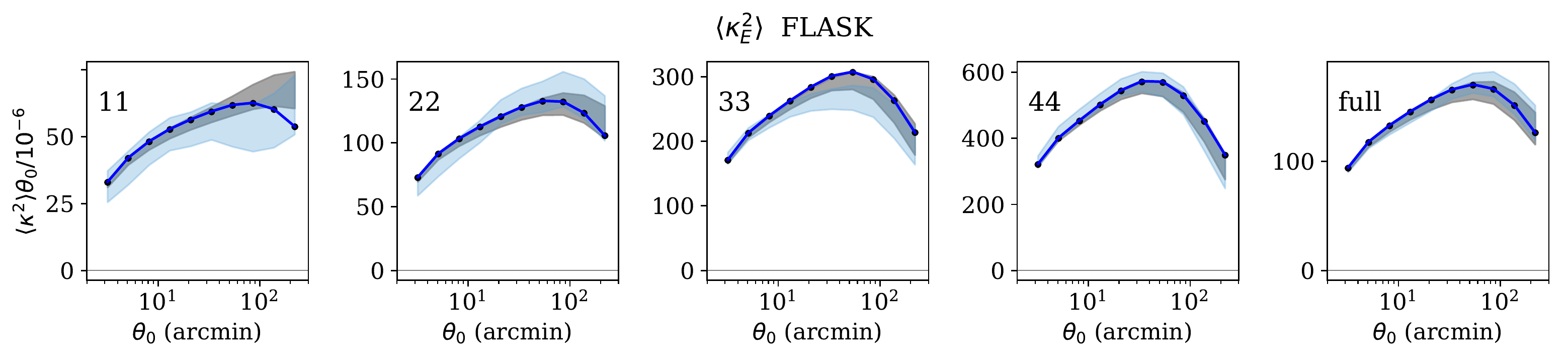}
\includegraphics[width=0.95 \textwidth]{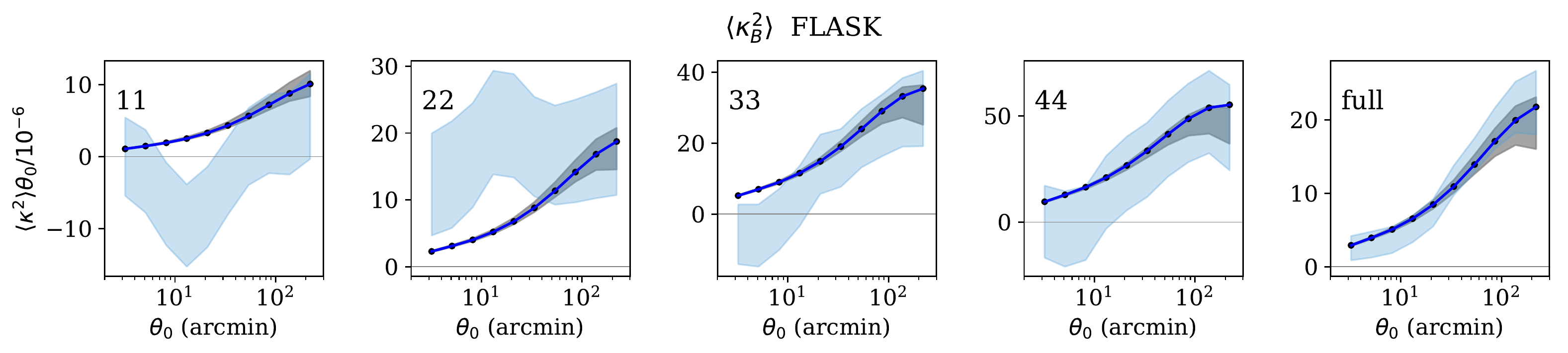}
\includegraphics[width=0.95 \textwidth]{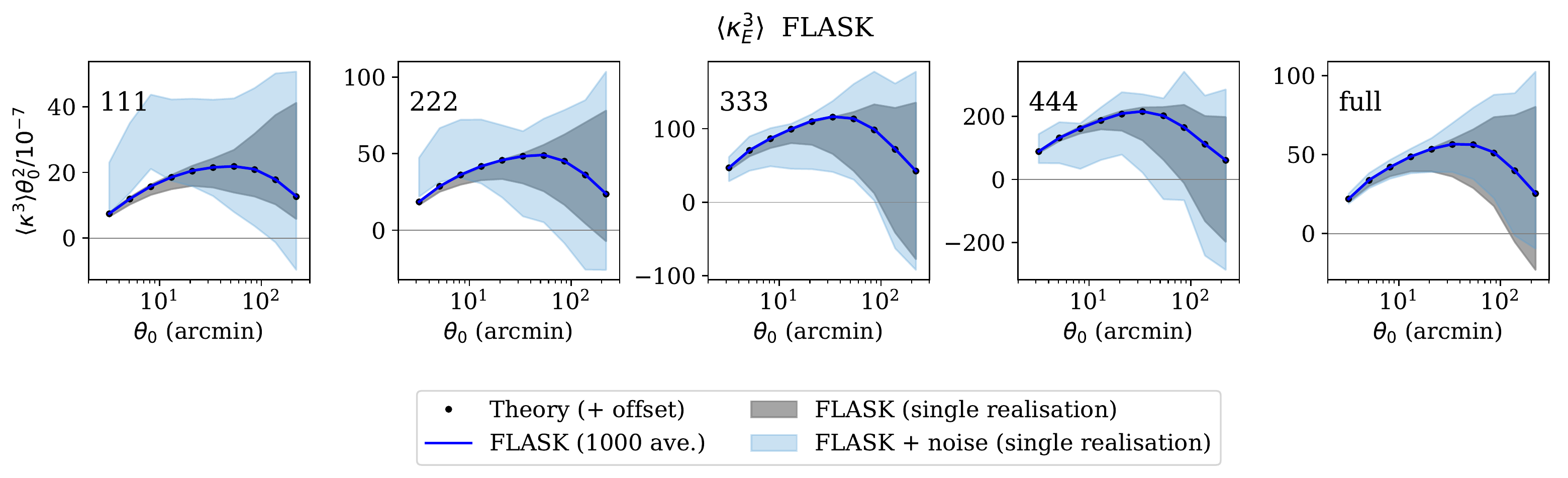}
\end{center}
\caption{Second moments (E and B modes) and third moments (E only) measured in \texttt{FLASK} simulations from partial-sky coverage realisations of the DES Y3 footprint. The convergence maps are obtained from the realisations of the \texttt{FLASK} shear fields configured as explained in \S\ref{sect:FLASK}. Mask effects are included in the theory modelling (black dots, Eqs.~\ref{eq:ksmM2}, \ref{eq:ksmM3}). Grey bands represent the measurement from one (taken at random) noiseless \texttt{FLASK} realisation, together with its uncertainty (measurements uncertainties are estimated in \S~\ref{sect:covariance}). Light blue bands also include shape noise. The average of the measurement over 1000 \texttt{FLASK} realisations are shown by the  blue lines (error bars are omitted). The numbers $11$, $22$, $33$ etc. in each plot refers to the combination of tomographic bins considered to compute the moments, while ``full'' refers to the non-tomographic case. Only auto-correlations are shown.
\textit{Upper panels:} second moments, E-mode of the convergence maps. 
\textit{Middle panels:} second moments, B-mode of the convergence maps. B-modes are much smaller than E-modes and are due to mask effects. 
\textit{Lower panels:} third moments, E-mode of the convergence maps. Third moments measured in \texttt{FLASK} simulations are not expected to match the input theory perfectly (see text for more details); here, the theoretical predictions for the third moments are replaced by the average measurement of third moments in many full-sky \texttt{FLASK} realisations. } 
\label{fig:FLASK_modeling}
\end{figure*}

\begin{figure*}
\begin{center}
\includegraphics[width=0.95 \textwidth]{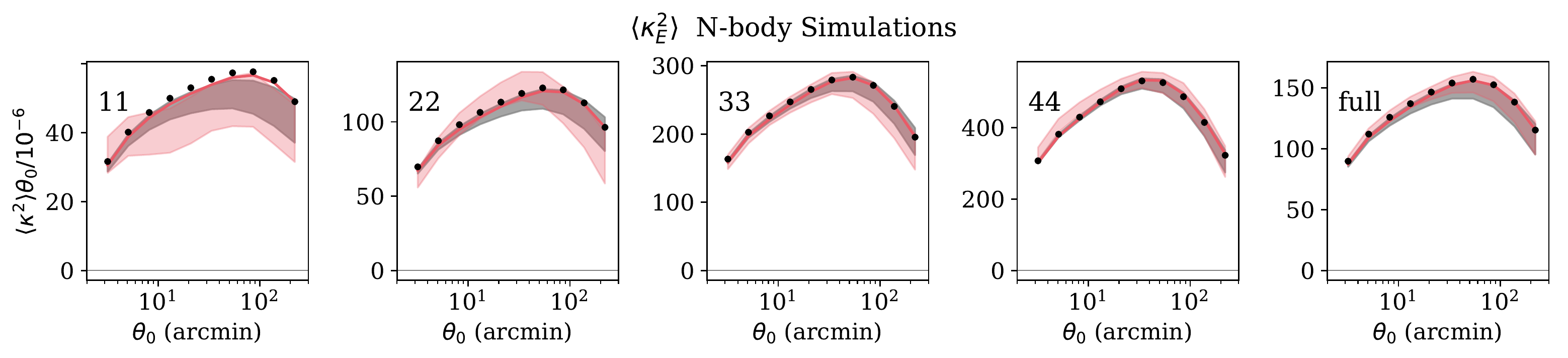}
\includegraphics[width=0.95 \textwidth]{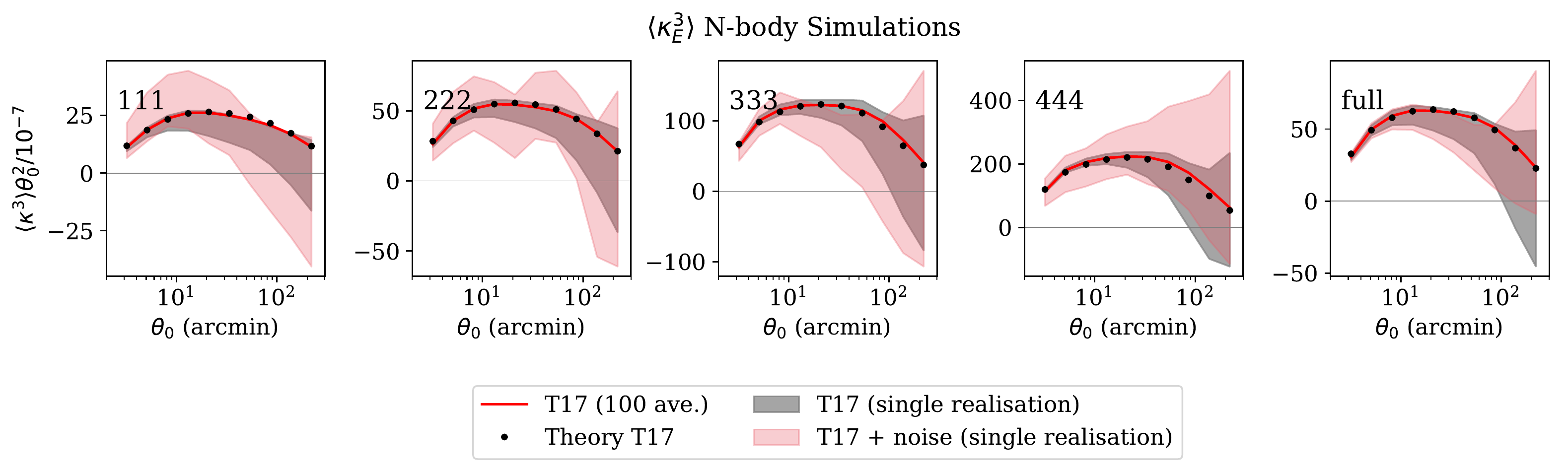}
\end{center}
\caption{Second moments and third moments (E-modes) measured in \citet{Takahashi2017} simulations from partial-sky coverage realisations of the DES Y3 footprint. The convergence maps have been obtained starting from a realisation of the DES Y3 shear field. Mask effects are included in the theory modelling (black dots, Eqs.~\ref{eq:ksmM2}, \ref{eq:ksmM3}). Grey bands represent the measurement from one (taken at random) noiseless T17 realisation, together with its uncertainty (measurements uncertainties are estimated in \S~\ref{sect:covariance}). Red bands also include shape noise.The average of the measurement over 100 T17 realisations are shown by the  red lines (error bars are omitted). The numbers $11$, $22$, $33$ etc. in each plot refers to the combination of tomographic bins considered to compute the moments, while ``full'' refers to the non-tomographic case. Only auto-correlations are shown. \textit{Upper panels:} second moments, E-modes of the convergence maps. \textit{Lower panels:} third moments, E-modes of the convergence maps.}
\label{fig:taka_modeling}
\end{figure*}
In this section we present a series of validation tests with simulations to show that our model presented in \S~\ref{sect:theo} does indeed model the second and third moments of the convergence maps. 
We first validate our model for the effect of masking (i.e. the mixing matrix approach) in \S~\ref{sect:validate_mask}, then validate the remaining components of the modelling of the second and third moments in \S~\ref{sect:validate_others}. In \S \ref{sect:HR1} we estimate the potential impact of baryonic feedback at smal scales; finally, in \S \ref{sect:HR2}, we assess the impact of higher-order lensing corrections (such as reduced shear or source crowding) not included in our modelling.

\subsection{Testing mask effects}
\label{sect:validate_mask}

We first considered the case of no shape noise. We used 1000 \texttt{FLASK} realizations of the DES Y3 footprint, and measured the convergence field starting from the shear field using the method explained in $\S \ref{sect:Map Making}$. This has been done for the four tomographic bins and the non-tomographic sample. We then smoothed the map with a top hat filter at different smoothing scales. We choose as an interval $\theta \in [3.2,220]$ arcmin, and we used 10 equally (logarithmic) spaced scales (even though we expect scales close to the pixels size, which is $\approx 3.4$ arcmin, to not contain much information).
To measure the moments of the smoothed map, we use a simple estimator:

\begin{equation}
\hat{\avg{\kappa^2_{\theta_0}}}^{i,j} = \frac{1}{N_{tot}}\sum_{pix}^{N_{tot}} k_{\theta_0,pix}^i k_{\theta_0,pix}^j,
\end{equation}
\begin{equation}
\hat{\avg{\kappa^3_{\theta_0}}}^{i,j,k} = \frac{1}{N_{tot}}\sum_{pix}^{N_{tot}} k_{\theta_0,pix}^i k_{\theta_0,pix}^j k_{\theta_0,pix}^k,
\end{equation}
where $i,j,k$ refers to different tomographic bins. The sum runs over all the pixel of the sky, also outside the footprint: the transformation from the shear field to the convergence field is non-local and some of the power is transferred outside the footprint, despite most of it remaining confined to the footprint. The lack of power outside of the footprint (due to the fact the shear field is not defined there) is taken into account by the mixing matrices (Eqs.~\ref{eq:ksmM2}, \ref{eq:ksmM3}). 

The (smoothed) second moments, both for the E and B modes, are shown in the top and middle panels of Fig.~\ref{fig:FLASK_modeling} and compared with theoretical predictions. In the figure, we just show auto-moments (i.e., moments obtained from maps of the same tomographic bin). We also show the average of the 1000 \texttt{FLASK} realisations, which agrees to better than $0.5$ per cent with the theoretical modelling.  We note that without the mixing matrices, we would have not been able to predict any B-modes. Moreover, our theoretical predictions for the E-modes would have been biased high, as no leakage of E-modes into B-modes would have occurred. 

The third moments are shown in the lower panel of Fig.~\ref{fig:FLASK_modeling}. We just show E-modes as B-modes are not measured at any statistical significance. As explained in $\S$~\ref{sect:FLASK}, \texttt{FLASK} simulations are expected to recover the input third moments only at no smoothing; for larger smoothing scales, we expect (and measure) an offset with respect to theoretical predictions, up to $\sim$ 40 per cent for $\theta_0 \sim 200$ \footnote{The relevance of this offset when using \texttt{FLASK} to produce covariance matrices for our measurements is described in \S~\ref{sect:covariance}.}. Here, however, we are interested in how well the mixing matrices deal with the mask effects. To check this, we just need to verify that applying our mixing matrix to a partial-sky simulation (with DES Y3 mask) recovers the same answer as a the full-sky simulation.

This is shown in the lower panel of Fig.~\ref{fig:FLASK_modeling} . For \texttt{FLASK} third moments the theory lines are replaced by the average of the third moments measured in many full-sky realisations. These agree with the average of 1000 DES Y3 (partial-sky) \texttt{FLASK} realisations within $3$ per cent, which is much smaller than the observational uncertainties. We conclude that our mixing matricies deal efficiently with mask effects also for the third moments.

We next consider a more realistic scenario in which shape noise is included. In this case we need to perform the de-noising procedure (eqs.~\ref{eq:deno2}, \ref{eq:deno3}), which subtracts the shape-noise contributions from the measured moments. For the second moments we first checked that the mixed terms ($\langle  \kappa_{\rm rand} \kappa\rangle^{i,j}$ and $\langle  \kappa_{\rm rand} \kappa\rangle^{j,i}$) averaged to zero, while the terms $\langle  \kappa^2_{\rm rand}\rangle^{i,i}$ (corresponding to the noise-only second moments) did not and needed to be subtracted. As for the third moments, we found out that mixed terms of the form $\langle  \kappa \kappa^2_{\rm rand}\rangle^{i,j,k}$ did not vanish for some choice of indices and needed to be subtracted. This is due to source galaxy density -- convergence field correlations that do not vanish at the third order. All the other terms, including $\langle  \kappa^3_{\theta_0,{\rm rand}}\rangle^{i,i,i}$, averaged to zero and did not need to be subtracted.

 The de-noised measurements are shown again in Fig.~\ref{fig:FLASK_modeling} (light blue  shaded regions). The measurements are clearly noisier than the previous case, but we verified that when the averages over the 1000 \texttt{FLASK} realizations are considered, the match with the theory shows the same level of agreement as the noiseless case.

\subsection{Testing second and third moments modelling}
\label{sect:validate_others}
To validate our modelling of the second and third moments we need a full N-body simulation. In particular, we need to validate the E-modes, as they will be used in the cosmological analysis (B-modes have a low signal-to-noise, and they will be mainly used as a diagnostic). To do this, we use 100 realisations of the shear field obtained using \cite{Takahashi2017} simulations. Such realisations do not include non-linear lensing corrections (which are probed in \S~\ref{sect:HR2}). The comparison with the theory (second and third moments, E-modes) is shown in Fig.~\ref{fig:taka_modeling}. In the same figure, we also show the average of the 100 realizations of the DES Y3 footprint. 
For the second moments, the match with the theory is better than 1 per cent at large scales (compatible with the uncertainties in the modelling of mask effects) and it is at the level of 2-3 per cent at small scales (compatible with the accuracy of the simulations at low redshift). The good match at large scales also justify the use of the Limber approximation in our modelling.

For the third moment, the theory matches the measurement to better than 5 per cent at all scales. The modelling at small scales is obtained including the analytical refinement based on N-body, cold dark matter only simulations  \citep{Scoccimarro2001,GilMarin2012}; predictions from the third moments from perturbation theory only would start departing from the T17 measurement at $\sim 30$ arcmin, reaching a disagreement of 80 per cent at 5 arcmin in the first tomographic bin. This analytical refinement comes with a modelling uncertainty \citep{VanWaerbeke2001,Semboloni2011,HarnoisDeraps2012,Simon2015}, which is taken into account in our covariance matrix (see \S~\ref{sect:covariance}). We test in \S~\ref{sect:scale_cuts} that this level of accuracy of our modelling is good enough for our cosmological analysis. 

\subsection{Baryonic effects}\label{sect:HR1} 

\begin{figure*}
\begin{center}
\includegraphics[width=0.95 \textwidth]{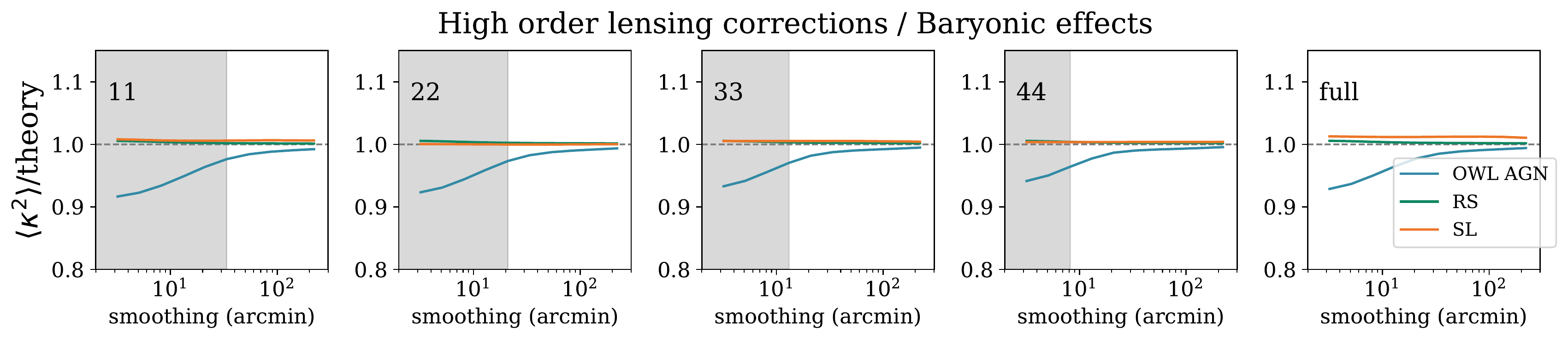}
\includegraphics[width=0.95 \textwidth]{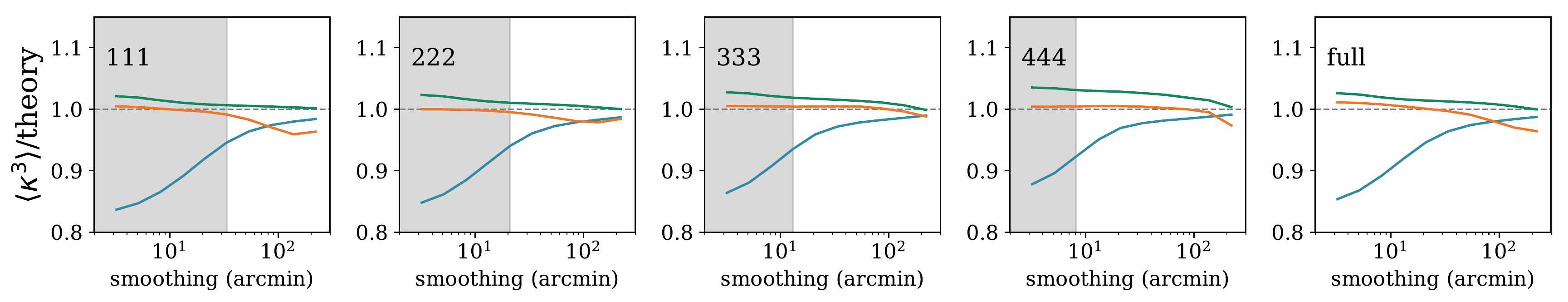}
\end{center}
\caption{Impact of baryonic effects (from OWLS simulations) and two non-linear lensing corrections to measured (E-modes) moments. The blue line (OWL) refers to the rescaled moments including baryonic contributions from AGN feedback. The orange line (label $\rm RS$) shows the contribution to reduced shear correction. The green line refers to source-lens clustering (label $\rm SL$). The grey shaded regions represent the angular scales cut out from the analysis (see \S ~\ref{sect:scale_cuts}; as the scales cut is determined only for the tomographic version of the data vector, we do not show any shaded region for the non-tomographic case).}
\label{fig:high_order}
\end{figure*}

We discuss in this and in the next subsections the impact of a number of effects not included in our modelling. Ultimately, the impact of these effects (together with the comparison with T17 sims from the previous section) will directly determine the scales to be used in the cosmological analysis. 

We consider here the possible contamination of our data vector by baryonic feedback effects at small scales. Baryonic effects are modelled using the results from the OWLS ``AGN'' simulations \citep{Schaye2010,vanDaalen2011} in similar fashion to what has been done in the DES Y1 cosmic shear analysis \citep{Troxel2017}. 
As done in DES Y1, we rescale the power spectrum so as to include contribution from the OWLS ``AGN'' sub-grid prescriptions:

\begin{equation}
\label{eq:owls}
P_{\rm NL} (k,z) \rightarrow \frac{P_{\rm DM + baryons}}{P_{\rm DM}}P_{\rm NL} (k,z),
\end{equation}
where $P_{\rm DM}$ is the OWLS power spectrum due to dark matter, and $P_{\rm DM + baryons}$ is the OWLS power spectrum including the ``AGN'' feedback prescription. Applying such correction to the power spectrum should account for most of the baryonic effects on the third moments as well (\citealt{Foreman2019} shows that baryonic contributions to the bispectrum goes as $P_{\rm DM + baryons}^2/P_{\rm DM}^2$, at least for the scales under study here).

Deviations from our theoretical modelling due to baryons are shown in Fig.~\ref{fig:high_order}. The OWLS power spectrum dampens the measured moments at small smoothing scales, whereas the effect is almost negligible at larger scales. From Fig.~\ref{fig:high_order} it is hard to compare the smoothing scales at which the OWLS power spectrum starts affecting the moments with the angular scales used in the DES Y1 cosmic shear analysis \citep{Troxel2017}, as the two probes get contributions from the high multipoles in harmonic space differently.

We note that OWLS suite is not the only set of simulations including baryonic effects  (see, e.g. EAGLE simulation \citealt{Hellwing2016}, IllustrisTNG simulations \citealt{Springel2018}, Horizon simulations \citealt{Chisari2018}), but due to its large impact on the DM power spectrum using it is generally considered a conservative choice. 

\subsection{Higher-order lensing corrections}\label{sect:HR2} 

We next verify the impact of a number of higher-order lensing corrections to our theoretical modelling \citep{Schneider1998,Schneider2002b,Schmidt2009,Krause2009}. We look at three different effects: reduced shear, source-lens clustering and magnification bias. The first is due to the fact that we cannot directly observe the shear field, but rather we observe the reduced shear (Eq. \ref{eq:reduced_mm}). Source-lens clustering is due to the correlation between source galaxies and lensing potentials along the line-of-sight. The convergence field traces the integrated density contrast up to the position where the sources are detected. Since we estimate the convergence field from an ensemble of sources at different redshifts, and the source galaxies are not uniformly distributed along the line-of-sight, this affects the estimated convergence values. The effect is enhanced in case of broad redshift distributions. We note that fluctuations in the density field are also caused by magnification effects (magnification bias).

In order to model the reduced shear contribution, we start from Eq.~\ref{eq:reduced_mm} and note that in the weak lensing limit $1/(1-\kappa) \sim 1+\kappa$. It follows that the observed shear has an additional contribution that can be modelled as:
\begin{equation}
\label{A1}
\gamma_{\rm obs} \rightarrow \gamma(1+\kappa).
\end{equation}
Source-lens clustering and magnification effects can be modelled by accounting for the effect of the density fluctuations along the line-of-sight when estimating the shear field:
\begin{equation}
\label{A2}
\gamma_{\rm obs} \rightarrow \gamma(1+\delta_{\rm obs}),
\end{equation}
where the $\delta_{\rm obs} \equiv 1-N_{\rm obs}/\avg{N}$ is the estimated density contrast ($N_{\rm obs}$ is the number of galaxies along the line-of-sight and $\avg{N}$ is the average number of galaxies). The fluctuations in the density field are due source galaxies overdensities and lensing magnification effects. Lensing magnification enhances the flux of  galaxies  and this can  locally increase the number density, as more galaxies pass the selection cuts/detection threshold of the sample; at the same time,  the  same  volume  of  space  appears  to  cover  a  different  solid  angle  on  the  sky, causing  the  observed number density to decrease. 
%
%
%
%
At first order, the impact of source galaxies overdensities and lensing magnification effects can be modelled as:
\begin{equation}
\delta_{\rm obs} =\delta_{\rm gal} + q\kappa,
\end{equation}
with $q$ expected to be the order of unity (see \citealt{Schmidt2009} for an approximate description of the term $q$).  Summing up Eqs. \ref{A1} and \ref{A2}:
\begin{equation}
\label{eq:highorder}
\gamma_{\rm obs} = \gamma[1+\delta_{\rm gal}+(1+q)\kappa].
\end{equation}

Reduced shear contributes as $\approx 1+\kappa$, magnification effects as $\approx 1+q\kappa$, lens-source clustering as $\approx 1+\delta_{\rm gal}$. To test the impact of these effects, we used \cite{Takahashi2017} simulations. Using the full-sky spherical harmonics approach laid out in \S~\ref{sect:Map Making}, we generated for every redshift layer of the simulations: 1) shear field $\gamma$ distributions starting from the convergence maps $\kappa$; 2) shear field distributions with $1+\kappa$ and $1+\delta_{\rm gal}$ contributions (Eqs.~\ref{A1} and \ref{A2}); 3)  density contrast field  distributions $1+\delta_{\rm obs}$. We then stacked the redshift layers together according to the redshift distributions of the WL tomographic bins, and generated the following maps:

\begin{equation}
\label{eq:no_map}
\avg{\gamma}_{\rm pix(\theta)} \approx \frac{\int dz n(z) \gamma(z,\theta)}{\int dz n(z)},
\end{equation}

\begin{equation}
\label{eq:rs_map}
\avg{\gamma}^{RS}_{\rm pix(\theta)} \approx \frac{\int dz n(z)(1+\kappa(z,\theta)) \gamma(z,\theta)}{\int dz n(z)},
\end{equation}

\begin{equation}
\label{eq:sl_map}
\avg{\gamma}^{SL}_{\rm pix(\theta)} \approx \frac{\int dz n(z)(1+\delta(z,\theta)) \gamma(z,\theta)}{\int dz n(z)(1+\delta(z,\theta))}.
\end{equation}

Eqs. \ref{eq:no_map}, \ref{eq:rs_map} and \ref{eq:sl_map} are, respectively, the shear fields with no non-linear lensing corrections, with reduced-shear contributions and source-lens clustering. As for the latter, we divided by the integrated density field to mimic the map making process, where each pixel contains the average of the shear field along the line of sight.

The impact of such corrections on E modes are shown in Fig.~\ref{fig:high_order}. We estimated the moments from a full-sky, noise-free realization of the simulation. For the reduced shear and source-lens clustering we considered as a ``theory'' the moments estimated from the same realization of the simulations using Eq.~\ref{eq:no_map} to estimate the shear field. We do not show error bars for the moments measurement as we expect them to be much smaller than DES Y3 uncertainties\footnote{First, since we are considering the full area, we expect the covariance of the moments measurement to be roughly $\approx 8$ times smaller. Second, as we are using the moments of the same realization with no non-linear lensing corrections as the ``theory'', we can expect the measurements to be highly correlated, and the uncertainties in their ratio should be very small.} . We also do not show magnification effects as they are of the same order as the reduced-shear correction (assuming $q$ of the order of unity). We find that these non-linear lensing corrections are much smaller then DES Y3 uncertainties and sub-dominant with respect to baryonic effects (except for source-lens clustering at very large scales of the third moments, but we do not expect this level of contamination to bias our cosmological inference).

\section{Covariance and Likelihood}\label{sect:like_tt}

\subsection{Covariance estimation}\label{sect:covariance}

\begin{figure}
\begin{center}
\includegraphics[width=0.45 \textwidth]{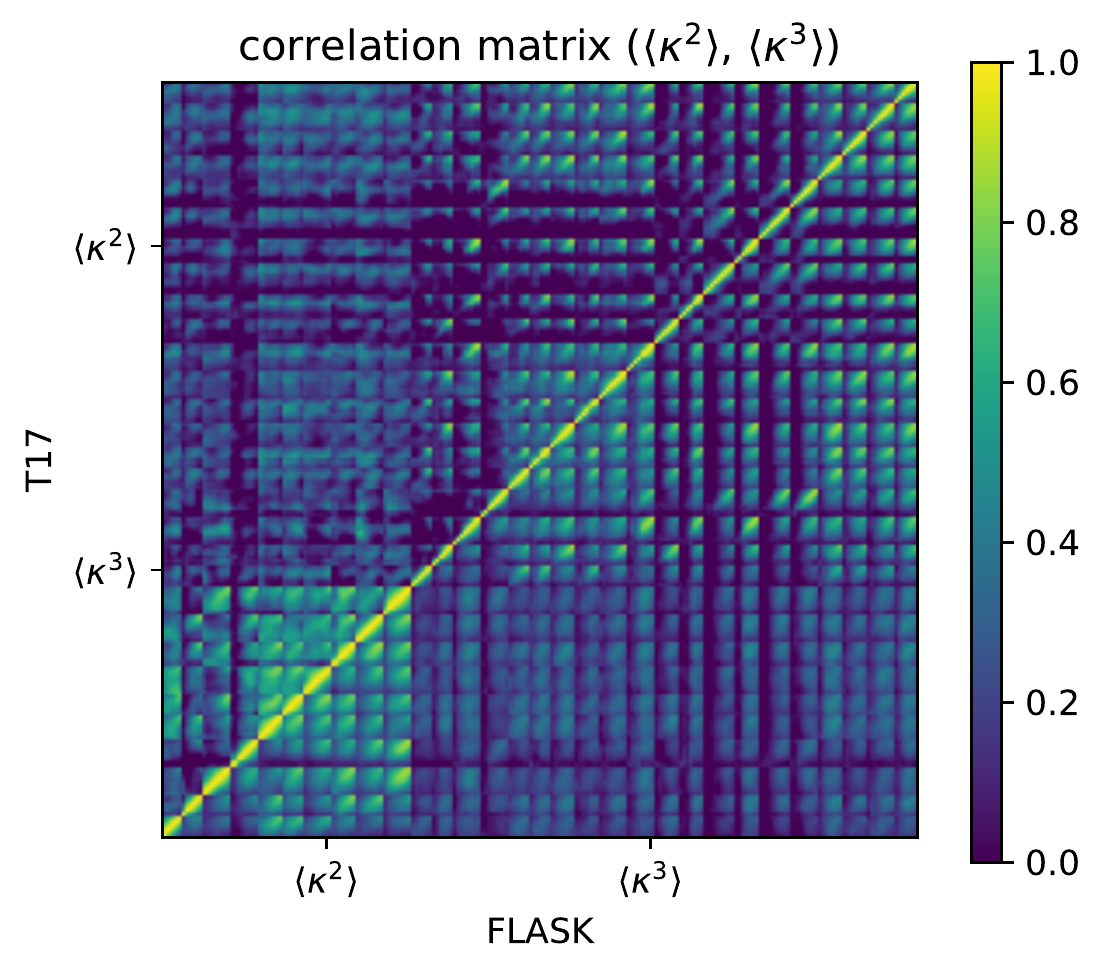}
\end{center}
\caption{Measured correlation matrix of second and third moments from 1000 \texttt{FLASK} simulations (lower right corner) and from 100 T17 simulations (upper left corner). A $12 h^{ - 1}$ Mpc scale cut has been applied (see \S~\ref{sect:scale_cuts} for a definition of the scale cuts).}
\label{fig:covariance}
\end{figure}

\begin{figure*}
\begin{center}
\includegraphics[width=0.9 \textwidth]{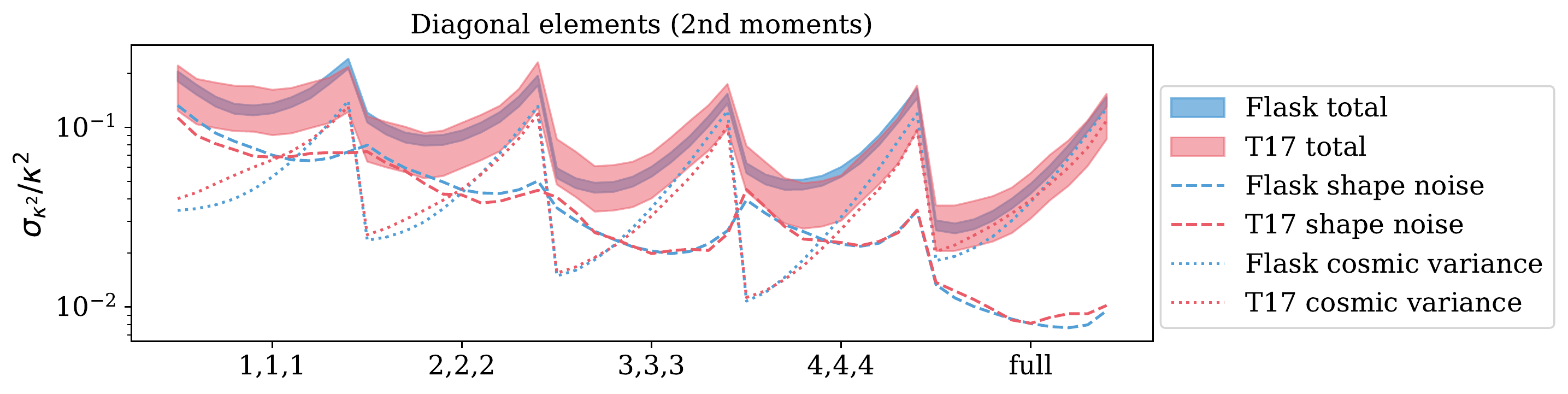}
\includegraphics[width=0.9 \textwidth]{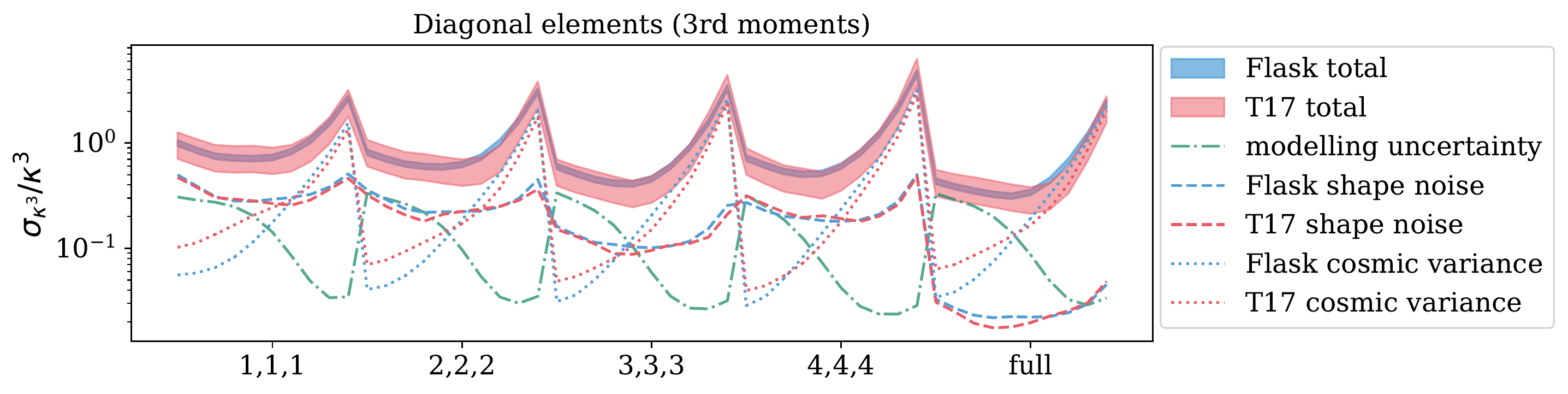}
\end{center}
\caption{Diagonal elements of the covariance of the second and third moments  estimated from \texttt{FLASK} simulations. The x-axis shows the corresponding data vector entries. We show  separate contributions due to shape noise and cosmic variance. For the third moments, we also show the modelling uncertainties related to the small scales analytical fitting formulae. For comparison, we also show the covariance estimated from \citet{Takahashi2017} simulations. 
For the total covariance, we show uncertainty due to the finite number of simulation realisations.}
\label{fig:covariance_diag}
\end{figure*}

To correctly infer cosmological parameters from our data, we need an accurate estimate of the measurement uncertainty. We estimate the covariance from 1000 independent realisations of the \texttt{FLASK} simulation. For each \texttt{FLASK} realisation, we measure the second and third moments of the smoothed convergence field as explained in \S~\ref{sect:validate_mask}. We then build our covariance matrix as:

\begin{equation}
\hat{{C}} = \frac{1}{\nu}{\sum_{i=1}}^{N_s} (\hat{{d}}_i-\hat{{d}}){(\hat{{d}}_i-\hat{{d}})}^{T},
\end{equation}
where $\nu = N_s-1$ with $N_s$ the number of realisations, $\hat{{d}}_i$ the data vector measured in the $i$-th simulation and $\hat{{d}}$ the sample mean. The data vector is made of a combination of second and third moments as measured at different smoothing scales. 

Within single realisations, variations in the measured moments among different simulations are mostly due to two different contributions: 1) a combination of galaxy intrinsic shape and measurement noise, or ``shape noise'', and 2) the cosmic density field inside the DES Y3 footprint is a random realisation of the underlying cosmology, or ``cosmic variance''. For third moments, we also include in our covariance a ``modelling uncertainty'' related to the analytical fitting formulae describing the third moments at small scales. The modelling uncertainty is taken to be equivalent to the scatter between \cite{Scoccimarro2001} and \cite{GilMarin2012} predictions at small scales, scatter that can reach $\sim 20$ per cent at $\sim 5 $ arcmin for the first tomographic bin (see Appendix~\ref{sect:Skewness}).

The measured correlation matrix is shown in Fig.~\ref{fig:covariance}. The matrix has been obtained applying a scale cut of $12 h^{ - 1}$ Mpc, so different tomographic bins have a different number of angular scales included in the data vector. Values at different smoothing scales but for the same moment are highly correlated. Fig.~\ref{fig:covariance} also shows that second and third moments are not very correlated. This is mostly due to shape noise and third moment modelling uncertainties at small scales that wash out existing correlations.

The values of the diagonal elements of the covariance matrix, relative to values of their data vector entries, are shown in Fig.~\ref{fig:covariance_diag}, for both \texttt{FLASK} and T17 simulations. We also show the errors due to the finite number of simulation realisations. One can see that for both second and third moments the intermediate scales are the ones with better signal-to-noise, and that in general second moments have a much better signal-to-noise than third moments.

The sample variance part of the covariance is cosmology-dependent and dominates at large scales. We do not expect this cosmological dependence to significantly impact the recovery of cosmological parameters (see discussion in \S~\ref{sect:scale_cuts}).  We also note here that the lognormal approximation assumed by \texttt{FLASK} needs to be checked for the sample variance part of the covariance for third moments. However, the scales dominated by sample variance have a smaller signal-to-noise smaller for the third moments; moreover, despite \texttt{FLASK} limitations, Fig.~\ref{fig:covariance_diag} shows that the \texttt{FLASK} and T17 covariances agree within uncertainties. In Appendix ~\ref{sect:data-compression} we provide further evidence that the uncertainties in the modelling of the third moments covariance have little effect on the cosmological inference.

We therefore decided to rely on \texttt{FLASK} simulations to build our fiducial covariance because the cosmological parameters can be easily changed and we can produce a large number  of simulations. The T17 simulations have a fixed cosmology, and, above all, are limited in numbers, causing the inverse of the covariance matrix to be extremely noisy. However, in the next section we show an implementation of a data-compression algorithm that greatly reduces the size of the data vector (and the noise in the covariance due to the paucity of simulations). The data compression algorithm is implemented in our fiducial analysis and in principle allows us to run our cosmological pipeline also using the T17 covariance (although it will be still noisier than the \texttt{FLASK} covariance). While we still use \texttt{FLASK} as our fiducial covariance, we show in Appendix ~\ref{sect:data-compression} that the differences in the recovered cosmological parameters between using the T17 covariance (in combination with the data compression algorithm) or \texttt{FLASK} covariance are small.

\subsection{Data-compression}\label{sect:data-compression_main}



To reduce the noise in our covariance matrix estimated from \texttt{FLASK} mocks, we implement the MOPED data-compression algorithm \citep{Tegmark1997,Heavens2000,Gualdi2018}. We follow \cite{Heavens2000} and include a data-compression scheme based on the Karhuned-Lo\`eve algorithm. The algorithm works by assigning  weights to each element of the data vector  that are proportional to the sensitivity of the element to the variation of a given model parameter. In such a way, it is possible to reduce the dimension of the data-vector to the number of model parameters considered. The compressed data vector can be written as:
\begin{equation}
\label{eq:data_compression}
    d^{\rm compr}_{i} = \avg{d}^{T}_{,i} \hat{{C}}^{-1} d \equiv b_i d ,
\end{equation}
where $x$ is the full-length data vector, $\hat{{C}}$ is the measurement covariance and $d^{\rm compr}_{i}$ is the $i$-th element of the compressed data vector. The index $i$ refers to the $i$-th model parameter $p$ considered, and $\avg{d}^{T}_{,i}$ is the derivative of the model data vector with respect that parameter.

The above equation assumes that the dependence of the covariance on cosmological parameters is mild ($\partial\ln C/\partial\ln p_i<<1$). While being reasonable, we do not explicitly prove the latter assumption as it would require producing many covariance matricies, which is computationally expensive. We also note that for the compression algorithm to be lossless, the likelihood of the non-compressed data vector must be Gaussian. We check this in Appendix~\ref{sect:data-compression}, and we show that the uncompressed data vector shows only mild deviations from Gaussianity. We note, however, that we expect the compressed data vector to have a more Gaussian distribution, due to the central limit theorem \citep{Heavens2017}. We show this in \S~\ref{sect:scale_cuts}.

In general, if one or more assumptions underlying the data-compression algorithm are violated, we can expect the compression to be not optimal. In this case the credible regions would  be larger than they could be \citep{Heavens2017,Alsing2018}, but the parameter inference would still be valid. 

To implement the algorithm described in Eq. \ref{eq:data_compression}, we use the \texttt{FLASK} covariance, and we estimate the derivative of the data vector using a 5-point stencil derivative centred on the true value of the simulation parameters. As the model parameters we use the five cosmological parameters and all the nuisance parameters as described in $\S$~\ref{systematics}. The compressed covariance can be easily obtained as:
\begin{equation}
 \hat{{C}}_{ij}^{\rm compr} = b_i^T \hat{{C}} b_j.
\end{equation}
%
%
We defer the validation of the compression algorithm to  Appendix~\ref{sect:data-compression}, where we compare the posterior distributions obtained with and without the data-compression algorithm. In general, we find smaller contours for the chains run with the compressed data vector, as expected by the lower noise in the covariance (we explain how we deal with the noise in the covariance in the next section).

\subsection{Data vector and likelihood}\label{sect:like}

The final data vector includes all the ``auto'' moments of different tomographic bins (e.g., $[1,1], [1,1,1], [2,2], [2,2,2]$) and the ``cross'' moments (e.g., $[1,2], [1,1,2], [1,2,2] $), for a total of 10 combinations for second moments and 20 combinations for third moments. The scale cuts are discussed in the next section.

We evaluate the posterior of the parameters conditional on the data by assuming a Gaussian likelihood for the data, i.e.
\begin{equation}
\label{eq:ll}
- 2 \ln \mathcal{L} = f_2 f_1 [\hat{d}-M(p)] \hat{{C}}^{-1}[\hat{d}-M(p)]^T
\end{equation}
 (see $\S$ \ref{sect:scale_cuts} for an investigation of this assumption). Here $M(p)$ is our theoretical model, $\hat{d}$ is the data vector, and $\hat C^{-1}$ is the inverse of our covariance estimate. The posterior is then the product of the likelihood and the priors. 
 Note that the quantities $M(p)$, $\hat{d}$ and $\hat C^{-1}$ in Eq.~\ref{eq:ll} are to be considered compressed quantities, and we have dropped the superscript ``${\rm compr}$'' for brevity. The terms $f_1$ and $f_2$ account for noise introduced when the covariance matrix  is estimated from random realisations of the data. Even if a covariance estimate $\hat C$ from $N_{sims}$ random realisations is an unbiased estimate of the true covariance of the data, its inverse $\hat C^{-1}$ is only a biased estimate of the true precision matrix $C^{-1}$ \citep{Hartlap2011}. This bias can be corrected with the multiplicative factor
\begin{equation}
\label{eq:hartlap}
f_1 = \frac{N_{sims}-N_{data}-2}{N_{sims} -1} , 
\end{equation}
where in our case the number of independent realisations used to estimate the covariance is $N_{sims} = 1000$, and $N_{data}$ is the length of the data vector. Note that this is just an approximate treatment of the noise in the covariance matrix, since the data likelihood depends on the precision matrix in a non-linear way. \citet{Sellentin2016} have devised a more accurate treatment, taking into account the impact of the covariance estimation noise on the entire likelihood. We investigate their alternative likelihood in Appendix \ref{sect:data-compression} and find that after our data compression it has a negligible effect.

There is a second - and often more severe - problem in estimating the likelihood of data from a finite number of random realisations that is not solved by the likelihood of \citet{Sellentin2016}. This problem is that the noise in a covariance estimate does not just change the width of parameter contours but also their location (\citealt{Dodelson2013}, see also figure 1 in \citealt{Friedrich2018b} for a simple demonstration of the effect). An approximate way to take this into account is to multiply our log-likelihood by
\begin{equation}
\label{eq:DS}
f_2 = \left[1 + \frac{(N_{\rm data}-N_{\rm par})(N_{\rm sims}-N_{\rm data}-2)}{(N_{\rm sims}-N_{\rm data}-1)(N_{\rm sims}-N_{\rm data}-4)}\right]^{-1}\ .
\end{equation}
This correction (dubbed Dodelson-Schneider-factor by \citealt{Friedrich2018b}) assumes the model to be linear in all the parameters and widens the contours to encompass the additional noise inhibited by the maximum likelihood parameter estimates due to the noise in the parameters \citep{Dodelson2013}. We note that as the data-compression greatly reduces the length of the data vectors, $f_1$ and $f_2$ become close to 1.

 To sample the posteriors of our parameters, we  use an EMCEE sampler \cite{Foreman-Mackey2013} and we test the convergence of our chains with the \cite{GelmanRubin1992} test. 
 
For the cosmological parameters, we assume a flat $\Lambda$CDM cosmology and vary five parameters: $\Omega_{\rm m}$ (the matter density in units of the critical density), $\Omega_{\rm b}$ (the baryonic density in units of the critical density), $\sigma_8$ (the amplitude of structure in the
present day Universe, parameterised as the standard deviation of the linear overdensity fluctuations on a 8$h^{ - 1}$ Mpc scale), $n_s$ (the spectral index of primordial density fluctuations) and $h$ (the dimensionless Hubble parameter). We assume wide flat priors on $\Omega_m$ and $\sigma_8$ and adopt the informative priors in $h$, $n_s$ and $\Omega_b$ that have been used the DES Y1 2-point function analysis (see Table \ref{table1}). 
When constraining cosmological parameters, we marginalise over nuisance parameters describing photo-$z$ uncertainties, shear biases and IA effects in our measurements. The modelling of our nuisance parameters is described in $\S$~\ref{systematics}. As at the time of finishing this work, the DES Y3 priors were not finalised yet, so we again assume DES Y1 priors for all the nuisance parameters (priors are summarised in Table \ref{table1}). Photo-$z$ uncertainties are parametrised by a shift in the mean of the distribution (one for each tomographic bin). Priors for the shifts come from redshift distributions of a matched sample of galaxies in the COSMOS survey and angular cross correlation with \redmagic galaxies \citep{Hoyle2018}. Multiplicative shear bias priors are described in \cite{shearcat}. We also assume wide flat priors for intrinsic alignment.

\begin{figure}
\begin{center}
\includegraphics[width=0.5 \textwidth]{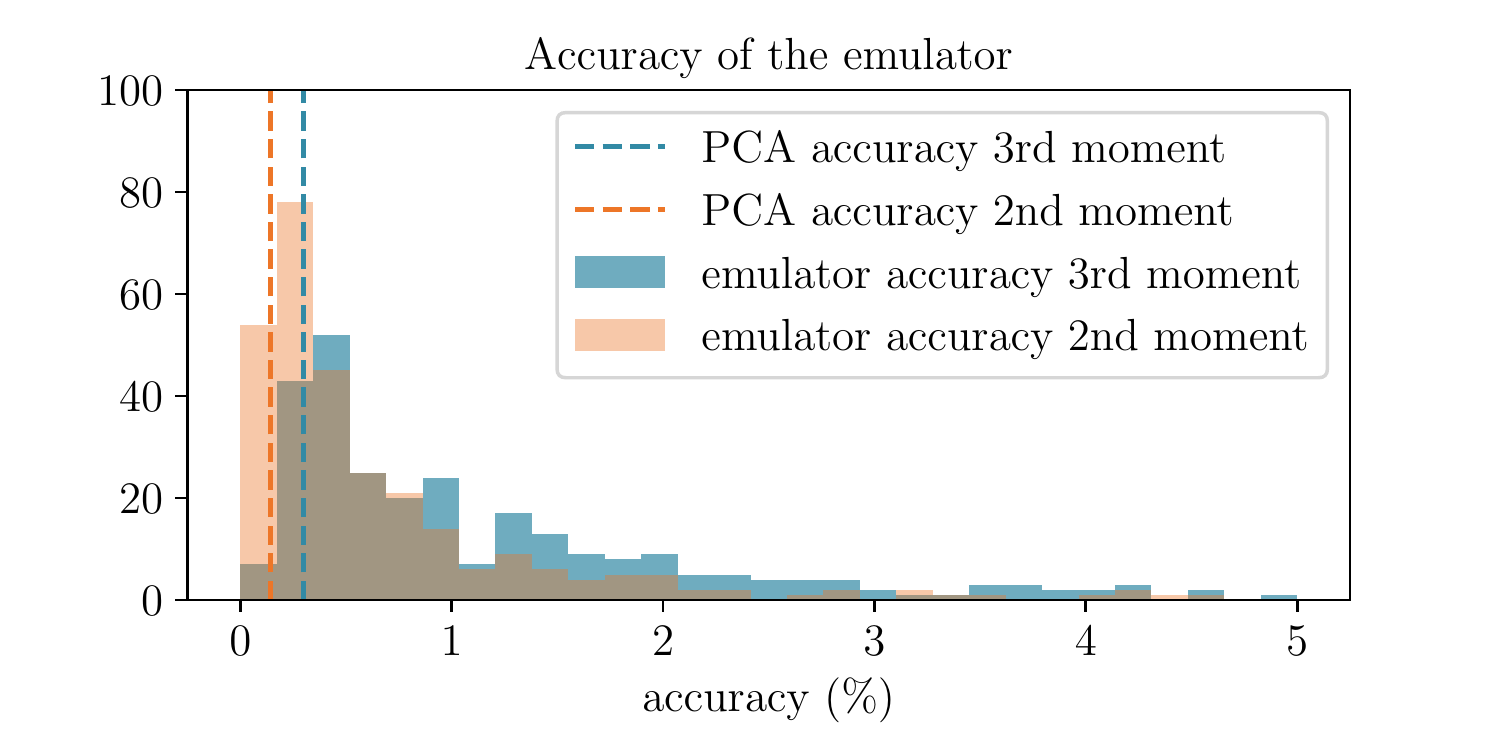}
\end{center}
\caption{Accuracy of the emulator for the second and third moments. We tested the emulator using a validation sample of 500 points. Each entry of the histogram refers to the maximum relative discrepancy between the emulator predictions and the validation model over all the smoothing scales and redshifts considered. The vertical dashed lines show the error introduced by selecting only 15 principal components.}
\label{fig:emu1}
\end{figure}

\begin{figure*}
\begin{center}
\includegraphics[width=0.9 \textwidth]{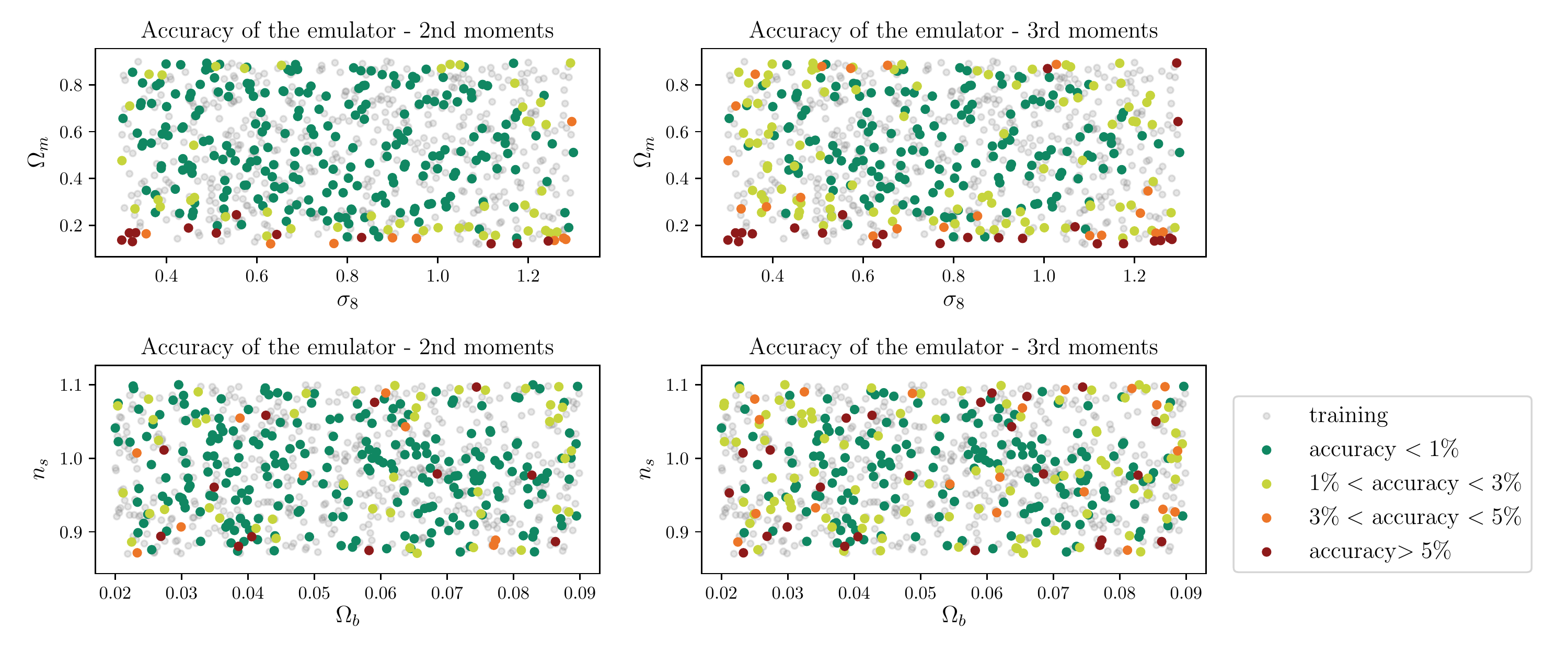}
\end{center}
\caption{Accuracy of the emulator for the second and third moments as a function of four cosmological parameters. We tested the emulator using a validation sample of 500 points. Each entry of the scatter plots refers to the mean relative discrepancy between the emulator predictions and the validation model over all the smoothing scales and redshifts considered. Training points are shown in grey.}
\label{fig:emu2}
\end{figure*}

\subsection{Fast theory predictions}
\label{sect:emu}
The theory prediction described in \S~\ref{sect:theo} can be quite time-consuming due to the large number of cross-correlations and integrations involved. In order to speed up this calculation, we implemented an emulator \citep{Heitmann2006,Habib2007}. Typically, emulators in the cosmology context are used when an expensive calculation is needed in a large parameter space, but the variation of the calculation over the parameter space is smooth. A primary example is predicting the dark matter power spectrum given cosmological parameters \citep{Heitmann2009,Kwan2015}. The power spectrum is computed to high accuracy at a given number of points in the cosmological parameter space and an interpolator is used to derive the power spectrum at some arbitrary point in the parameter space. 

In our case, the quantities we wish to emulate are the second and third moments of the matter power spectrum once the mask edges are accounted for, namely $\langle \delta^2_{\theta_0,NL}\rangle^{EE/BB}  (\chi)$ and $\langle \delta^3_{\theta_0,NL}\rangle^{EE/BB}  (\chi)$. We first compute their values at specific points in our parameter space, and then we build an interpolator that provides fast prediction at any point of the parameter space.

To decide which points to use for build the interpolator, we sampled our parameter space using a Latin hypercube \citep{McKay1979}, which is a scheme that provides good space-filling properties. We sampled the space delimited by the priors defined in Table \ref{table1}, and chose 500 points. For each point of the Latin hypercube, we predicted the second and third moments of the dark matter density field (Eqs. \ref{eq:dd2}, \ref{eq:dd3}) with a resolution of $\delta z = 0.01$ up to redshift 4, for 12 equally logarithmic spaced smoothing scales between $\theta_0 = 0$ arcmin and $\theta_0 = 220$ arcmin. For each smoothing scale, we organised the predictions of our second and third moments in a matrix of dimensionality $n_z \times n_{points} = 400 \times 500$. Since interpolating a $400 \times 500$ matrix as a function of cosmological parameters would be impractical, we further reduce the dimensionality using the singular value decomposition. We define $\eta  = \textbf{UBV}^T$, where $\textbf{U}$ has dimensionality $n_z\times n_z$ and $\textbf{V}$ $n_z \times n_{points} $. $\textbf{B}$ is a diagonal matrix of singular values. We defined the basis vectors $\Phi = \frac{1}{\sqrt{n_z}} \textbf{UB}$ and weights $\omega = \sqrt{n_z} \textbf{V}^T$. Then, we kept only the first $p < n_z$ principal components of our basis vectors:
%
%
\begin{multline}
\langle \delta^2_{\theta_0,NL}\rangle^{EE/BB}  (\chi(z),\Omega_m,\Omega_b,\sigma_8,n_s,h_{100}) = \\\sum_{i=0}^{p} \omega_i^{\delta^2,\theta_0} (\Omega_m,\Omega_b,\sigma_8,n_s,h_{100}) \Phi_{i}^{\delta^2,\theta_0}(\chi(z)),
\end{multline}
\begin{multline}
\langle \delta^3_{\theta_0,NL}\rangle^{EE/BB}  (\chi(z),\Omega_m,\Omega_b,\sigma_8,n_s,h_{100}) = \\\sum_{i=0}^{p} \omega_i^{\delta^3,\theta_0} (\Omega_m,\Omega_b,\sigma_8,n_s,h_{100}) \Phi_{i}^{\delta^3,\theta_0}(\chi(z)),
\end{multline}
where the basis and weights are different for the second and third moments and depends on the smoothing scale. We found that setting $p=15$ and $p=45$ retains most of the information in the moments (99.9 per cent and 99.7 per cent for second and third moments respectively), so we can neglect the other components. The third moments require more components due to the complex dependence on cosmological parameters at small scales.

After the singular value decomposition, we are left to interpolate, as a function of five cosmological parameters, 60 weight functions in total between  $\omega_i^{\delta^2,\theta_0}$ and $\omega_i^{\delta^3,\theta_0}$ measured at 500 different points in our parameter space. We opted for a Gaussian process \citep{Rasmussen2006} interpolation scheme. A Gaussian process is a stochastic process where any finite subset forms a multivariate Gaussian distribution. At each reconstruction point $x = (\Omega_m,\Omega_b,\sigma_8,n_s,h_{100})$ of our parameter space, the weights $\omega_i^{\delta^2,\theta_0}$,$\omega_i^{\delta^3,\theta_0}$ are modelled as multivariate Gaussian distributions with a given mean value and Gaussian errors. The latter is determined by a covariance function $k(x; x')$ that correlates the function at different points. The covariance function depends on only two hyper-parameters (the amplitude and the typical scale of the correlation) which are fixed during the training phase.

We tested the accuracy of our interpolation scheme by training and validating over two different sets of 500 points determined using two different Latin hypercubes. The resulting accuracy is shown in Figs~\ref{fig:emu1} and \ref{fig:emu2}. The performance of the emulator is generally better than $1$ per cent. The recovery gets worse close the the edges of the priors. This is particularly evident for $\sigma_8$ and $\Omega_{\rm m}$ (Fig.~\ref{fig:emu2}) as these two are the parameters to which our measurement is most sensitive. The emulator performs slightly worse for the third moment, due to a more complex dependence on the cosmological parameters. We note that Figs.~\ref{fig:emu1}, \ref{fig:emu2} report the mean accuracy of the emulator across smoothing scales and redshifts. While for the second moments the accuracy does not strongly depend on the smoothing scales or redshift, we found that the emulator for the third moments performs slightly worse at low redshift and intermediate scales, where the accuracy is around $\sim 3$ per cent, still well below observational uncertainties. The speedup achieved by using the emulator is of two orders of magnitudes.

\begin{figure}
\begin{center}
\includegraphics[width=0.4 \textwidth]{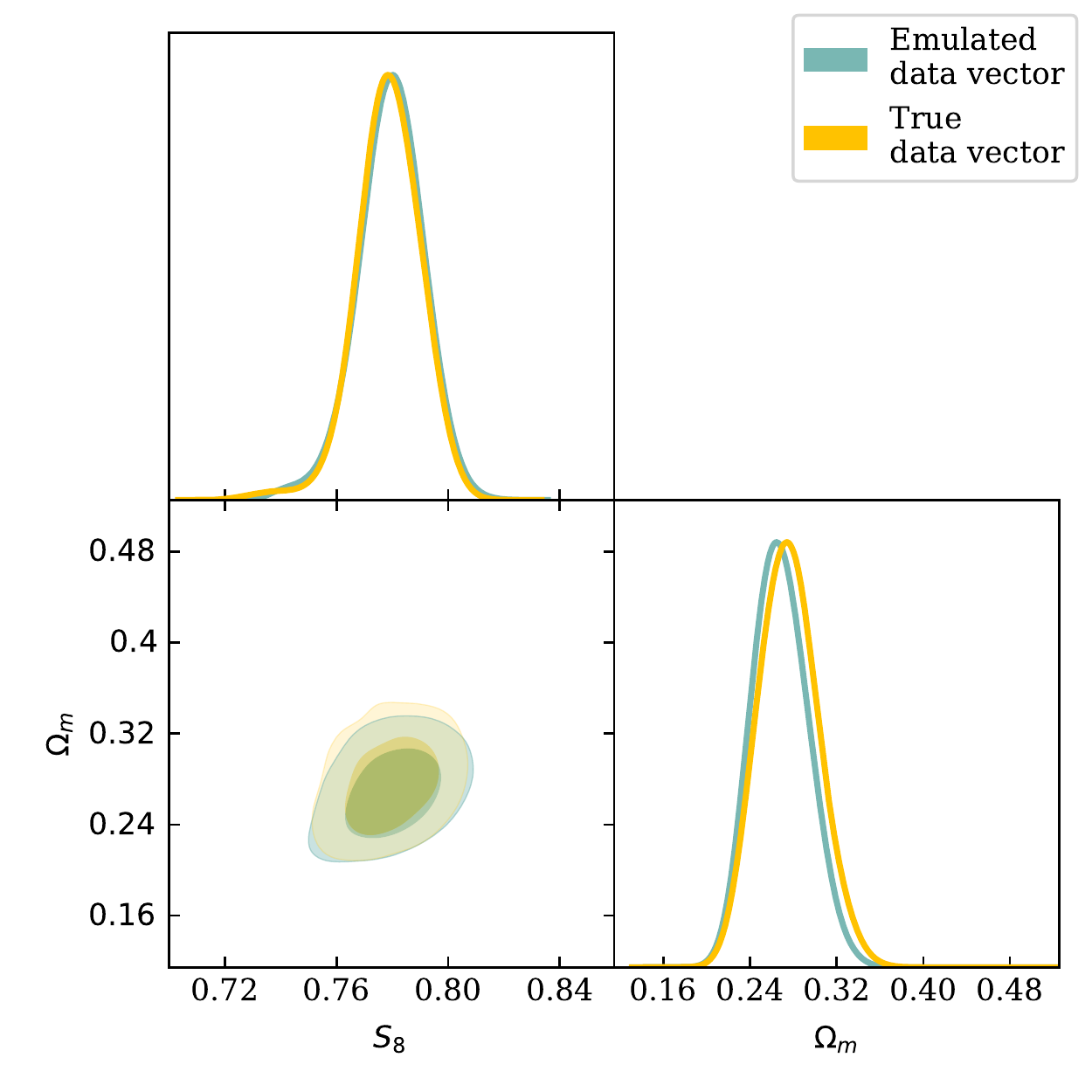}
\end{center}
\caption{Forecast posteriors for cosmological parameters, obtained with a theory data vector and an emulated data vector (see \S~\ref{sect:emu}). We marginalise over nuisance parameters as explained in \S \ref{sect:like}. Constraints with the second and third moments combined are shown in the $S_8 - \Omega_{\rm m}$ plane.}
\label{fig:emu_chain}
\end{figure}

After predicting the masked second and third moments of the dark matter density field with the emulator, we took into account the lensing kernel of the samples and the nuisance parameters as described in $\S$~\ref{sect:theo}. We checked that the emulated theory data vector causes small variations in the  $\chi^2$ with respect to a theory data vector obtained without approximations. For the fiducial cosmology, such variations are of the order of $\Delta \chi^2 \sim 0.2 - 0.4$, the exact value depending on the particular scale cut combination of second and third moments considered. We also verified that the difference between the maximum of the 1-D marginalised posterior of the cosmological parameters obtained running a MCMC chain on an emulated theory data vector and on a non-approximated one are much smaller than the parameters' 1-$\sigma$ confidence intervals. This is shown in Fig.~\ref{fig:emu_chain}, and the differences are at the level of $<1.5$ per cent for $\Omega_{\rm m}$ and $<0.3$ per cent for $S_8 = \sigma_8 (\Omega_{\rm m}/0.3)^{0.5}$.

\section{Cosmological constraints from moments of the convergence field}\label{sect:cosmology}

\subsection{Fiducial scale cuts}\label{sect:scale_cuts}

\begin{figure*}
\begin{center}
\includegraphics[width=0.85 \textwidth]{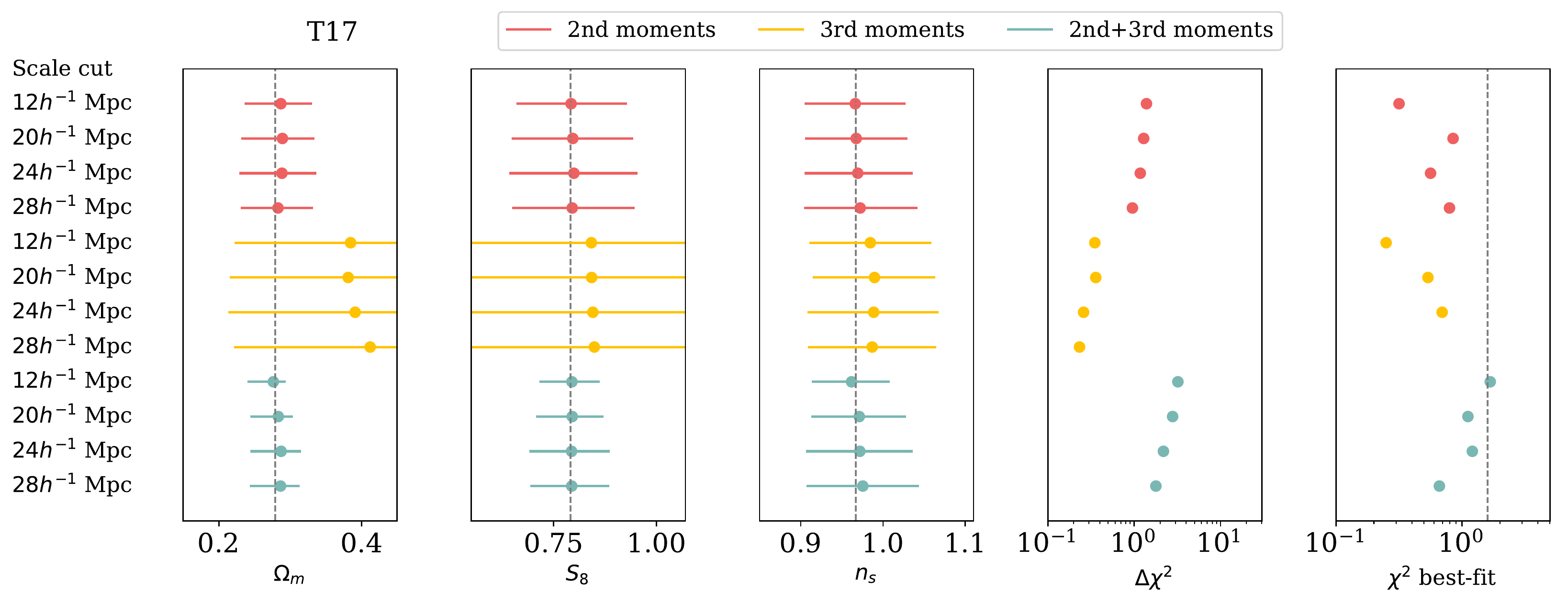}
\includegraphics[width=0.85 \textwidth]{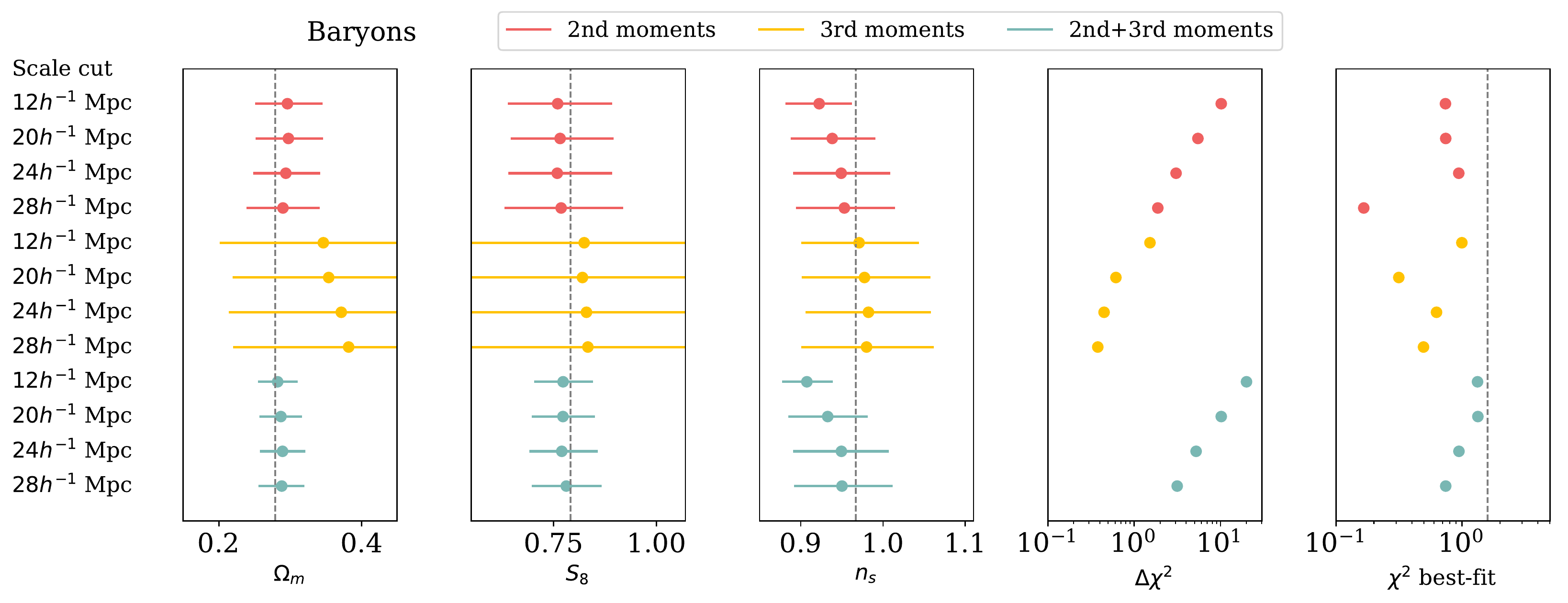}
\end{center}
\caption{The 1-$\sigma$ marginalised constraints on cosmological parameters for a number of different scale cuts.  In the upper plot, the average of 100 T17 simulations have been used as the data vector. In the lower plot, the constraints are obtained by using a theory data vector contaminated with the OWLS AGN power spectrum. Data points represents the mean of the 1-D marginalised posterior, while for the confidence interval we show the two-tail symmetric intervals. The column $\Delta \chi^2$ represents the $\chi^2$ of the data vector contaminated with baryonic effects or from the average of T17 sims with respect to a theory data vector. The  $\chi^2$ best-fit column represents the $\chi^2$ of the best-fitting cosmology from the MCMC chain. The vertical line in the last column marks the $\Delta \chi^2 = 1.6$ boundary.}
\label{fig:scale cuts}
\end{figure*}

\begin{figure*}
\begin{center}
\includegraphics[width=0.85 \textwidth]{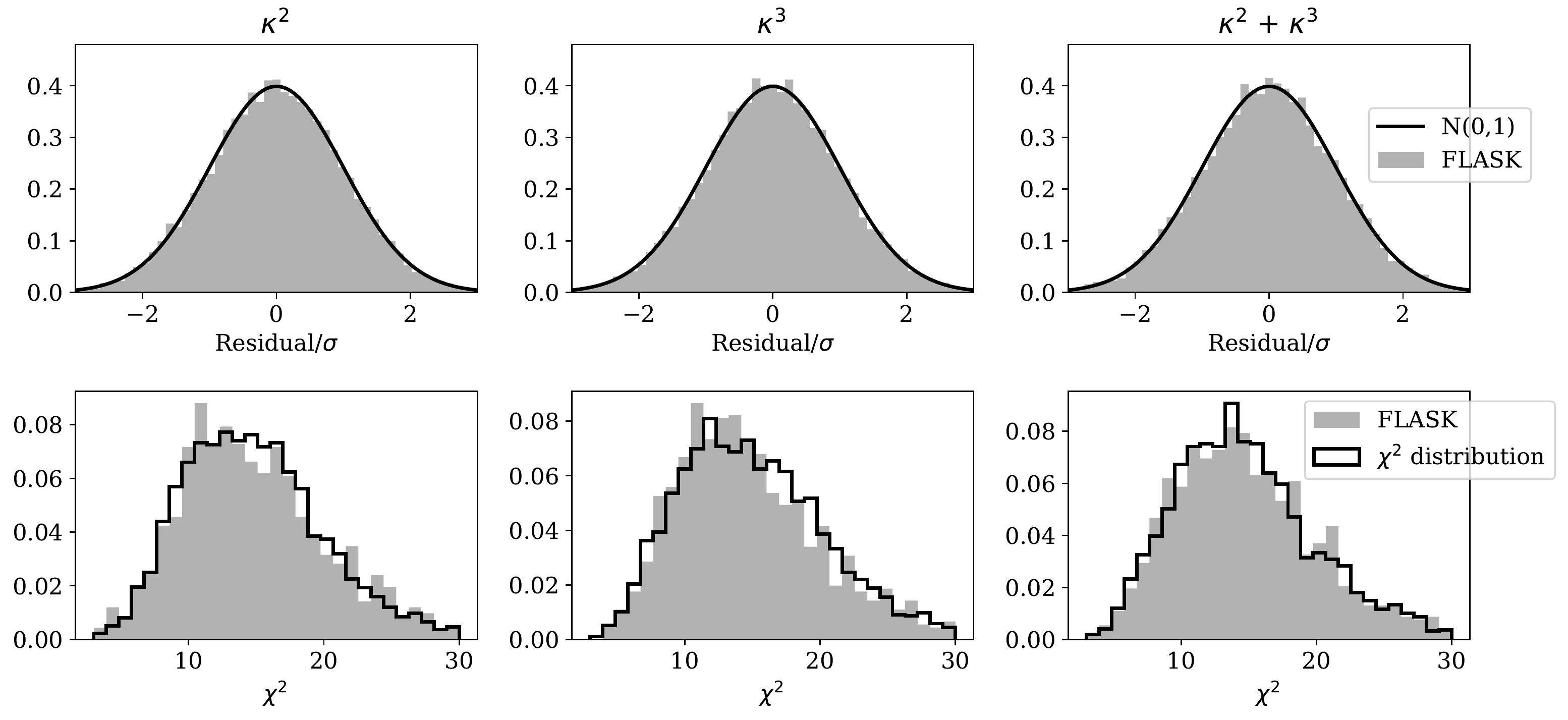}
\end{center}
\caption{\textit{Top panels}: residuals of individual data points
in units of their expected standard deviation for a compressed data vector. We compare to a Gaussian with zero mean and unit standard deviation. \textit{Bottom panels}: Distribution of the $\chi^2$ of each realisation of the FLASK simulations, compared to a theoretical $\chi^2$ distribution.}
\label{fig:residual}
\end{figure*}

The last analysis choice to make before presenting the final cosmological constraints of the second and third moments of the convergence field is which scales are to be used for the analysis. The scale cuts we use are determined based on two tests. First we check that our theoretical modelling is adequate to describe the data vectors as obtained from the average of many N-body simulations from T17; second, we check that the impact of baryons on our data vector is not significant. For both tests we run MCMC chains for different combinations of scale cuts using a data vector from T17 simulations or a baryons-contaminated data vector and require the resulting constraints on cosmological parameters not to be biased against the truth. 

For a combination of scales to be acceptable, we require the mean of the marginalised 1-D posterior of $\Omega_{\rm m}$, $S_8 = \sigma_8 (\Omega_{\rm m}/0.3)^{0.5}$ to be within 0.3 $\sigma$ of the values obtained with a ``theory'' data vector. As we partially constrain $n_s$, we also require the posterior of $n_s$ to be within 0.5 $\sigma$ of the baseline value. 

We also adopt a second criterion on the $\chi^2$ of the best-fit cosmology. When analysing the data, the best-fit $\chi^2$ is used for hypothesis testing and as a proxy of the adequacy of the data vector modelling. A bad best-fit $\chi^2$ implies that either our covariance or the parametrisation of the data vector is not adequate to describe the measurement. Since we do not model baryonic effects or the small discrepancies between our theoretical predictions and the data vector from T17 simulations, we should expect the best-fit $\chi^2$ from the data to be biased. By adopting a criteria on the $\chi^2$ of the best-fit cosmology of the contaminated data vector we make sure the biases from these two effects are small. In particular, we require the $\chi^2$ of the best-fit cosmology obtained from a contaminated data vector to be within $0.3$ of the expected spread of the $\chi^2$ distribution. Therefore, since the length of the compressed data vector is 15, we require the best-fit $\chi^2 < 1.6$. Ideally, for negligible contamination we  expect a best-fit  $\chi^2 = 0$, as we are using a theory data vector as a baseline (whereas using a noisy data vector would give, on average,  $\chi^2 \sim $ d.o.f. ).


In this section, scale cuts are expressed in terms of a specific comoving scale $R_0$; the relation with the smoothing scale $\theta_0$ is given by $\theta_0 = R_0/ \chi(\avg{z})$, where $\chi(\avg{z})$ is the comoving distance of the mean redshift of a given tomographic bin. In the case of moments from different tomographic bins, we took the average of the $\avg{z}$ of the bins.

The tests run on the data vectors obtained from the average of the T17 simulations are shown in the upper plot of Fig.~\ref{fig:scale cuts}. As we estimated the bias in the cosmological parameters induced by the emulator in the previous section, we re-scaled the measured data vector by the ratio between an emulated theory data vector and a non-approximated one predicted at the T17 cosmology. This assumes the emulator uncertainties propagate linearly to the data vector; this is justified as at the T17 cosmology the emulator accuracy is below the per cent level.

Besides emulator inaccuracies (which are handled by the re-scaling), there are different known reasons why the data-vector from the average of T17 might differ from our theoretical predictions: inaccuracies of the simulations or in the modelling of the third moments at small scales (\S~\ref{sect:sims} and \S~\ref{sect:validate_others}), inaccuracies in accounting for mask effects (\S~\ref{sect:validate_mask}), inaccuracies in the covariance modelling (\S~\ref{sect:covariance}), etc. In the past sections we showed (or discussed) these differences to be small at the level of the data vector, but here we want to assess the impact on the cosmological parameters posteriors.

The plots show the marginalised 1-D posterior for three out of five cosmological parameters under study. We do not show constraints for $\Omega_b$ and $h_{100}$ because the posteriors are heavily prior dominated. For each parameter, we show the mean of the posterior and the symmetric 1-$\sigma$ confidence interval. We note that $n_{\rm s}$ is mildly constrained and its posterior is partially dominated by the prior (which is assumed to be flat with $n_s \in [0.87,1.07]$; see Table~\ref{table1}). The constraints from second moments and from the combination of second and third moments are close to the input cosmology, and pass our $0.3 \sigma$ criteria at all scales. We note that the values of $\Omega_{\rm m}$ from the third moments are biased. This is due to the fact that the posterior is strongly asymmetric. We checked that the posterior of a theory data vector shows the same level of shifts in the mean value of $\Omega_{\rm m}$ for the third moments, and the difference with respect to the results from the T17 data vector are much smaller than $0.3 \sigma$. 

In Fig.~\ref{fig:scale cuts} we show both the difference $\Delta \chi^2$ of the T17 data vector and the theory data vector, and the $\chi^2$ of the best-fit cosmology. The former quantity gives a rough idea of the discrepancy of the data vector with respect to the truth: a variation of $\Delta \chi^ 2 = 1$ could, in the worst case possible, cause a 1-$\sigma$ shift in the marginalised 1-D posterior of one of the parameters probed. Usually the difference is absorbed and shared across all the parameters probed (and this is the case). The values of best fit $\chi^2$ for the T17 data vectors also pass our $0.3\sigma$ criteria, being always $\chi^2<1.6$.

We next test the impact of baryonic effects, by contaminating a theory data vector with the effects from the OWLS AGN simulation, as described in \S~\ref{sect:HR1}. The results are shown in the lower panel of Fig.~\ref{fig:scale cuts}. The impact on the data vector from baryons is more pronounced than from the T17 data vector, as shown by the larger $\Delta \chi^2$ values, and it is more important at small scales. This translates in a bias in $n_s$ at small scales. Second moments pass our scale cuts criteria starting from $20 h^{-1}$ Mpc, while the combination of second and third moments from $24 h^{-1}$ Mpc. As for the third moments, they pass our criteria at all the scales probed here (similarly to the T17 data vector test, the values of the mean of the $\Omega_{\rm m}$ posteriors show a negligible shift with respect to the values obtained using a theory data vector). At all scales and for the the combinations of second and third moments, our criteria on the best-fit $\chi^2$ is passed.

We note that we performed these tests adopting a FLASK covariance, which has a slightly different cosmology with respect to the T17 data vector. This, however, did not significantly bias our posteriors, as shown in the upper panel of Fig.~\ref{fig:scale cuts}.

Based on these tests, we adopt the following fiducial scale cuts: $20 h^{-1}$ Mpc as a minimum smoothing scale for second moments, $12 h^{-1}$ Mpc for third moments, and $24 h^{-1}$ Mpc when second and third moments are combined. We note that the scale $24 h^{-1}$ Mpc translates into a cut at $\approx 33$ $(8)$ arcmin for the first (fourth) tomographic bin, while $12 h^{-1}$ Mpc translates into a cut at $\approx 16$ $(4)$ arcmin for the first (fourth) tomographic bin. As there is no significant information below twice the pixel size (i.e., $<7$ arcmin) and most of the constraining power comes from the two high redshift tomographic bins, we have not considered scales smaller than $12 h^{-1}$ Mpc in the above tests.

With the final scale cuts determined, we perform extra checks on the covariance and data vector. We checked that the mean $\chi^2$ of the 1000 FLASK realisations agreed within errors with the number of degree of freedom of our data vector. The distributions of the measured $\chi^2$ are shown in the bottom panels of Fig.~\ref{fig:residual}. We also verified that the distribution of the residuals for each entry of our data vector followed a Gaussian distribution. This is shown in the top panels of Fig.~\ref{fig:residual}. We note that the data-compression algorithm surely helps in giving the compressed data a more Gaussian distribution, due to the central limit theorem \citep{Heavens2017}.

\subsection{Forecast}
\begin{figure*}
\begin{center}
\includegraphics[width=0.7 \textwidth]{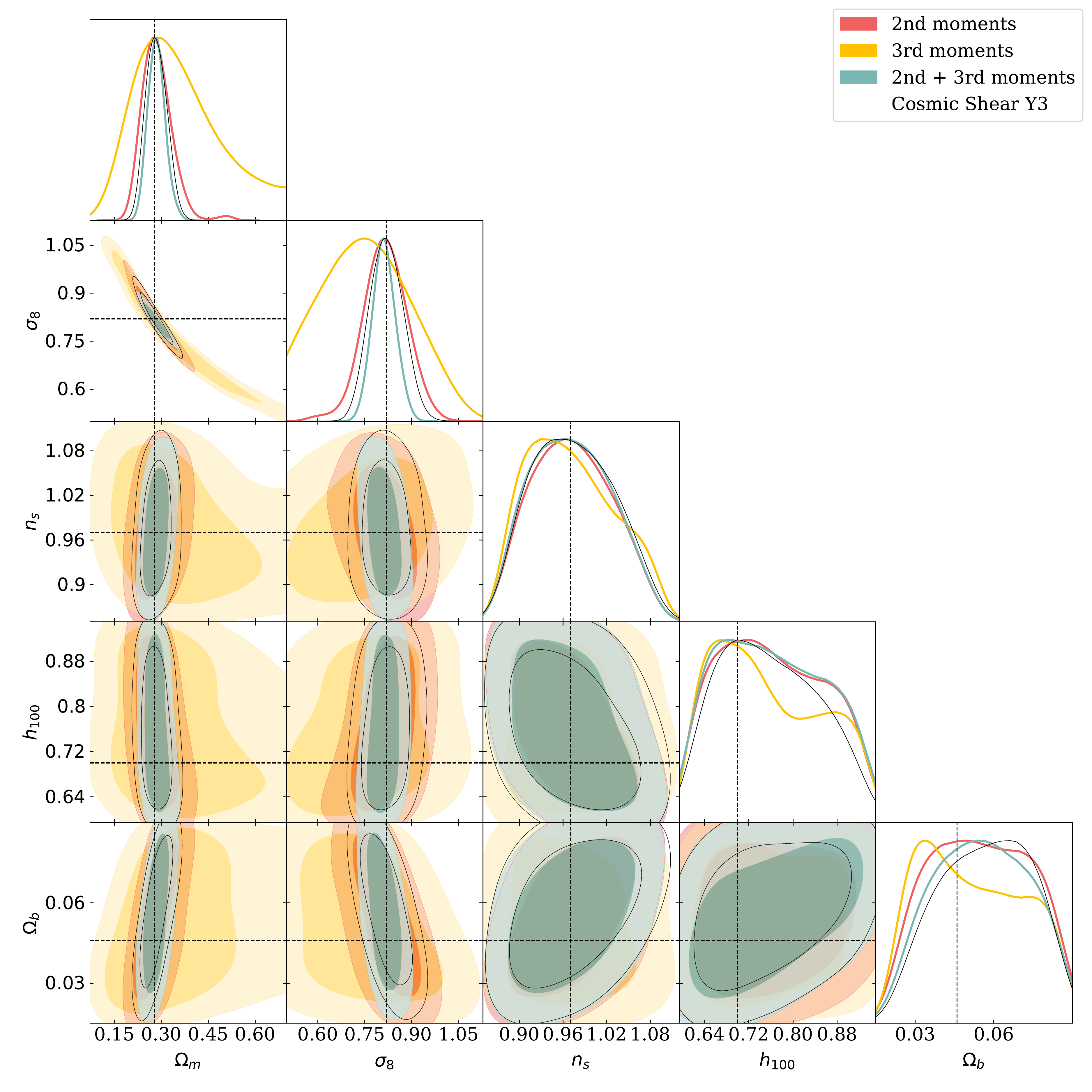}
\end{center}
\caption{Forecast posteriors for cosmological parameters. We marginalise over nuisance parameters as explained in \S \ref{sect:like}. We show constraints from second moments, third moments and second and third moments combined, along with constraints from a shear 2-point correlation function analysis.}
\label{fig:buzzard}
\end{figure*}

\begin{figure}
\begin{center}
\includegraphics[width=0.4 \textwidth]{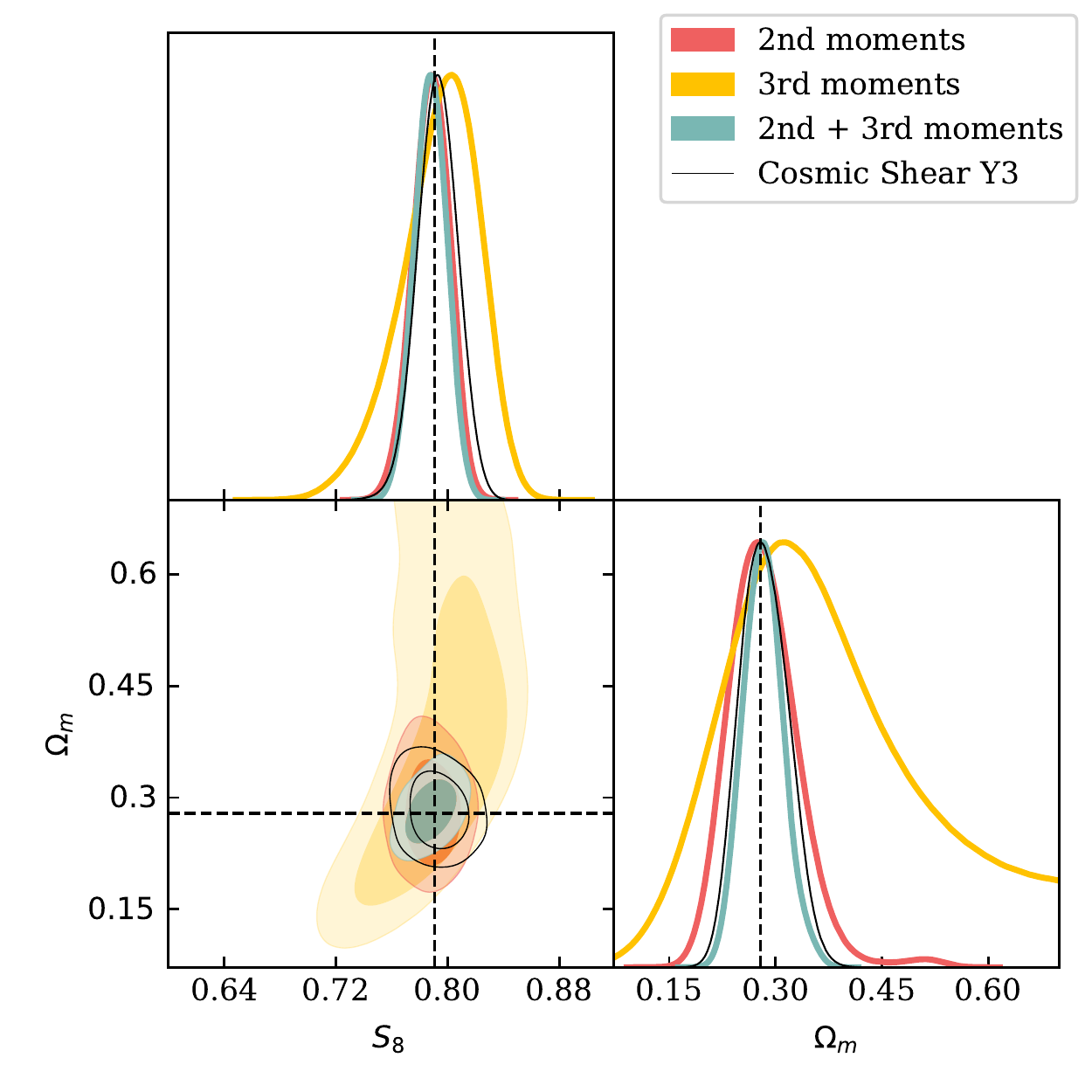}

\end{center}
\caption{Same as Fig.~\ref{fig:buzzard}, but now in the $S_8 - \Omega_{\rm m}$ plane.}
\label{fig:buzzard2}
\end{figure}

\begin{table}
\tiny
\caption {DES Y3/Y5 forecast: fractional accuracy (1-$\sigma$ marginalised posterior confidence intervals over input value) for $\Omega_m$, $S_8$ and $n_s$.  DES Y5 forecast is obtained with the expected DES Y5 number density and DES Y3 scale cuts and tomographic binning.}

\centering
\begin{adjustbox}{width=0.45\textwidth}
\begin{tabular}{|c|c|c|c|}
 \hline
 & $\Omega_{\rm m}$ & $S_8$ &  $n_s$ \\
 \hline
2 moments (Y3) &17 \%& 1.8 \%& 6.9 \%\\
3 moments (Y3) &66 \%& 3.6 \%& 7.9 \%\\
2 + 3 moments (Y3) &10 \%& 1.5 \%& 6.5 \%\\
cosmic shear (Y3) &12 \%& 1.8 \%& 6.4 \%\\
 \hline
2 moments (Y5) &14 \%& 1.5 \%& 6.0 \%\\
3 moments (Y5) &48 \%& 2.8\% & 7.9 \%\\
2 + 3 moments (Y5) & 8 \%& 1.4 \%& 5.7 \%\\
 \hline
\end{tabular}
\end{adjustbox}
\label{tablec}
\end{table}

\begin{table}
\tiny
\caption {DES Y3/Y5 forecast: fractional improvements of the 1-$\sigma$ marginalised posterior confidence intervals for $\Omega_m$, $S_8$ and $n_s$. DES Y5 forecast is obtained with the expected DES Y5 number density and the same DES Y3 scale cuts and tomographic binning.}

\centering
\begin{adjustbox}{width=0.45\textwidth}
\begin{tabular}{|c|c|c|c|}
 \hline
{$24 h^{-1}$ Mpc}  & $\Omega_{\rm m}$ C.I. & $S_8$ C.I. &  $n_s$ C.I. \\
\textbf{scale cuts} & fimprovement  & improvement &  improvement \\
 \hline
2 (Y3) $\to$ 2 + 3 (Y3) &0.57 & 0.79 & 0.97\\
2 (Y5) $\to$ 2 + 3 (Y5) &0.58 & 0.89 & 1.01\\
2 (Y3) $\to$ 2 (Y5) &0.85 & 0.86 & 0.83\\
3 (Y3) $\to$ 3 (Y5) &0.78 & 0.80 & 0.99\\
2 + 3 (Y3) $\to$ 2 + 3 (Y5) &0.87 & 0.97  & 0.89\\
 \hline
\end{tabular}
\end{adjustbox}
\label{table2}
\end{table}

With the scale cuts finalised, we now show the forecast constraints for DES Y3 in Fig.\ref{fig:buzzard}, for 5 cosmological parameters. We adopted the fiducial scale cuts determined in the previous section and the \texttt{FLASK} covariance described in \S~\ref{sect:covariance}. We compressed our data vector following \S~\ref{sect:data-compression_main}. We further marginalise over nuisance parameters as explained in \S ~\ref{sect:theo}. In the modelling of the theory data vector we assumed perfect knowledge of the shape of the redshift distributions. 

As we commented in the previous section, second and third moments mostly constrain $\Omega_{\rm m}$ and $\sigma_8$, while $n_s$ is partially affected by the prior and $h_{100}$ and $\Omega_{\rm b}$ are prior dominated. In general, third moments are less constraining than second moments; however, they contain additional non-Gaussian information and they have a slightly different degeneration axis in the $\Omega_{\rm m}-\sigma_8$ plane compared to second moments. This helps breaking the degeneracy when the two are combined, delivering tighter constraints. This is also shown in Fig.\ref{fig:buzzard2}, where we show results in the $\Omega_{\rm m}-S_8$ plane.

We report in Table~\ref{tablec} the constraining power of moments for $\Omega_{\rm m}$, $S_8$ and $n_s$; the level of improvement when the moments are combined is reported in Table~\ref{table2}. Second, third moments and their combination constrain $\Omega_{\rm m}$ to 17 per cent, 66 per cent and 10 per cent respectively, and $S_8$ to 1.8 per cent, 3.6 per cent and  1.5 per cent respectively. These particular values are obtained specifically for DES Y3 and depends on the particular scales and the noise properties of sample considered.

We also forecast in Table~\ref{table2} how much we expect to improve our constraints when moving to the final DES release, which will include all the data from the five years (Y5) of observations. The values have been obtained by assuming the expected DES Y5 number density (which should roughly double DES Y3 one) and the same DES Y3 scale cuts and tomographic binning. We did not take into account the possibility of having more than four tomographic bins, which would be possible having a deeper sample. In general, we can expect to further improve our constraints by $10-20$ per cent with respect to DES Y3. 

We overlay in Figs.~\ref{fig:buzzard}, \ref{fig:buzzard2} the expected posteriors from the DES Y3 shear 2-point correlation function analysis. Scale cuts for the cosmic shear analysis have been chosen by contaminating a shear 2-point data vector with the effect of baryons and looking at the bias in the parameters' posteriors, in a fashion similar to what has been done in \S~\ref{sect:scale_cuts}. The measurement covariance has been obtained using jackknife resampling and a fiducial DES Y3 simulation \citep{DeRose2018}. The shear 2-point analysis delivers slightly tighter posteriors than second moments alone, but is less constraining than the combination of second and third moments. Indeed, we find it to constrain $\Omega_{\rm m}$ and $S_8$ at the level of 12 per cent and 1.8 per cent; the combined second and third moments result is 20 per cent more constraining. %
%
Without measuring the cross-covariance between moments and shear 2-point correlation function, it is hard to quantitatively explain why the latter is more constraining than second moments alone. One reason could be that they have access to the same information (the power spectrum), but they probe scales differently (cosmic shear is more localised in harmonic space, whereas moments get contributions from a broader range of multipoles, being prone to baryonic effects at all smoothing scales). A different sensitivity to the effects that drive the scale selection can limit the constraining power of a probe compared to others (see, e.g., \citealt{Asgari2019}).  More in general, a different sensitivity to angular scales might cause different observables to be only weakly correlated, even if they belong to the same category of 2-point statistics (see, e.g. \citealt{Hamana2019}). Future works will investigate further the correlation between cosmic shear and second and third moments.

One relevant feature that can be observed from Fig.~\ref{fig:buzzard} and Fig.~\ref{fig:buzzard2}, is that shear 2-point correlation function has a similar degeneracy direction compared to second moments only. Combining shear 2-point correlation function with any other probes sensitive to the bispectrum (such as the third moments) is likely to significantly improve the constraints due to the different degeneracy direction of their constraints.


\section{Summary}\label{sect:summary}
In this paper, we have presented a simulated cosmology analysis using the second and third moments of the weak lensing mass (convergence) maps, targeted at the third year (Y3) data from the Dark Energy Survey (DES).  The second moments, or variances, of the maps as a function of smoothing scale contains similar information as the standard two-point correlations. The third moment, or the skewness, contains additional non-Gaussian information of the field. We described how the convergence maps are constructed starting from the shear catalogue using the Kaiser-Squires formalism. We obtain analytical predictions for the second and third moments using perturbation theory. We included the effects of partial sky coverage in the theoretical modelling of the moments using the pseudo-$C_\ell$ formalism. We validated the modelling of the convergence moments using a large suite of simulations, including the effects of the survey mask and non-linear lensing corrections (such as reduced-shear and source crowding). We used the same simulations to estimate the covariance. We furthermore showed how the computation of theoretical predictions can be sped up without introducing biases in the cosmological analysis by implementing a 5-parameter emulator. 

We tested our pipeline through simulated likelihood analyses varying five cosmological parameters ($\Omega_{\rm m}$, $\sigma_8$, $n_{\rm s}$, $\Omega_{\rm b}$, $h_{100}$) and 10 nuisance parameters (modelling redshift uncertainties, shear biases, and intrinsic alignments). We determined the scale cuts based on the impact of baryonic physics and modelling inaccuracies of the third moments at small scales.  

 We then forecast the constraints achievable with a DES Y3 analysis. We found that second moments, third moments, and their combination constrain $\Omega_{\rm m}$ to 17 per cent, 66 per cent and 10 per cent respectively, and $S_8$ to 1.8 per cent, 3.6 per cent and 1.5 per cent respectively. The combination of second and third moments provides improved constraints with respect to second moments due to the extra non-Gaussian information probed by the third moments and the different inclination of the degeneracy axis in the $\sigma_8 - \Omega_{\rm m}$ plane of the two probes. For DES Y5, where we expect to have a data set with higher galaxy density, we forecast a further improvement in the constraining power at the level of $10-20$ per cent. 
 
 We also compared with a forecast shear 2-point analysis for DES Y3, which yields constraints at the level of 12 per cent and 1.8 per cent for $\Omega_{\rm m}$ and $S_8$. The combined second and third moments result is about 20 per cent more constraining. This analysis shows the importance of including in the analysis probes of higher order statistics to improve on the cosmological constraints. 

This paper has been geared towards the DES Y3 analysis; the application to DES Y3 data will follow. The application of our methodology to DES Y3 data will necessarily require extra checks of the data level, especially concerning potential systematic effects such as modelling errors in the point spread function, inhomogeneities in the noise, and spurious dependencies of shear with observing conditions.  

The methods developed here are general and can be applied to other datasets. We note that as the upcoming survey data becomes deeper, the constraints from higher moments relative to second-moments are expected to improve. This is especially encouraging going forward on the DES Y5 and LSST data.

\section*{Acknowledgements}
Funding for the DES Projects has been provided by the U.S. Department of Energy, the U.S. National Science Foundation, the Ministry of Science and Education of Spain, 
the Science and Technology Facilities Council of the United Kingdom, the Higher Education Funding Council for England, the National Center for Supercomputing 
Applications at the University of Illinois at Urbana-Champaign, the Kavli Institute of Cosmological Physics at the University of Chicago, 
the Center for Cosmology and Astro-Particle Physics at the Ohio State University,
the Mitchell Institute for Fundamental Physics and Astronomy at Texas A\&M University, Financiadora de Estudos e Projetos, 
Funda{\c c}{\~a}o Carlos Chagas Filho de Amparo {\`a} Pesquisa do Estado do Rio de Janeiro, Conselho Nacional de Desenvolvimento Cient{\'i}fico e Tecnol{\'o}gico and 
the Minist{\'e}rio da Ci{\^e}ncia, Tecnologia e Inova{\c c}{\~a}o, the Deutsche Forschungsgemeinschaft and the Collaborating Institutions in the Dark Energy Survey. 

The Collaborating Institutions are Argonne National Laboratory, the University of California at Santa Cruz, the University of Cambridge, Centro de Investigaciones Energ{\'e}ticas, 
Medioambientales y Tecnol{\'o}gicas-Madrid, the University of Chicago, University College London, the DES-Brazil Consortium, the University of Edinburgh, 
the Eidgen{\"o}ssische Technische Hochschule (ETH) Z{\"u}rich, 
Fermi National Accelerator Laboratory, the University of Illinois at Urbana-Champaign, the Institut de Ci{\`e}ncies de l'Espai (IEEC/CSIC), 
the Institut de F{\'i}sica d'Altes Energies, Lawrence Berkeley National Laboratory, the Ludwig-Maximilians Universit{\"a}t M{\"u}nchen and the associated Excellence Cluster Universe, 
the University of Michigan, the National Optical Astronomy Observatory, the University of Nottingham, The Ohio State University, the University of Pennsylvania, the University of Portsmouth, 
SLAC National Accelerator Laboratory, Stanford University, the University of Sussex, Texas A\&M University, and the OzDES Membership Consortium.

Based in part on observations at Cerro Tololo Inter-American Observatory, National Optical Astronomy Observatory, which is operated by the Association of 
Universities for Research in Astronomy (AURA) under a cooperative agreement with the National Science Foundation.

The DES data management system is supported by the National Science Foundation under Grant Numbers AST-1138766 and AST-1536171.
The DES participants from Spanish institutions are partially supported by MINECO under grants AYA2015-71825, ESP2015-66861, FPA2015-68048, SEV-2016-0588, SEV-2016-0597, and MDM-2015-0509, 
some of which include ERDF funds from the European Union. IFAE is partially funded by the CERCA program of the Generalitat de Catalunya.
Research leading to these results has received funding from the European Research
Council under the European Union's Seventh Framework Program (FP7/2007-2013) including ERC grant agreements 240672, 291329, and 306478.
We  acknowledge support from the Brazilian Instituto Nacional de Ci\^encia
e Tecnologia (INCT) e-Universe (CNPq grant 465376/2014-2).

This manuscript has been authored by Fermi Research Alliance, LLC under Contract No. DE-AC02-07CH11359 with the U.S. Department of Energy, Office of Science, Office of High Energy Physics




\bibliographystyle{mnras}		  
\bibliography{bibliography}



\appendix

\section{Skewness parameter}\label{sect:Skewness}
In perturbation theory, the Fourier space equations of motion for the matter density contrast $\delta$ and the divergence of the velocity field $\theta = \nabla \textbf{v}$ are \citep{Bernardeau2002}:

\begin{multline}
 \frac{\partial \delta(\textbf{k},\tau)}{\partial \tau}  + \theta(\textbf{k},\tau) = \\-\int d^3k_1 d^3k_2 \delta_D(\bk-\bk_{12}),\alpha(\bk_1,\bk_2)\delta(\bk_1,\tau)\theta(\bk_2,\tau) \equiv \alpha[\delta,\theta,\bk],
\end{multline}

\begin{multline}
 \frac{\partial \theta(\textbf{k},\tau)}{\partial \tau}  + H \theta(\textbf{k},\tau) + \frac{3\Omega_m H_0^2}{2a} \delta(\bk,\tau)=\\ -\int d^3k_1 d^3k_2 \delta_D(\bk-\bk_{12}),\beta(\bk_1,\bk_2)\theta(\bk_1,\tau)\theta(\bk_2,\tau)  \equiv \beta[\delta,\theta,\bk],
\end{multline}
with $\tau$ being the conformal time, $a$ the scale factor, $H=\frac{d}{d\tau} $ln $a$, $\bk_{12} = \bk_1+\bk_2$ and $\alpha$ and $\beta$ defined by:
\begin{equation}
\alpha(\bk_1,\bk_2) = 1+\frac{1}{2}\frac{\bk_1  \bk_2}{k_1k_2}(\frac{k_1}{k_2}+\frac{k_2}{k_1}),
\end{equation}

\begin{equation}
\beta(\bk_1,\bk_2) = \frac{1}{2}\frac{\bk_1  \bk_2}{k_1k_2}(\frac{k_1}{k_2}+\frac{k_2}{k_1})+\frac{(\bk_1  \bk_2)^2}{k_1^2k_2^2}.
\end{equation}

The matter density contrast and the divergence of the velocity field can be expanded as:

\begin{equation}
 \delta(\bk,\tau) = \sum_{n=1} \delta_n(\bk,\tau),
\end{equation}

\begin{equation}
 \theta(\bk,\tau) = -\frac{\partial ln D_{+}(\tau)}{\partial \tau}\sum_{n=1} \theta_n(\bk,\tau),   
\end{equation}
where $n$ indicates the order at which the fields are approximated and $D_+$ is the linear growth factor. At linear order, $\delta_1(\bk,\tau) = \theta_1(\bk,\tau) = D_+(\tau) \delta_1(\bk)$.

At second order the Fourier equations of motions are solved by:

\begin{equation}
\delta_2(\bk,\tau) = D_+^2(\tau)\alpha[\delta_1,\delta_1,\bk] + D_2(\tau)(\beta[\delta_1,\delta_1,\bk]-\alpha[\delta_1,\delta_1,\bk]),
\end{equation}
with $D_2$ the solution of the following differential equation:

\begin{equation}
\frac{\partial^2D_2(\tau)}{\partial^2\tau} + H \frac{\partial D_2(\tau)}{\partial\tau}-\frac{3\Omega_mH_0^2}{2a}D_2(\tau) = (\frac{\partial D_+(\tau)}{\partial\tau})^2
\end{equation}

Lastly, we define the following quantity $\mu$, as it will enter in the modeling of the third moment:

\begin{equation}
\mu \equiv 1 - D_2/D_+^2 
\end{equation}

At leading order in perturbation theory, one can compute the variance of the dark matter density field smoothed by a top hat filter as:
\begin{equation}
\langle \delta^2_{\theta_0,lin} \rangle (\tau) = \frac{1}{2 \pi} \int dk k W(k,\theta_0)^2 P_{lin} (k,\tau);
\end{equation}
while the skewness will be described by the following equation:
\begin{multline}
\label{eq:3rdmoment}
\langle \delta^3_{\theta_0,lin} \rangle (\tau) = \frac{6}{(2\pi)^3} \int d^2k_1d^2k_2  W(\bk_1,\theta_0)W(\bk_2,\theta_0)W(\bk_1+\bk_2,\theta_0) \\ \times P_{lin} (\bk_1,\tau),P_{lin} (\bk_2,\tau) F_2(\bk_1,\bk_2,\tau),
\end{multline}
where $P_{lin} (k,\tau)$ is the linear power spectrum and $W(k,\theta_0)$ is the top hat filter described in Eq. \ref{eq:filter}. The term $F_2(\bk_1,\bk_2,\tau)$ reads:
\begin{equation}
\label{eq:F2}
F_2(\bk_1,\bk_2,\tau) = \frac{1}{2}[(1+\frac{k_1}{k_2}cos\phi)+(1+\frac{k_2}{k_1}cos\phi)] + [1-\mu(\tau)](cos^2\phi-1).
\end{equation}

We implement here a refinement of the term $F_2$ based on N-body simulations (while Eq.~\ref{eq:F2} has been obtained, so far, exclusively relying on perturbation theory). The refinement we are implementing here has been first obtained by \cite{Scoccimarro2001} and later on by \cite{GilMarin2012} fitting an analytical formula to the non-linear evolution of the bispectrum based on a suite of cold dark matter N-body simulations.
Implementing such corrections, Eq.~\ref{eq:F2} becomes:
\begin{multline}
F_2(\bk_1,\bk_2,\tau) = \frac{1}{2}b_1b_2[(1+\frac{k_1}{k_2}cos\phi)+(1+\frac{k_2}{k_1}cos\phi)]\\ + [1-\mu(\tau)]c_1c_2(cos^2\phi-1) + [a_1a_2\mu(\tau)-b_1b_2+[1-\mu(\tau)]c_1c_2].
\end{multline}
The terms $a$, $b$, $c$ are taken from \cite{GilMarin2012}; their subscripts in the above equations indicate if they refer to $k_1$ or $k_2$. In particular:
\begin{equation}
\label{eq:a}
a(n,k,\tau) = \frac{1+(\sigma_8D_+)^{a_6}[0.7 (4-2^n)/(1+2^{2n+1})]^{1/2}(qa_1)^{n+a_2}}{1+(qa_1)^{n+a_2}},
\end{equation}
\begin{equation}
\label{eq:b}
b(n,k,\tau) = \frac{1+0.2a_3(n+3)(qa_7)^{n+3+a_8}}{1+(qa_7)^{n+3.5+a_8}},
\end{equation}
\begin{equation}
\label{eq:c}
c(n,k,\tau) = \frac{1+4.5a_4/[1.5+(n+3)^4{(qa_5)^{n+3+a_9}}}{1+(qa_5)^{n+3.5+a_9}}.
\end{equation}
In the above equations $q\equiv k/k_{\rm NL}$, where $k_{\rm NL}$ is the scale where non-linearities start to be important and it is defined so that $k_{\rm NL}^3 P(k,\tau)/2\pi^2 = 1$.
We report in table~\ref{table_coefficients} the values of the coefficients $a_1, ..., a_9$ as from \cite{Scoccimarro2001} and \cite{GilMarin2012}. Implementing these corrections in Eq.~\ref{eq:3rdmoment} leads to:
\begin{multline}
\label{eq:3rdmoment_2}
\langle \delta^3_{\theta_0,lin} \rangle (\tau) = \frac{6}{(4\pi^2)} \int dk_1dk_2  W(k_1,\theta_0)W(k_2,\theta_0) \\ \times  P_{lin} (k_1,\tau),P_{lin} (k_2,\tau) \int d\phi W(\sqrt{k_1^2+k_2^2+2k_1k_2cos\phi},\theta_0) \\ \times  F_2(k_1,k_2,\phi,\tau).
\end{multline}
The integral on the angle $\phi$ can be written as:
\begin{multline}
\label{eq:delirio}
\int d\phi W(\sqrt{k_1^2+k_2^2+2k_1K_2cos\phi},\theta_0) F_2(k_1,k_2,\phi,\tau) \\ = \frac{1}{2}b_1b_2 \int d\phi W(\sqrt{k_1^2+k_2^2+2k_1k_2cos\phi},\theta_0)[2+(\frac{k_1}{k_2}+\frac{k_2}{k_1})cos\phi]\\
+  \int d\phi W(\sqrt{k_1^2+k_2^2+2k_1k_2cos\phi},\theta_0) [(1-\mu) c_1c_2(cos^2\phi-1)]\\
+  \int d\phi W(\sqrt{k_1^2+k_2^2+2k_1k_2cos\phi},\theta_0) [a_1a_2\mu-b_1b_2+(1-\mu)c_1c_2].
\end{multline}
For brevity, we omitted the dependence on $\tau$ from $\mu$.
The three integrals in Eq.~\ref{eq:delirio} can be solved as:
\begin{multline}
b_1b_2[2\pi W(k_1,\theta_0)W(k_2,\theta_0) + \frac{\pi}{2}\frac{\partial}{\partial \theta_0}(W(k_1,\theta_0)W(k_2,\theta_0))]\\
-c_1c_2[\pi(1-\mu)W(k_1,\theta_0)W(k_2,\theta_0)]\\
+2\pi [a_1a_2\mu-b_1b_2+(1-\mu)c_1c_2] W(k_1,\theta_0)W(k_2,\theta_0) = \\ \frac{\pi}{2}b_1b_2\frac{\partial}{\partial \theta_0}[W(k_1,\theta_0)W(k_2,\theta_0)]\\+\pi[2a_1a_2-(1-\mu) c_1c_2] W(k_1,\theta_0)W(k_2,\theta_0).
\end{multline}
After some algebra, one can express Eq. \ref{eq:3rdmoment_2} as:
\begin{multline}
\langle \delta^3_{\theta_0,lin} \rangle (\tau) =  6\left[\int dk k W(k,\theta_0)^2 P_{lin} (k,\tau)\right]^2\\ - 3\left[\int dk k (1-\mu) c W(k,\theta_0)^2 P_{lin} (k,\tau)\right]^2\\ + \frac{3}{4}\frac{\partial}{\partial ln \theta_{,0}} \left[\int dk k b W(k,\theta_0)^2 P_{lin} (k,\tau)\right]^2,
\end{multline}
\begin{multline}
\langle \delta^3_{\theta_0,lin} \rangle (\tau) = 3[2( \langle \delta^2_{\theta_0,lin,a} \rangle (\tau))^2 - (1-\mu) ( \langle \delta^2_{\theta_0,lin,c} \rangle (\tau))^2 + \\\frac{3}{2}\frac{\partial \langle \delta^2_{\theta_0,lin,b} \rangle (\tau) }{\partial ln \theta_0}.
\end{multline}
In the above equation we have defined
\begin{equation}
\langle \delta^2_{\theta_0,lin,X} \rangle (\tau) = \frac{1}{2 \pi} \int dk k X(k,\tau) W(k,\theta_0)^2 P_{lin} (k,\tau),
\end{equation}
with $X$ that can be either $a$, $b$ or $c$. We finally define the reduced skewness parameter as
\begin{equation}
S_3 \equiv  \frac{\langle\delta^3_{\theta_0,lin} \rangle (\tau)}{\langle\delta^2_{\theta_0,lin} \rangle (\tau)^2}.
\end{equation}
The original perturbation theory result can be obtained noting that in the limit of $a,b,c \rightarrow 1$ we have $\delta^2_{\theta_0,lin,a}$,  $\delta^2_{\theta_0,lin,b}$, $\delta^2_{\theta_0,lin,c}  \rightarrow \delta^2_{\theta_0,lin}$; in this case, the reduced skewness parameter assumes the following form:
\begin{equation}
S_3 \equiv  \frac{\langle\delta^3_{\theta_0,lin} \rangle (\tau)}{\langle\delta^2_{\theta_0,lin} \rangle (\tau)^2} = 3(1+\mu)  + \frac{3}{2}\frac{\partial ln \langle \delta^2_{\theta_0,lin} \rangle (\tau) }{\partial ln \theta_0}.
\end{equation}

The equations above for the third moments hold in the linear regime, but they are usually extrapolated to the mild non-linear regime using predictions of the non-linear power spectrum.

\begin{table}
\tiny
\caption {Values of the coefficients for the fitting formula described in Eqs.~\ref{eq:a}, \ref{eq:b} and \ref{eq:c} from \citet{Scoccimarro2001} (SC01) and \citet{GilMarin2012} (GM12).}
\centering
\begin{adjustbox}{width=0.25\textwidth}
\begin{tabular}{|c|c|c|}

 \hline
\textbf{coefficient}  & SC01  & GM12 \\
 \hline
$\alpha_1$ & 0.25 & 0.484\\
$\alpha_2$ & 3.5 & 3.740\\
$\alpha_3$ & 2 & -0.849\\
$\alpha_4$ & 1 & 0.392\\
$\alpha_5$ & 2 & 1.013\\
$\alpha_6$ & -0.2 & -0.575\\
$\alpha_7$ & 1 & 0.128\\
$\alpha_8$ & 0 & -0.722\\
$\alpha_9$ & 0 & -0.926\\
\hline
\end{tabular}
\end{adjustbox}
\label{table_coefficients}
\end{table}

We note that there is up to a $20\%$ difference between \cite{Scoccimarro2001} and \cite{GilMarin2012} fitting formulae at small scales ($\sim 5 $ arcmin for the first tomographic bin). In our main analysis we use the values for the coefficients from \cite{Scoccimarro2001} because they provide a better fit to our simulations, but we include the difference between the \cite{Scoccimarro2001} and \cite{GilMarin2012} models in our covariance in order to grasp the small scales modeling uncertainty of the skeweness.

\section{Mode-mode coupling matrices}\label{sect:MM}

We provide here mathematical recipes for the mode-mode coupling matrices $\boldsymbol{M}$ used in $\S$ \ref{sect:theo} to account for masking effects. Such matrices are developed in the contest of pseudo power spectrum estimators (e.g, \citealt{Wandelt2001,Brown2005,Hikage2011,Hikage2016}). In particular, we strictly follow here $\S 2.1$ of \cite{Hikage2011}.

In the presence of a window function (in our case, the DES Y3 footprint) $K(\theta,\phi)$, the shear field assumes the following expression:

\begin{equation}
\bar{\gamma}_1(\theta,\phi) + \bar{\gamma}_2(\theta,\phi) = K(\theta,\phi)(\gamma_1(\theta,\phi) + \gamma_2(\theta,\phi)).
\end{equation}

When the shear field is transformed into its spherical harmonic counterpart (Eq. \ref{eq:gammanomask}), it obtains an additional contribution due to the convolution with the footprint mask:
\begin{equation}
\hat{\bar{\gamma}}_{E,lm} \pm i \hat{\bar{\gamma}}_{B,lm}  = \int d\Omega [K(\theta,\phi)(\gamma_1(\theta,\phi) + \gamma_2(\theta,\phi))]_{\pm2}Y^{*}_{lm}(\theta,\phi).
\end{equation}

The quantities $\hat{\bar{\gamma}}_{E,lm}$ and $\hat{\bar{\gamma}}_{B,lm}$ are called pseudo E and B modes (as they are convolved with the footprint mask) and their relation with the true E and B modes can be written as:
\begin{equation}
\hat{\bar{\gamma}}_{E,lm} \pm i \hat{\bar{\gamma}}_{B,lm}  = \sum_{l'm'} (\hat{{\gamma}}_{E,lm} \pm i \hat{{\gamma}}_{B,lm})_{\pm2}W_{ll'mm'},
\end{equation}
where $_{\pm2}W_{ll'mm'}$ is a convolution kernel

\begin{multline}
_{\pm2}W_{ll'mm'} = \int d\Omega _{\pm2}Y_{l'm'}(\theta,\phi) C (\theta,\phi) _{\pm2}Y^{*}_{lm}(\theta,\phi) =\\
\sum_{l''m''} K_{l''m''} (-1)^{m}\sqrt{\frac{(2l+1)(2l'+1)(2l"+1)}{4\pi}} \times \\ \tj{l}{l'}{l''}{\pm2}{\mp2}{0}\tj{l}{l'}{l''}{m}{m'}{m''},
\end{multline}
with $\tj{l}{l'}{l''}{m}{m'}{m''}$ Wigner 3$j$ symbols and $K_{lm} = \int d\Omega K(\theta,\phi)Y^*_{lm}(\theta,\phi) $ the harmonic transform of the window function. Defining

\begin{equation}
C_{l}^{EE} = \frac{1}{2l+1}\sum_m | \hat{\gamma}_{E,lm} |^2,
\end{equation}
\begin{equation}
C_{l}^{EB} = \frac{1}{2l+1}\sum_m  \hat{\gamma}_{E,lm}  \hat{\gamma}_{B,lm}^* ,
\end{equation}
\begin{equation}
C_{l}^{BB} = \frac{1}{2l+1}\sum_m | \hat{\gamma}_{B,lm} |^2,
\end{equation}
we can write the masked (pseudo) spectra as the convolution of the true spectra with a mode-mode coupling matrix:
\begin{equation}
\textbf{C}_{\ell} = \sum_{\ell'} \textbf{M}_{\ell \ell'} \textbf{C}_{\ell'},
\end{equation}
where we introduced the vector $\textbf{C}_{\ell}  (C_{\ell}^{EE},C_{\ell}^{EB},C_{\ell}^{BB})$. The mode-mode coupling matrix $\textbf{M}$ is expressed in terms of $M^{EE,EE}_{\ell \ell'}$, $M^{BB,BB}_{\ell \ell'}$, $M^{EB,EB}_{\ell \ell'}$, $M^{EE,BB}_{\ell \ell'}$:

\begin{multline}
M^{EE,EE}_{ll'} = M^{BB,BB}_{ll'} \\
= \frac{2l'+1}{8\pi} \sum_{l''}(2l''+1)K_{l''}[1+(-1)^{l+l'+l''}]  \times \\
\tj{l}{l'}{l''}{2}{-2}{0}^2,
\end{multline}

\begin{multline}
M^{EE,BB}_{ll'} = M^{BB,EE}_{ll'} \\
= \frac{2l'+1}{8\pi} \sum_{l''}(2l''+1)K_{l''}[1-(-1)^{l+l'+l''}]  \times \\
\tj{l}{l'}{l''}{2}{-2}{0}^2,
\end{multline}

\begin{multline}
M^{EB,EB}_{ll'} = \frac{2l'+1}{4\pi} \sum_{l''}(2l''+1)K_{l''}
\tj{l}{l'}{l''}{2}{-2}{0}^2,
\end{multline}
with $K_{l} = \frac{1}{2l+1}\sum_m K_{lm}K^*_{lm}$.

\section{Constraints with data-compression and alternative covariance matrix}\label{sect:data-compression}

\begin{figure}
\begin{center}
\includegraphics[width=0.45 \textwidth]{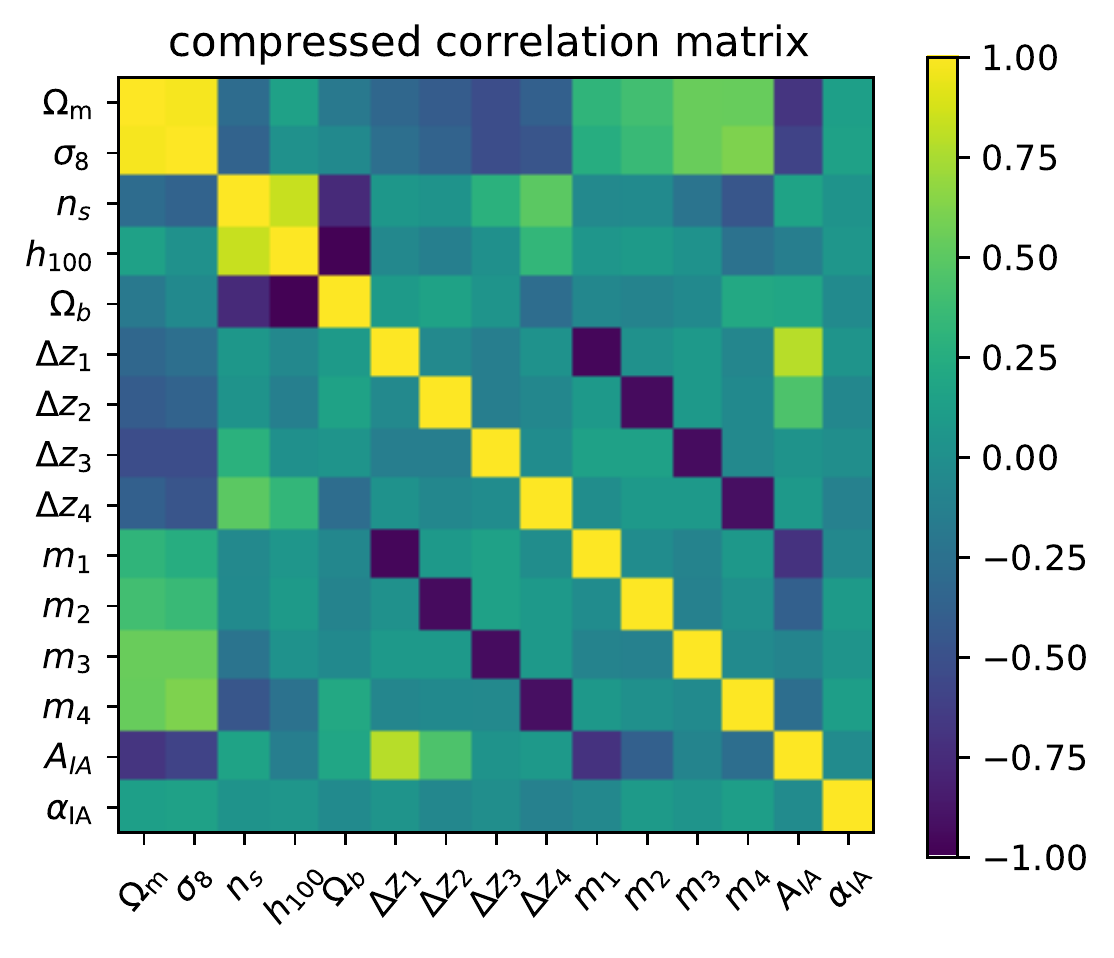}
\end{center}
\caption{Measured compressed correlation matrix of second and third moments from 1000 \texttt{FLASK} simulations. A $12 h^{ - 1}$ Mpc scale cut has been applied (see \S~\ref{sect:scale_cuts} for a definition of the scale cuts). The entries of the correlation matrix are shown with respect to the parameter used to compress the data vector.}
\label{fig:covariance_compressed}
\end{figure}

\begin{figure*}
\begin{center}
\label{fig:svd}
\includegraphics[width=0.4 \textwidth]{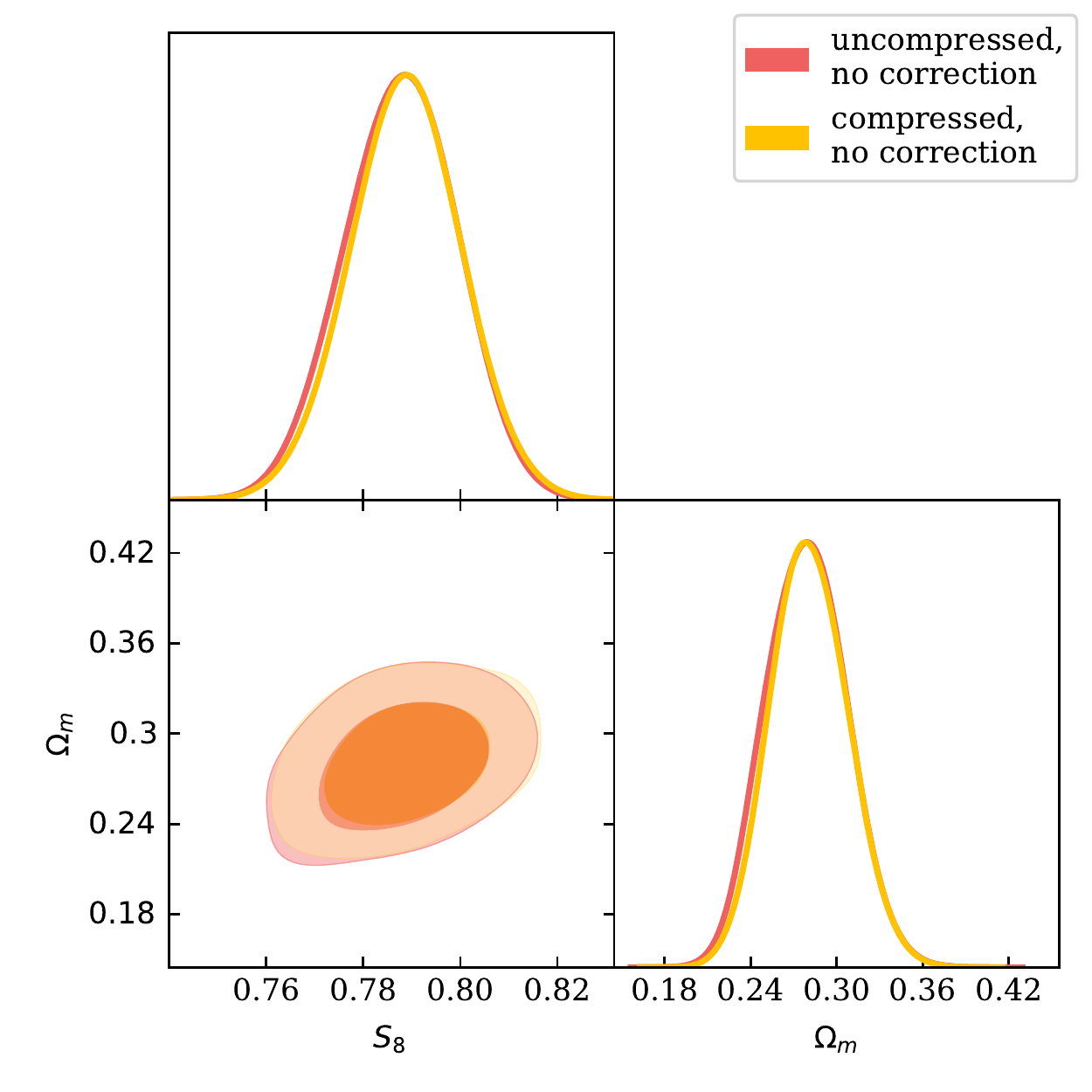}
\includegraphics[width=0.4 \textwidth]{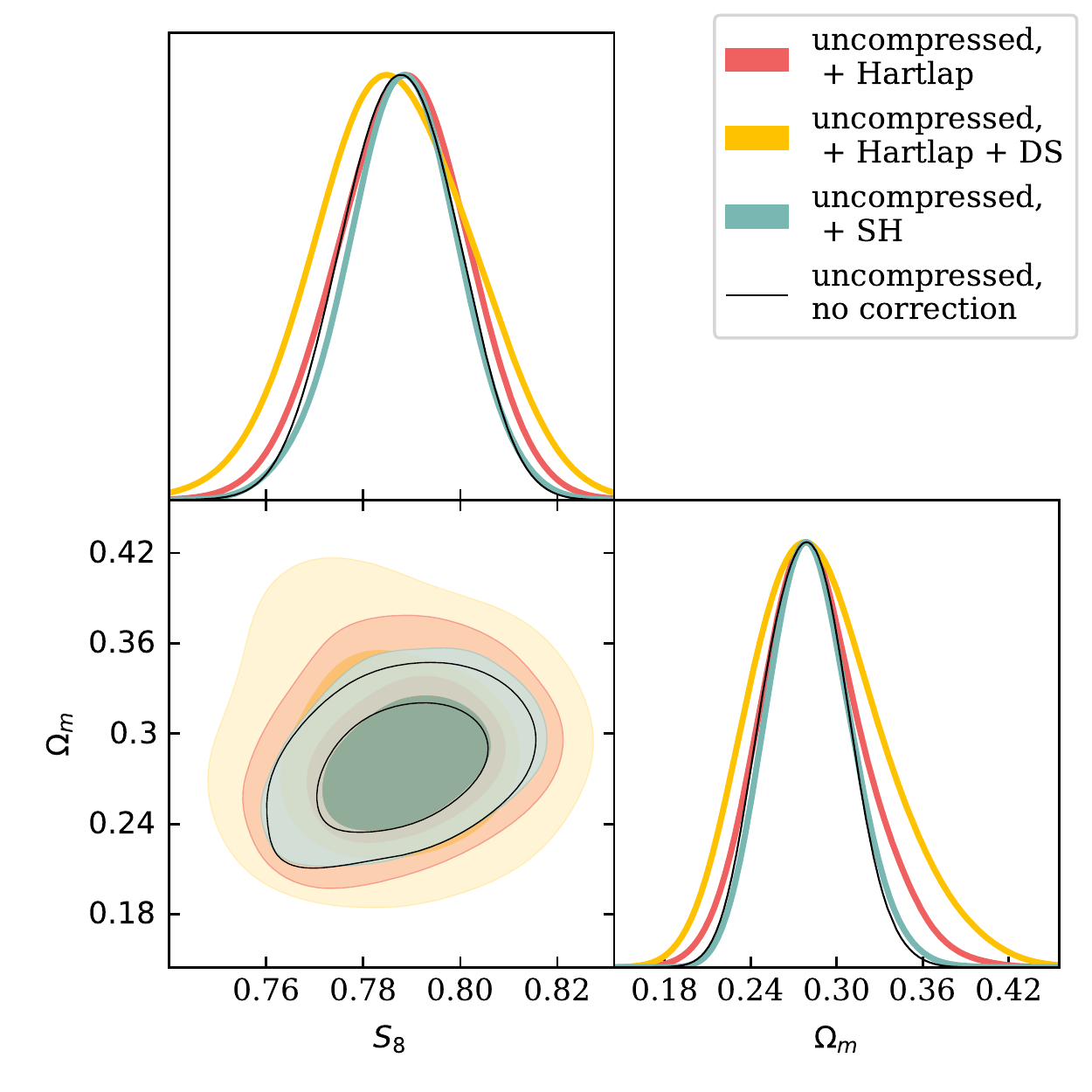}
\includegraphics[width=0.4 \textwidth]{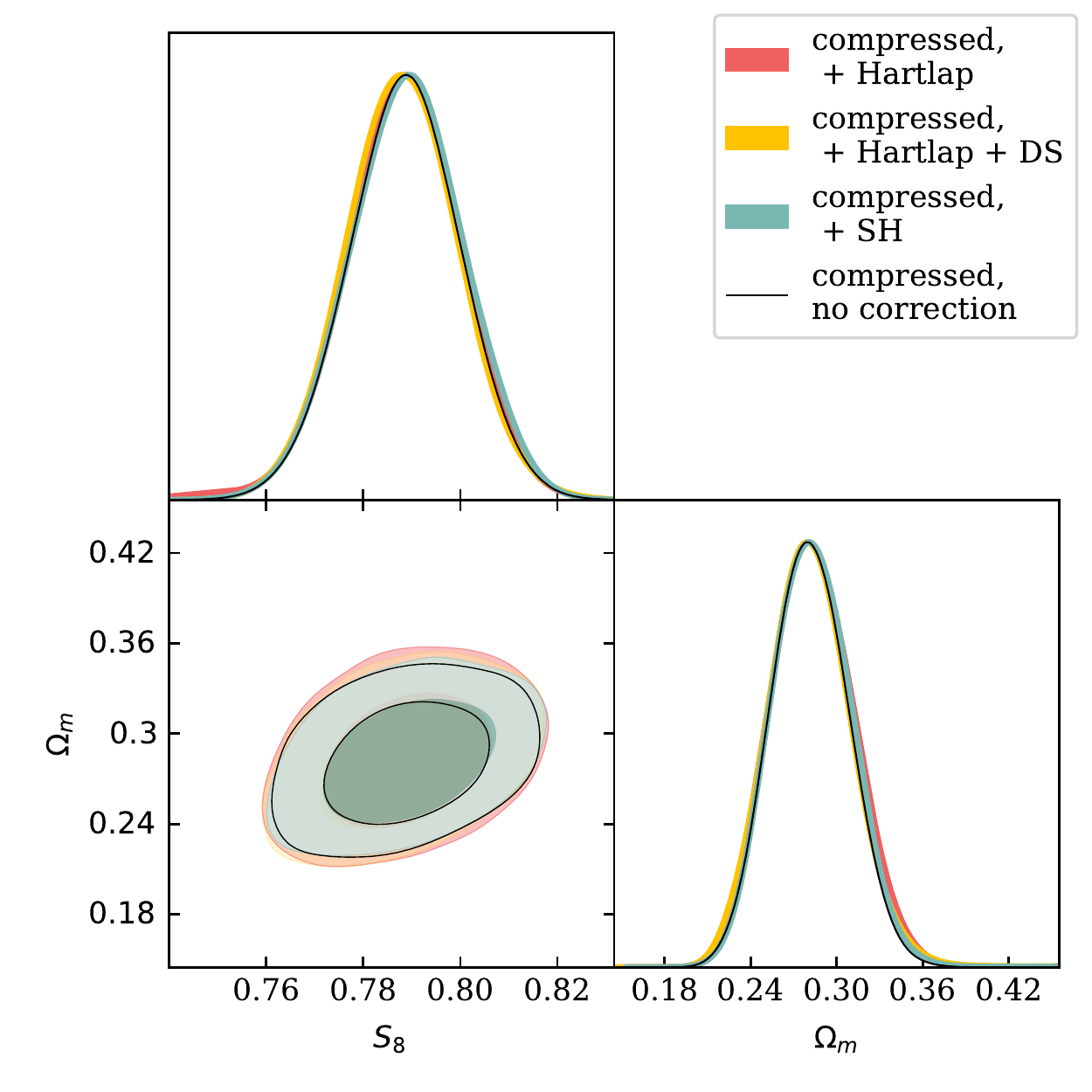}
\includegraphics[width=0.4 \textwidth]{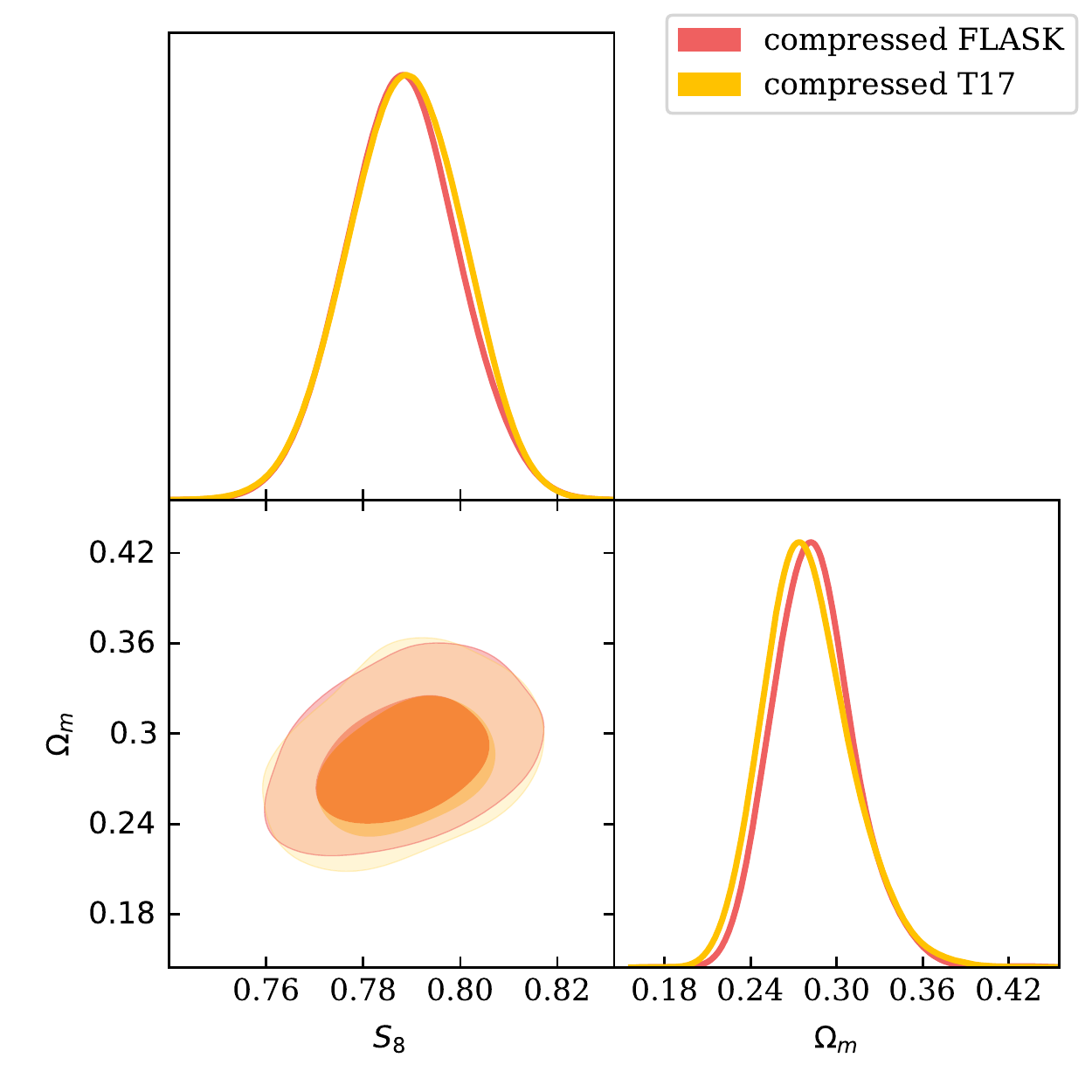}
\end{center}
\caption{Posterior of  $\Omega_{\rm m}$ and $S_8$ for four different cases. \textit{Top left}: posteriors obtained using the uncompressed and compressed \texttt{FLASK} covariance, without applying any corrections due to the noise (Eqs.~\ref{eq:hartlap} and \ref{eq:DS}). \textit{Top right}: posteriors obtained using uncompressed FLASK covariance, with a number of corrections to account for the noise in the estimated covariance matrix. ``Hartlap'' refers to the \citet{Hartlap2011} correction (Eq.~\ref{eq:hartlap}), ``DS'' refers to the \citet{Dodelson2013} correction (Eq.~\ref{eq:DS}), while ``SH'' refers to the \citet{Sellentin2016} likelihood (see text in Appendix~\ref{sect:data-compression} for more details). \textit{Bottom left}: same as the top right panel, but for compressed data vectors. \textit{Bottom right}: posteriors obtained using the compressed  \texttt{FLASK} and T17 covariances. In all the cases above, to highlight the effects of the noise in the estimate of the covariance matrix, only 400 N-body simulations have been used (instead of 1000).}
\label{fig:compression}
\end{figure*}

\begin{figure*}
\begin{center}
\label{fig:svd}
\includegraphics[width=0.8 \textwidth]{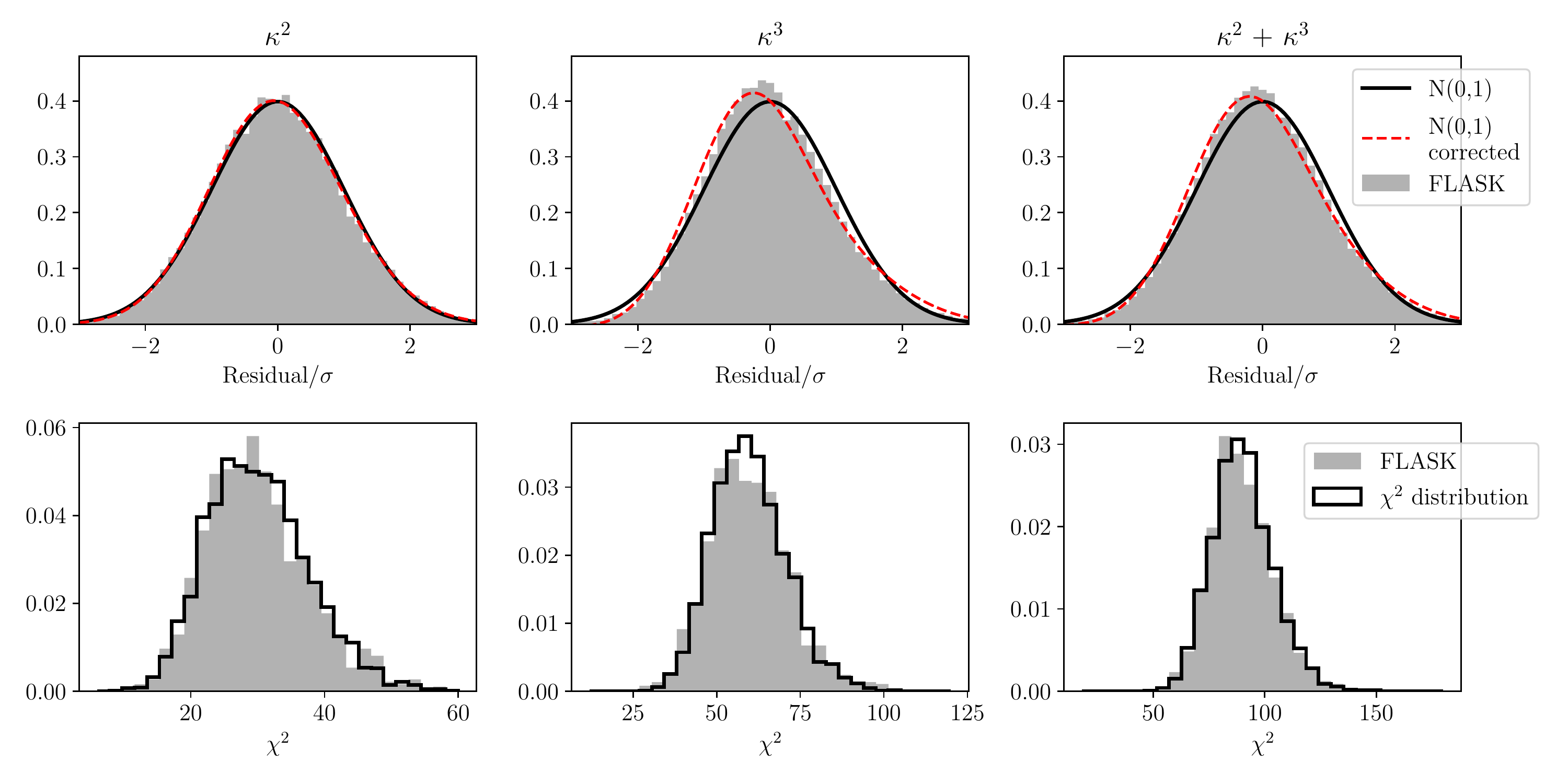}
\end{center}
\caption{This figure is the same as Fig.~\ref{fig:residual} but for a uncompressed data vector. \textit{Upper panels}: residuals of individual data points in units of their expected standard deviation. We compare to a Gaussian with 0 mean and unit standard deviation; we also compare to a Gaussian corrected by the first term of the Edgeworth expansion of the likelihood (see text for more details). \textit{Bottom panels}: Distribution of the $\chi^2$ of each realization of the \texttt{FLASK} simulations, compared to a theoretical $\chi^2$ distribution.}
\label{fig:compression2}
\end{figure*}

Our fiducial analysis has been carried out using a covariance matrix obtained from multiple \texttt{FLASK} realisations (see \S~\ref{sect:FLASK}). \texttt{FLASK} is a log-normal simulation, where the only required inputs are the desired auto and cross power spectra of the convergence fields and the so-called log-normal shift parameters \citep[which effectively set the skewness of the simulated fields at one scale, see e.g.][]{Friedrich2018, Gruen2018}. No additional physics is encoded in the \texttt{FLASK} maps. This means that our \texttt{FLASK} realisations reproduce the correct 2nd moments set by our $\Lambda$CDM input spectra, but has only limited accuracy in its 3rd moments.
We have shown that this does not strongly bias the recovery of input cosmological parameters once applied to N-body simulations (see $\S$~\ref{sect:scale_cuts}).

In this section, we show how to obtain cosmological constraints from our pipeline using the T17 covariance and compare them to the ones obtained from the \texttt{FLASK} covariance, using a data compression algorithm (described in \S~\ref{sect:data-compression_main}). We also validate the efficiency of the data compression algorithm and show how it helps to reduce the noise in the inferred parameters caused by the paucity of simulations used to estimate the covariance matrix.

We show the compressed correlation matrix in Fig.~\ref{fig:covariance_compressed}. The correlation matrix has now 15 entries, as many as as the number of parameters we constrain in our analysis. Interestingly, the correlation between the different elements of the compressed data vector reflects the correlation between parameters (e.g., $\Omega_{\rm m}$ and $\sigma_8$ show a significant correlation, as expected from Fig.~\ref{fig:buzzard}).

We next perform here several tests to validate our compression algorithm. First, we run two forecast chains using the compressed and uncompressed \texttt{FLASK} covariance and compare the contours. This is shown in the top left panel of Fig.~\ref{fig:compression}, for the $\Omega_{\rm m}$ and $S_8$ parameters. In this first test, we did not apply any correction for the noise in the inverse of the covariance (Eqs.~\ref{eq:hartlap} and \ref{eq:DS}), as we are interested in validating the compression algorithm only. The marginalised 1-D posteriors of $\Omega_{\rm m}$ and $S_8$ have similar width, showing that the data compression implemented is basically lossless. As a caveat, we remind the reader that we assume the likelihood to be Gaussian, which in the case of the uncompressed data vector is only an approximation (see below).

Second, we show in the top right panel of Fig.~\ref{fig:compression} how the constraints degrade once the uncertainties in the inverse of the covariance matrix are taken into account. The \cite{Hartlap2011} and \cite{Dodelson2013} corrections (Eqs.~\ref{eq:hartlap} and \ref{eq:DS}) noticeably enlarge the contours, the net effect depending on the number of simulations used to estimate the covariance matrix. We also show, for comparison purposes, how the posteriors would look if the likelihood from \cite{Sellentin2016} was used. \cite{Sellentin2016} argue that when the covariance matrix is estimated from simulations, the likelihood is no longer Gaussian but rather is described by an adapted version of a multivariate \textit{t}-distribution, a fact not taken into account by the \cite{Hartlap2011} correction. They suggest that marginalising over the true covariance improves over the simple \cite{Hartlap2011} correction, and this is confirmed by the top right panel of Fig.~\ref{fig:compression}. We note, however, that the additional scatter in the parameters posterior encoded by the \cite{Dodelson2013} correction is not accounted for in the \cite{Sellentin2016} framework.

The lower left panel of Fig.~\ref{fig:compression} is the same as the top right panel but for the compressed data vector. The compression greatly reduces the noise in the estimated covariance matrix and Eqs.~\ref{eq:hartlap} and \ref{eq:DS} approaches $\sim1$. Also the \cite{Sellentin2016} likelihood approaches a multivariate Gaussian, becoming almost indistinguishable from the no correction case.

Lastly, in the lower right panel of Fig.~\ref{fig:compression} we show the contours obtained using the compressed T17 covariance matrix. We expect the shape of the posterior to be different when using the compressed T17 covariance and the \texttt{FLASK} covariance in two ways. First, the cosmology of the T17 simulations is slightly different from \texttt{FLASK} one. Second, third moments should be more accurately modelled in the T17 simulations as \texttt{FLASK} does not contain the physics to model the third moments beyond the log-normal shift. Differences in the widths between the two compressed covariances are smaller than 2 per cent, suggesting that the two factors considered above have a modest impact.

Finally, we comment on the more Gaussian nature of the compressed data vector compared to the uncompressed one. This is shown in  Fig.~\ref{fig:compression2}. The residuals of the uncompressed data vector appear much less Gaussian for the third moments and the combination of second and third moments compared to what we found for the compressed data vector in Fig.~\ref{fig:residual} (no significant difference in the distribution of the residuals is seen for when only second moments are used). We compute how the distribution of residuals would look if the likelihood were not purely Gaussian, by means of a multivariate Edgeworth expansion of the likelihood (e.g., \citealt{Amendola1996}):
\begin{equation}
\mathcal{L} = G(x,C)[1+\frac{1}{6}k_{x}^{ijk}h_{ijk} + ...],
\end{equation}
with
\begin{equation}
  h_{ijk} = (-1)^3 G^{-1} (x,C) \partial_{ijk}G (x,C),
\end{equation}
where $G(x,C)$ is the Gaussian part of the likelihood, $x$ and $C$ are the data vector and its covariance respectively, and $k_{x}^{ijk} = \avg{x^ix^jx^k}$ is the third order cumulant of the data vector (which can be measured in simulations). The predicted distribution of residuals in Fig.~\ref{fig:compression2} obtained with the first term of the Edgeworth expansion is in better agreement with the one measured in \texttt{FLASK} simulations.

\section*{Affiliations}

$^{1}$ Institut de F\'{\i}sica d'Altes Energies (IFAE), The Barcelona Institute of Science and Technology, Campus UAB, 08193 Bellaterra (Barcelona) Spain\\	
$^{2}$ Department of Astronomy and Astrophysics, University of Chicago, Chicago, IL 60637, USA	0000-0002-7887-0896\\
$^{3}$ Kavli Institute for Cosmological Physics, University of Chicago, Chicago, IL 60637, USA	0000-0002-7887-0896\\
$^{4}$ Kavli Institute for Cosmology, University of Cambridge, Madingley Road, Cambridge CB3 0HA, UK	\\
$^{5}$ Department of Physics and Astronomy, University of Pennsylvania, Philadelphia, PA 19104, USA	\\
$^{6}$ Institute of Cosmology and Gravitation, University of Portsmouth, Portsmouth, PO1 3FX, UK	\\
$^{7}$ 	Institut d'Estudis Espacials de Catalunya (IEEC), 08034 Barcelona, Spain	\\
$^{8}$ Institute of Space Sciences (ICE, CSIC),  Campus UAB, Carrer de Can Magrans, s/n,  08193 Barcelona, Spain	\\
$^{9}$ Department of Physics, Stanford University, 382 Via Pueblo Mall, Stanford, CA 94305, USA	\\
$^{10}$ Kavli Institute for Particle Astrophysics \& Cosmology, P. O. Box 2450, Stanford University, Stanford, CA 94305, USA	\\
$^{11}$Institute of Theoretical Astrophysics, University of Oslo.
P.O. Box 1029 Blindern, NO-0315 Oslo, Norway\\
$^{12}$ SLAC National Accelerator Laboratory, Menlo Park, CA 94025, USA	\\
$^{13}$ Jodrell Bank Center for Astrophysics, School of Physics and Astronomy, University of Manchester, Oxford Road, Manchester, M13 9PL, UK\\
$^{14}$ Department of Physics \& Astronomy, University College London, Gower Street, London, WC1E 6BT, UK	0000-0003-2927-1800\\
$^{15}$ Center for Cosmology and Astro-Particle Physics, The Ohio State University, Columbus, OH 43210, USA	\\
$^{16}$ Department of Physics, The Ohio State University, Columbus, OH 43210, USA	\\
$^{17}$ Department of Physics, University of Arizona, Tucson, AZ 85721, USA	\\
$^{18}$Cerro Tololo Inter-American Observatory, National Optical Astronomy Observatory, Casilla 603, La Serena, Chile	\\
$^{19}$Fermi National Accelerator Laboratory, P. O. Box 500, Batavia, IL 60510, USA	0000-0002-7069-7857\\
$^{20}$Instituto de Fisica Teorica UAM/CSIC, Universidad Autonoma de Madrid, 28049 Madrid, Spain	\\
$^{21}$Centro de Investigaciones Energ\'eticas, Medioambientales y Tecnol\'ogicas (CIEMAT), Madrid, Spain	0000-0003-3044-5150\\
$^{22}$Laborat\'orio Interinstitucional de e-Astronomia - LIneA, Rua Gal. Jos\'e Cristino 77, Rio de Janeiro, RJ - 20921-400, Brazil	0000-0003-3044-5150\\
$^{23}$Department of Astronomy, University of Illinois at Urbana-Champaign, 1002 W. Green Street, Urbana, IL 61801, USA	0000-0002-4802-3194\\
$^{24}$National Center for Supercomputing Applications, 1205 West Clark St., Urbana, IL 61801, USA	0000-0002-4802-3194\\
$^{25}$Physics Department, 2320 Chamberlin Hall, University of Wisconsin-Madison, 1150 University Avenue Madison, WI  53706-1390\\	
$^{26}$Observat\'orio Nacional, Rua Gal. Jos\'e Cristino 77, Rio de Janeiro, RJ - 20921-400, Brazil	\\
$^{27}$Department of Physics, IIT Hyderabad, Kandi, Telangana 502285, India	0000-0002-0466-3288,\\
$^{28}$Department of Astronomy/Steward Observatory, University of Arizona, 933 North Cherry Avenue, Tucson, AZ 85721-0065, USA	0000-0002-1894-3301\\
$^{29}$Jet Propulsion Laboratory, California Institute of Technology, 4800 Oak Grove Dr., Pasadena, CA 91109, USA	0000-0002-1894-3301\\
$^{30}$Santa Cruz Institute for Particle Physics, Santa Cruz, CA 95064, USA\\
$^{31}$Department of Astronomy, University of Michigan, Ann Arbor, MI 48109, USA	0000-0002-4876-956X\\
$^{32}$Department of Physics, University of Michigan, Ann Arbor, MI 48109, USA	0000-0002-4876-956X\\
$^{33}$Australian Astronomical Optics, Macquarie University, North Ryde, NSW 2113, Australia	0000-0003-0120-0808\\
$^{34}$Lowell Observatory, 1400 Mars Hill Rd, Flagstaff, AZ 86001, USA	0000-0003-0120-0808\\
$^{35}$Departamento de F\'isica Matem\'atica, Instituto de F\'isica, Universidade de S\~ao Paulo, CP 66318, S\~ao Paulo, SP, 05314-970, Brazil\\
$^{36}$George P. and Cynthia Woods Mitchell Institute for Fundamental Physics and Astronomy, and Department of Physics and Astronomy, Texas A\&M University, College Station, TX 77843,  USA	0000-0003-0710-9474\\
$^{37}$Department of Astrophysical Sciences, Princeton University, Peyton Hall, Princeton, NJ 08544, USA\\
$^{38}$Instituci\'o Catalana de Recerca i Estudis Avan\c{c}ats, E-08010 Barcelona, Spain	0000-0002-6610-4836\\
$^{39}$School of Physics and Astronomy, University of Southampton,  Southampton, SO17 1BJ, UK	0000-0002-3321-1432\\
$^{40}$Brandeis University, Physics Department, 415 South Street, Waltham MA 02453	0000-0001-6082-8529\\
$^{41}$Computer Science and Mathematics Division, Oak Ridge National Laboratory, Oak Ridge, TN 37831	0000-0002-7047-9358\\
$^{42}$Department of Physics, Duke University Durham, NC 27708, USA	\\
$^{43}$Institute for Astronomy, University of Edinburgh, Edinburgh EH9 3HJ, UK\\

\bsp	
\label{lastpage}
\end{document}